\def\ftmagnification{1200}
\def\spacingNumerator{5}
\def\spacingDenominator{4}

\def\ifundefined#1{\expandafter\ifx\csname#1\endcsname\relax}
\ifundefined{ftmagnification}  \def\ftmagnification{1200} \fi
\ifundefined{spacingNumerator}  \def\spacingNumerator{5} \fi
\ifundefined{spacingDenominator}  \def\spacingDenominator{4} \fi


\magnification\ftmagnification
\tolerance=10000
\hsize=17truecm\vsize=23truecm

\parindent=40pt
\mathsurround=0pt
     \multiply\baselineskip by \spacingNumerator
     \divide \baselineskip by \spacingDenominator 

%
%
\def\today{\ifcase\month\or January\or February\or March\or April\or
     May\or June\or July\or August\or September\or October\or November\or
     December\fi\space\number\day, \number\year}
%
%
\def\dst{\displaystyle}
\def\sst{\scriptstyle}
\def\tst{\textstyle}
%
%
\def\frac#1#2{\dst {#1\over#2}}     
\def\sfrac#1#2{{\tst{#1\over#2}}}   

\def\deqalign#1{\vcenter{\openup1\jot \mathsurround=0pt \ialign{
                \strut\hfil$\displaystyle{##}$&&$\displaystyle{{}##}$\hfil
                \crcr
                #1\crcr}}}         

\def\meqalign#1{\vcenter{\openup1\jot \mathsurround=0pt \ialign{
                &\strut\hfil$\displaystyle{##}$&$\displaystyle{{}##}$\hfil&
                \quad$##$\crcr
                #1\crcr}}}         

%
%
\def\al{\alpha}
\def\be{\beta}
\def\ga{\gamma}
\def\de{\delta}
\def\ep{\epsilon}
\def\ze{\zeta}
\def\et{\eta}

\def\ka{\kappa}
\def\la{\lambda}

\def\si{\sigma}

\def\up{\upsilon}

\def\om{\omega}

\def\La{\Lambda}
\def\Si{\Sigma}

\def\Om{\Omega}   
%
%
\def\pmb#1{\setbox0=\hbox{#1}       
     \kern-.025em\copy0\kern-\wd0
     \kern.05em\copy0\kern-\wd0
     \kern-.025em\box0}             
\def\0{{\bf 0}}

\def\k{{\bf k}}

\def\x{{\bf x}}

\def\cB{{\cal B}}
\def\cE{{\cal E}}
\def\cF{{\cal F}}
\def\cG{{\cal G}}

\def\cL{{\cal L}}
\def\cO{{\cal O}}

%
%
\font\tenfrak                 = eufm10
\font\sevenfrak               = eufm7
\font\fivefrak                = eufb5
\newfam\frakfam
     \textfont\frakfam=\tenfrak
     \scriptfont\frakfam=\sevenfrak   
     \scriptscriptfont\frakfam=\fivefrak
\def\frak{\fam\frakfam\tenfrak}
\font \tensans                = cmss10
\font \fivesans               = cmss10 at 5pt
\font \sevensans              = cmss10 at 7pt
\newfam\sansfam
     \textfont\sansfam=\tensans
     \scriptfont\sansfam=\sevensans
     \scriptscriptfont\sansfam=\fivesans
\def\sans{\fam\sansfam\tensans}
%
%
\def\bbbr{{\rm I\!R}}  
\def\bbbn{{\rm I\!N}}

\def\bbbc{{\mathchoice {\setbox0=\hbox{$\displaystyle\rm C$}\hbox{\hbox 
to0pt{\kern0.4\wd0\vrule height0.9\ht0\hss}\box0}}
{\setbox0=\hbox{$\textstyle\rm C$}\hbox{\hbox
to0pt{\kern0.4\wd0\vrule height0.9\ht0\hss}\box0}}
{\setbox0=\hbox{$\scriptstyle\rm C$}\hbox{\hbox
to0pt{\kern0.4\wd0\vrule height0.9\ht0\hss}\box0}}
{\setbox0=\hbox{$\scriptscriptstyle\rm C$}\hbox{\hbox
to0pt{\kern0.4\wd0\vrule height0.9\ht0\hss}\box0}}}}
\def\bbbq{{\mathchoice {\setbox0=\hbox{$\displaystyle\rm               
Q$}\hbox{\raise
0.15\ht0\hbox to0pt{\kern0.4\wd0\vrule height0.8\ht0\hss}\box0}}
{\setbox0=\hbox{$\textstyle\rm Q$}\hbox{\raise
0.15\ht0\hbox to0pt{\kern0.4\wd0\vrule height0.8\ht0\hss}\box0}}
{\setbox0=\hbox{$\scriptstyle\rm Q$}\hbox{\raise
0.15\ht0\hbox to0pt{\kern0.4\wd0\vrule height0.7\ht0\hss}\box0}}
{\setbox0=\hbox{$\scriptscriptstyle\rm Q$}\hbox{\raise
0.15\ht0\hbox to0pt{\kern0.4\wd0\vrule height0.7\ht0\hss}\box0}}}}
\def\bbbz{{\mathchoice {\hbox{$\sans\textstyle Z\kern-0.4em Z$}}       
{\hbox{$\sans\textstyle Z\kern-0.4em Z$}}
{\hbox{$\sans\scriptstyle Z\kern-0.3em Z$}}
{\hbox{$\sans\scriptscriptstyle Z\kern-0.2em Z$}}}}
%
%
\def\const{{\rm const}\,}

\def\half{\sfrac{1}{2}}

\def\optbar#1{\vbox{\ialign{##\crcr\hfil${\scriptscriptstyle(}\mkern -1mu
         \vrule height 1.2pt width 3pt depth -.8pt
         {\scriptscriptstyle)}$\hfil\crcr
          \noalign{\kern-1pt\nointerlineskip}$\hfil\displaystyle{#1}\hfil$\crcr}}}
\def\<{\left<}
\def\>{\right>}

\def\smprod{\mathop{\textstyle\prod}}
\def\smsum{\mathop{\textstyle\sum}}
\def\set#1#2{\big\{ \ #1\ \big|\ #2\ \big\}}
\def\eval#1{\big|\lower4pt\hbox{$\displaystyle\sst #1$}}
%
%
\font \tafontt                = cmbx10 scaled\magstep2
\font \tbfontt                = cmbx10 scaled\magstep1
\def\titlea#1{\centerline{\tafontt #1 }\vskip.5truein}
\def\titleb#1{\removelastskip\vskip.3truein%
\noindent{\tbfontt #1 }\vskip.25truein}

%
%
\def\newenvironment#1#2#3#4{\long\def#1##1##2{%
\removelastskip\penalty-100\vskip\baselineskip%
\noindent{#3#2\if!##1!.\else\unskip\ \ignorespaces
##1\unskip\fi\ }{#4\ignorespaces##2\vskip\baselineskip}}}
\newenvironment\lemma{Lemma}{\bf}{\it}
\newenvironment\proposition{Proposition}{\bf}{\it}
\newenvironment\theorem{Theorem}{\bf}{\it}
\newenvironment\corollary{Corollary}{\bf}{\it}
\newenvironment\example{Example}{\bf}{\rm}
\newenvironment\problem{Problem}{\bf}{\rm}
\newenvironment\definition{Definition}{\bf}{\rm}
\newenvironment\remark{Remark}{\bf}{\rm}
\newenvironment\hypothesis{Hypothesis}{\bf}{\it}
\newenvironment\convention{Convention}{\bf}{\it}

\def\Item{\vskip.1in\noindent}

%
%
\long\def\proof#1{\removelastskip\penalty-100\vskip\baselineskip\noindent{\bf
            Proof\if!#1!\else\ \ignorespaces#1\fi:\ }\ \ \ignorespaces}
\long\def\prf{\removelastskip\penalty-100\vskip\baselineskip\noindent{\bf
            Proof:\ }\ \ \ignorespaces}
\def\endproof{\hfill\vrule height .6em width .6em depth 0pt\goodbreak\vskip.25in }

\ifundefined{warnForwardRef}  \def\warnForwardRef{n} \fi
\newcount\chapno
\newcount\sectno
\newcount\equano
\newcount\theono
\newcount\probno

\def\IgNoRe#1{}

\chapno=0
\sectno=0
\equano=0
\theono=0
\probno=0
\def\eqhead{}
\def\frefwarning{\if\warnForwardRef y\immediate\write16{   Forward reference on line \the\inputlineno}\fi}
\def\qqqrefwarning{\immediate\write16{   ??? reference on line \the\inputlineno}}

\def\chap#1{\equano=0\sectno=0\theono=0\probno=0\global\advance\chapno by 1%
\def\eqhead{\ifcase\chapno\or I\or II\or III\or IV\or V\or VI\or VII\or
VIII\or IX\or X\or XI\or XII\or XIII\or XIV\or XV\or XVI\or XVII\or XVIII\or
XIX\or XX\or XXI\or XXII\or XXIII\or XXIV\or XXV\or XXVI\or XXVII\or XXVIII\or XXIX\or XXX\or XXXI\or XXXII\or XXXIII\or XXXIV\or XXXV\or XXXVI\or XXXVII\or XXXVIII\or XXXIX\fi.}%
\titlea{\eqhead \hglue 5pt #1}%
}

\def\sect#1{\global\advance\sectno by 1%
\titleb{\eqhead\number\sectno  \hglue 5pt #1}%
}%

\def\appendix#1#2{\equano=0\sectno=0\theono=0\probno=0\def\eqhead{#1.}
\titlea{Appendix #1: #2}%
}

\def\:#1{\def\temp{\expandafter\IgNoRe\string#1}%
\expandafter\ifx\csname\temp\endcsname\relax%
\expandafter\gdef#1{\qqqrefwarning ???}\fi#1}

\def\Eqn{{\hbox{\global\advance\equano by 1}}%
\eqno ({\rm \eqhead\number\equano})}%

\def\Eqno{{\hbox{\global\advance\equano by 1}}%
 ({\rm \eqhead\number\equano})}%

\def\EQN#1{\Eqn\edef\Zwi{\eqhead\number\equano}%
\global\let #1=\Zwi
}

\def\EQNO#1{\Eqno\edef\Zwi{\eqhead\number\equano}%
\global\let #1=\Zwi
}

\def\STM#1{{\global\advance \theono by 1}%
\eqhead\number\theono
\edef\Zwi{\eqhead\number\theono }
\global\let#1=\Zwi
}

\def\PRB#1{{\global\advance \probno by 1}%
\eqhead\number\probno
\edef\Zwi{\eqhead\number\probno }
\global\let#1=\Zwi
}

\def\PG#1{\def\Zwi{\number\pageno }
\global\let#1=\Zwi
}

\def\Stm{{\global\advance \theono by 1}%
\eqhead\number\theono
}

\def\Prb{{\global\advance \probno by 1}%
\eqhead\number\probno
}

\def\EDEF#1#2{
\def\tEmP{#1}\expandafter\gdef\tEmP{#2}
}



\def\suffix{ps}
\newcount\system
\global\system=3   

\def\ifundefined#1{\expandafter\ifx\csname#1\endcsname\relax}
\ifundefined{figdir}\def\figdir{}\fi
%
%
\newcount\firstline
\newdimen\pswidth  \newdimen\xleft
\newdimen\psheight \newdimen\ytop \newdimen\ybot
\newcount\justx \newcount\justy
\global\justx=0 \global\justy=0
\newdimen\vpos \newtoks\labeL 
\newread\labeLfile \newdimen\xcoord \newdimen\ycoord
\newif\ifdoit 
\newbox\labox
\newdimen\xdvikwid 
\newdimen\xdvikht
\newdimen\pspoints
\newdimen\rwi
\pspoints=1bp
\newcount\temp
\def\readdim#1{\global\read\labeLfile to \temp
\global #1=\temp pt}
%
%
%
%
\def\figcrop#1{\par
\openin\labeLfile=\figdir#1.lbl                                              
\global\read\labeLfile to\firstline\message{#1}               
\global\read\labeLfile to\temp
\readdim{\ybot}
\readdim{\xleft}
\readdim{\ytop}
\global\read\labeLfile to\justx
\global\read\labeLfile to\justy
\global\read\labeLfile to\labeL
\readdim{\pswidth}
\global\advance\pswidth by -\xleft
\readdim{\psheight}
\global\advance\ybot by -\psheight
\global\advance\psheight by -\ytop
\global\read\labeLfile to\justx
\global\read\labeLfile to\justy
\global\read\labeLfile to\labeL
\vbox to\psheight{\vfill
\ifnum\system=1
\fi 
\ifnum\system=2
\fi 
\ifnum\system=3
                                                 \fi         
\ifnum\system=4
\fi         
\ifnum\system=1
\hbox to \pswidth{\kern-\xleft\special{postscriptfile \figdir#1.\suffix }\hfil}\fi
\ifnum\system=2
\hbox to \pswidth{\kern-\xleft\special{ps: plotfile \figdir#1.\suffix }\hfil}\fi
\ifnum\system=3
\hbox to \pswidth{\kern-\xleft\includegraphics{\figdir#1.\suffix}\hfil}\fi
\ifnum\system=4
\hbox to \pswidth{\kern-\xleft\includegraphics{\figdir#1.\suffix}\hfil}\fi
\ifnum\system=5
\hbox to \pswidth{\kern-\xleft\includegraphics{\figdir#1.\suffix}\hfil}\fi 
\ifnum\system=6
   \xdvikwid=\pswidth
   \xdvikht=\psheight
   {\global\divide\xdvikwid by \pspoints}
   {\global\divide\xdvikht by \pspoints}
   \rwi=\xdvikwid
    {\global\multiply\rwi by 10}
\hbox to \pswidth{\kern-\xleft\includegraphics{\figdir#1.\suffix\space}\hfil}\fi                   
\vskip -\baselineskip
\vskip -\ybot 
\vskip-\psheight %
\hbox to\pswidth  {\hss}%
\parindent=0pt\offinterlineskip                                       
\vpos=0 pt%
\loop\readdim{\xcoord}                                 
\ifdim \xcoord < -999pt \doitfalse\else\doittrue\fi                        
\ifdoit \advance \xcoord by -\xleft
\readdim{\ycoord}
\advance \ycoord by -\ytop                              
\global\read\labeLfile to\justx                                       
\global\read\labeLfile to\justy                                       
\global\read\labeLfile to\labeL
\global\setbox\labox=\hbox{\labeL\hskip-0.3em}%
\advance\vpos by-\ycoord                                              
\vskip-\vpos \vpos=\ycoord                                         
\hbox to\pswidth{\hskip\xcoord %
\hbox to 0pt{\ifnum\justx>0\hss\fi%
\vbox to0pt{%
\ifnum\justy<2\vss\fi%
\copy\labox\kern0pt%
\ifnum\justy>0\vss\fi}%
\ifnum\justx<2\hss\fi}%
\hss}%
\repeat%
\advance\vpos by-\psheight%
\vskip-\vpos %
}\closein\labeLfile}
%
%
%
\def\figplace#1#2#3{
\openin\labeLfile=\figdir#1.lbl
\ifeof \labeLfile
       \immediate\write16{***Can't find \figdir#1.lbl; Skipping it.***}
\else  \closein\labeLfile
       \null\hskip#2\raise #3 \hbox{\figcrop{#1}}
\fi
}
%
%
%
%
\def\figput#1{
\openin\labeLfile=\figdir#1.lbl
\ifeof \labeLfile
       \immediate\write16{***Can't find \figdir#1.lbl; Skipping it.***}
\else  \closein\labeLfile
       \hbox{\figcrop{#1}}
\fi
}


\font\tenscript                 = pzcmi at 10pt
\font\sevenscript               = pzcmi at 7pt
\font\fivescript                = pzcmi at 5pt
\newfam\scriptfam
     \textfont\scriptfam=\tenscript
     \scriptfont\scriptfam=\sevenscript   
     \scriptscriptfont\scriptfam=\fivescript
\def\script{\fam\scriptfam\tenscript}

    \newenvironment\notation{Notation}{\bf}{\rm}

          \def\stoday{\number\day\space\ifcase\month\or Jan\or Feb\or 
                      Mar\or Apr\or May\or Jun\or Jul\or Aug\or Sep\or 
                      Oct\or Nov\or Dec\fi, \number\year}

         \def\squiggle{\raise2pt\hbox{${\scriptstyle\sim}$}}

    \def\form{{\ssst\script form}}
    \def\il{\jbar}
    \def\IR{{\rm\ssst IR}}
    \def\ren{{\rm ren}}
    \def\dunion{\cup\kern-0.7em\cdot\kern0.45em}
    \def\cb{{\frak c}}
    \def\ib{{\rm b}}
    
    \def\IB{{\rm\sst B}}
    \def\vi{\fl^{\raise2pt\hbox{$\scriptscriptstyle{1/n_0}$}}}

    \def\out{{\rm out}}
    \def\rg{{\rm rg}}


    \def\cst#1#2{{\rm const}^{#1}_{#2}\,}
    \def\abcst{{\sst const}}

     \def\veps{\varepsilon}
     \def\ssst{\scriptscriptstyle}
     \def\bde{{\mathchoice{\pmb{$\de$}}{\pmb{$\de$}}
                              {\pmb{$\sst\de$}}{\pmb{$\ssst\de$}}}}
    \def\jbar{{\mathchoice
                   {{\smash{\lower1ex\hbox{$\mathchar'26$}}\mkern-9mu j}}
                   {{\smash{\lower1ex\hbox{$\mathchar'26$}}\mkern-9mu j}}
                   {{\smash{\lower1.2ex\hbox{$\mathchar'26$}}\mkern-10.2mu j}}
                   {{\smash{\lower1.2ex\hbox{$\mathchar'26$}}\mkern-10.2mu j}}}}

    \def\cC{{\cal C}}
    \def\cD{{\cal D}}
    \def\cK{{\cal K}}
    \def\cL{{\cal L}}

    \def\cV{{\cal V}}
    \def\cW{{\cal W}}
    
    \def\fd{{\frak d}}
    \def\fe{{\frak e}}
    
    \def\fl{{\frak l}}
    
    \def\fN{{\frak N}}
    \def\fK{{\frak K}}

    \def\fX{{\frak X}}

    \def\rv{{\rm v}}

    \def\sQ{{\sst Q}}

    \def\sv{\pmb{$\sst\vert$}}
    \def\v{\pmb{$\vert$}}
    \def\V{\pmb{$\big\vert$}}
    \def\VV{\pmb{$\Big\vert$}}

     \def\tv{\kern8pt\tilde{\kern-8pt\pmb{$\vert$}}}
     \def\tV{\kern8pt\tilde{\kern-8pt\pmb{$\big\vert$}}}
     \def\tVV{\kern8pt\tilde{\kern-8pt\pmb{$\Big\vert$}}}

    \def\tn{|\kern-1pt|\kern-1pt|}
    \def\TN{\big|\kern-1.5pt\big|\kern-1.5pt\big|}
    \def\TTN{\Big|\kern-2pt\Big|\kern-2pt\Big|}

    \def\trn{|\kern-1pt|\kern-1pt|^{\,\tilde{\,}}}
    \def\TRN{\big|\kern-1.5pt\big|\kern-1.5pt\big|^{\,\tilde{\,}}}
    \def\TTRN{\Big|\kern-2pt\Big|\kern-2pt\Big|^{\,\tilde{\,}}}

     \def\tnorm{\kern8pt\tilde{\kern-8pt\|}}
     \def\Tnorm{\kern8pt\tilde{\kern-8pt\big\|}}
     \def\TNorm{\kern8pt\tilde{\kern-8pt\Big\|}}
     \def\TNOrm{\kern8pt\tilde{\kern-8pt\bigg\|}}

    \def\rw{\mathclose{:}}
    \def\lw{\mathopen{:}}
    \def\lW{\mathopen{{\tst{\hbox{.}\atop\raise 2.5pt\hbox{.}}}}}
    \def\rW{\mathclose{{\tst{{.}\atop\raise 2.5pt\hbox{.}}}}}
    \def\lww{\mathopen{{\tst{\raise 1pt\hbox{.}\atop\raise 1pt\hbox{.}}}}}
    \def\rww{\mathclose{{\tst{\raise 1pt\hbox{.}\atop\raise 1pt\hbox{.}}}}}

   \font\sixrm=cmr6   \font\eightrm=cmr8  
   \font\sixi=cmmi6   \font\eighti=cmmi8  
  \font\sixsy=cmsy6  \font\eightsy=cmsy8 
  \font\sixbf=cmbx6  \font\eightbf=cmbx8 
                     \font\eightit=cmti8 
                     \font\eightsl=cmsl8 
                     \font\eighttt=cmtt8 

\font\eightfrak=eufm7 at 8pt

\def\eightpoint{\def\rm{\fam0\eightrm}
 \textfont0=\eightrm \scriptfont0=\sixrm \scriptscriptfont0=\fiverm
 \textfont1=\eighti \scriptfont1=\sixi \scriptscriptfont1=\fivei
 \textfont2=\eightsy \scriptfont2=\sixsy \scriptscriptfont2=\fivesy
 \textfont3=\tenex \scriptfont3=\tenex \scriptscriptfont3=\tenex
 \textfont\itfam=\eightit \def\it{\fam\itfam\eightit}%
 \textfont\slfam=\eightsl \def\sl{\fam\slfam\eightsl}%
 \textfont\ttfam=\eighttt \def\tt{\fam\ttfam\eighttt}%
 \textfont\frakfam=\eightfrak \def\frak{\fam\frakfam\tenfrak}%
 \textfont\bffam=\eightbf \scriptfont\bffam=\sixbf
 \scriptscriptfont\bffam=\fivebf \def\bf{\fam\bffam\eightbf}%
 \normalbaselineskip=9pt
 \setbox\strutbox=\hbox{\vrule height7pt depth2pt width0pt}%
 \let\sc=\sixrm \let\big=\eightbig \normalbaselines\rm}
\catcode`@=11
\def\footnote#1{\edef\@sf{\spacefactor\the\spacefactor}#1\@sf
     \insert\footins\bgroup\eightpoint
     \interlinepenalty100 \let\par=\endgraf
     \leftskip=0pt \rightskip=0pt
     \splittopskip=10pt plus 1pt minus 1pt \floatingpenalty=20000
     \smallskip\item{#1}\bgroup\strut\aftergroup\@foot\let\next}
\skip\footins=12pt plus 2pt minus 4pt
\dimen\footins=30pc
\catcode`@=12


  \IgNoRe{PG}
  \IgNoRe{STM Assertion }
  \IgNoRe{PG}
  \IgNoRe{PG}
  \IgNoRe{STM Assertion }
  \IgNoRe{PG}
  \IgNoRe{STM Assertion }
  \IgNoRe{STM Assertion }
  \IgNoRe{EQN}
  \IgNoRe{STM Assertion }
  \IgNoRe{STM Assertion }
  \IgNoRe{PG}
  \IgNoRe{STM Assertion }
  \IgNoRe{STM Assertion }
  \IgNoRe{EQN}
  \IgNoRe{STM Assertion }
  \IgNoRe{STM Assertion }
  \IgNoRe{STM Assertion }
  \IgNoRe{STM Assertion }
  \IgNoRe{STM Assertion }
  \IgNoRe{STM Assertion }
  \IgNoRe{PG}
  \IgNoRe{STM Assertion }
  \IgNoRe{STM Assertion }
  \IgNoRe{STM Assertion }
  \IgNoRe{STM Assertion }
  \IgNoRe{STM Assertion }
  \IgNoRe{STM Assertion }
  \IgNoRe{STM Assertion }
  \IgNoRe{STM Assertion }
  \IgNoRe{STM Assertion }
  \IgNoRe{STM Assertion }
  \IgNoRe{STM Assertion }
  \IgNoRe{STM Assertion }
  \IgNoRe{PG}
  \IgNoRe{EQN}
  \IgNoRe{STM Assertion }
  \IgNoRe{STM Assertion }
  \IgNoRe{STM Assertion }
  \IgNoRe{PG}
 \def\scaleIntConsts{\frefwarning II.30} \IgNoRe{STM Assertion }
  \IgNoRe{STM Assertion }
 \def\corwicknorm{\frefwarning II.32} \IgNoRe{STM Assertion }
  \IgNoRe{STM Assertion }
  \IgNoRe{EQN}
  \IgNoRe{EQN}
  \IgNoRe{STM Assertion }
  \IgNoRe{PG}
  \IgNoRe{PG}
  \IgNoRe{STM Assertion }
  \IgNoRe{EQN}
  \IgNoRe{STM Assertion }
  \IgNoRe{STM Assertion }
  \IgNoRe{STM Assertion }
  \IgNoRe{EQN}
  \IgNoRe{STM Assertion }
  \IgNoRe{STM Assertion }
  \IgNoRe{STM Assertion }
  \IgNoRe{PG}
  \IgNoRe{STM Assertion }
  \IgNoRe{STM Assertion }
  \IgNoRe{PG}
  \IgNoRe{STM Assertion }
  \IgNoRe{STM Assertion }
  \IgNoRe{STM Assertion }
  \IgNoRe{PG}
  \IgNoRe{STM Assertion }
  \IgNoRe{STM Assertion }
  \IgNoRe{STM Assertion }
  \IgNoRe{STM Assertion }
  \IgNoRe{STM Assertion }
  \IgNoRe{STM Assertion }
  \IgNoRe{STM Assertion }
 \def\propBII{\frefwarning A.2} \IgNoRe{STM Assertion }
  \IgNoRe{PG}
  \IgNoRe{STM Assertion }
  \IgNoRe{STM Assertion }
  \IgNoRe{STM Assertion }
  \IgNoRe{STM Assertion }
  \IgNoRe{STM Assertion }
  \IgNoRe{STM Assertion }
  \IgNoRe{STM Assertion }
  \IgNoRe{STM Assertion }
  \IgNoRe{PG}
  \IgNoRe{STM Assertion }
  \IgNoRe{STM Assertion }
  \IgNoRe{PG}
  \IgNoRe{PG}
  \IgNoRe{STM Assertion }
  \IgNoRe{STM Assertion }
  \IgNoRe{PG}
  \IgNoRe{PG}
  \IgNoRe{STM Assertion }
  \IgNoRe{STM Assertion }
  \IgNoRe{EQN}
  \IgNoRe{STM Assertion }
  \IgNoRe{PG}
  \IgNoRe{STM Assertion }
  \IgNoRe{STM Assertion }
  \IgNoRe{STM Assertion }
  \IgNoRe{PG}
  \IgNoRe{STM Assertion }
  \IgNoRe{STM Assertion }
  \IgNoRe{STM Assertion }
  \IgNoRe{EQN}
  \IgNoRe{STM Assertion }
  \IgNoRe{PG}
  \IgNoRe{EQN}
  \IgNoRe{STM Assertion }
  \IgNoRe{STM Assertion }
  \IgNoRe{STM Assertion }
  \IgNoRe{STM Assertion }
  \IgNoRe{EQN}
  \IgNoRe{EQN}
  \IgNoRe{PG}
  \IgNoRe{PG}
  \IgNoRe{STM Assertion }
  \IgNoRe{STM Assertion }
  \IgNoRe{STM Assertion }
  \IgNoRe{STM Assertion }
  \IgNoRe{STM Assertion }
  \IgNoRe{STM Assertion }
  \IgNoRe{STM Assertion }
  \IgNoRe{STM Assertion }
  \IgNoRe{PG}
  \IgNoRe{STM Assertion }
  \IgNoRe{STM Assertion }
  \IgNoRe{STM Assertion }
  \IgNoRe{STM Assertion }
  \IgNoRe{STM Assertion }
  \IgNoRe{STM Assertion }
  \IgNoRe{STM Assertion }
  \IgNoRe{STM Assertion }
  \IgNoRe{STM Assertion }
  \IgNoRe{STM Assertion }
  \IgNoRe{PG}
  \IgNoRe{STM Assertion }
  \IgNoRe{STM Assertion }
  \IgNoRe{STM Assertion }
  \IgNoRe{PG}
  \IgNoRe{STM Assertion }
  \IgNoRe{STM Assertion }
  \IgNoRe{STM Assertion }
  \IgNoRe{STM Assertion }
  \IgNoRe{PG}
  \IgNoRe{STM Assertion }
  \IgNoRe{STM Assertion }
  \IgNoRe{PG}
  \IgNoRe{STM Assertion }
  \IgNoRe{PG}
  \IgNoRe{STM Assertion }
  \IgNoRe{PG}
  \IgNoRe{STM Assertion }
  \IgNoRe{STM Assertion }
  \IgNoRe{STM Assertion }
  \IgNoRe{STM Assertion }
  \IgNoRe{PG}
  \IgNoRe{PG}
  \IgNoRe{STM Assertion }
  \IgNoRe{STM Assertion }
  \IgNoRe{EQN}
  \IgNoRe{EQN}
  \IgNoRe{STM Assertion }
  \IgNoRe{STM Assertion }
  \IgNoRe{STM Assertion }
  \IgNoRe{STM Assertion }
  \IgNoRe{PG}
  \IgNoRe{STM Assertion }
  \IgNoRe{STM Assertion }
  \IgNoRe{STM Assertion }
  \IgNoRe{STM Assertion }
  \IgNoRe{STM Assertion }
  \IgNoRe{STM Assertion }
  \IgNoRe{STM Assertion }
  \IgNoRe{PG}
  \IgNoRe{STM Assertion }
  \IgNoRe{EQN}
  \IgNoRe{EQN}
  \IgNoRe{PG}
  \IgNoRe{STM Assertion }
  \IgNoRe{EQN}
  \IgNoRe{STM Assertion }
  \IgNoRe{STM Assertion }
  \IgNoRe{STM Assertion }
  \IgNoRe{PG}
  \IgNoRe{STM Assertion }
  \IgNoRe{EQN}
  \IgNoRe{STM Assertion }
  \IgNoRe{PG}
  \IgNoRe{PG}


  \IgNoRe{PG}
  \IgNoRe{PG}
  \IgNoRe{STM Assertion }
  \IgNoRe{EQN}
  \IgNoRe{STM Assertion }
  \IgNoRe{PG}
  \IgNoRe{STM Assertion }
  \IgNoRe{EQN}
  \IgNoRe{STM Assertion }
  \IgNoRe{EQN}
  \IgNoRe{STM Assertion }
  \IgNoRe{STM Assertion }
  \IgNoRe{STM Assertion }
  \IgNoRe{STM Assertion }
  \IgNoRe{PG}
  \IgNoRe{STM Assertion }
  \IgNoRe{STM Assertion }
  \IgNoRe{STM Assertion }
  \IgNoRe{STM Assertion }
  \IgNoRe{PG}
  \IgNoRe{STM Assertion }
  \IgNoRe{STM Assertion }
  \IgNoRe{EQN}
  \IgNoRe{STM Assertion }
  \IgNoRe{STM Assertion }
  \IgNoRe{STM Assertion }
  \IgNoRe{PG}
  \IgNoRe{PG}
  \IgNoRe{STM Assertion }
  \IgNoRe{EQN}
  \IgNoRe{STM Assertion }
  \IgNoRe{STM Assertion }
  \IgNoRe{PG}
  \IgNoRe{STM Assertion }
  \IgNoRe{STM Assertion }

  \IgNoRe{STM Assertion }
  \IgNoRe{STM Assertion }
  \IgNoRe{PG}
  \IgNoRe{PG}
  \IgNoRe{STM Assertion }
  \IgNoRe{STM Assertion }
  \IgNoRe{STM Assertion }
  \IgNoRe{STM Assertion }
  \IgNoRe{STM Assertion }
  \IgNoRe{STM Assertion }
  \IgNoRe{PG}
  \IgNoRe{STM Assertion }
  \IgNoRe{STM Assertion }
  \IgNoRe{STM Assertion }
  \IgNoRe{STM Assertion }
  \IgNoRe{STM Assertion }
  \IgNoRe{STM Assertion }
  \IgNoRe{EQN}
  \IgNoRe{PG}
  \IgNoRe{STM Assertion }
  \IgNoRe{STM Assertion }
  \IgNoRe{STM Assertion }
  \IgNoRe{EQN}
  \IgNoRe{STM Assertion }
  \IgNoRe{STM Assertion }
  \IgNoRe{STM Assertion }
  \IgNoRe{PG}
  \IgNoRe{STM Assertion }
  \IgNoRe{STM Assertion }
  \IgNoRe{STM Assertion }
  \IgNoRe{STM Assertion }
  \IgNoRe{EQN}
  \IgNoRe{EQN}
  \IgNoRe{EQN}
  \IgNoRe{EQN}
  \IgNoRe{STM Assertion }
  \IgNoRe{STM Assertion }
  \IgNoRe{EQN}
  \IgNoRe{STM Assertion }
  \IgNoRe{EQN}
  \IgNoRe{EQN}
  \IgNoRe{EQN}
  \IgNoRe{EQN}
  \IgNoRe{EQN}
  \IgNoRe{PG}
  \IgNoRe{STM Assertion }
  \IgNoRe{STM Assertion }
  \IgNoRe{EQN}

  \IgNoRe{EQN}
  \IgNoRe{EQN}
  \IgNoRe{PG}
  \IgNoRe{EQN}
  \IgNoRe{EQN}
  \IgNoRe{STM Assertion }
  \IgNoRe{STM Assertion }
  \IgNoRe{EQN}
  \IgNoRe{EQN}
  \IgNoRe{EQN}
  \IgNoRe{EQN}
  \IgNoRe{STM Assertion }
  \IgNoRe{STM Assertion }
  \IgNoRe{STM Assertion }
  \IgNoRe{STM Assertion }
  \IgNoRe{STM Assertion }
  \IgNoRe{STM Assertion }
  \IgNoRe{STM Assertion }
  \IgNoRe{STM Assertion }
  \IgNoRe{STM Assertion }
  \IgNoRe{STM Assertion }
  \IgNoRe{STM Assertion }
  \IgNoRe{STM Assertion }
  \IgNoRe{EQN}
  \IgNoRe{EQN}
  \IgNoRe{EQN}
  \IgNoRe{EQN}
  \IgNoRe{STM Assertion }
  \IgNoRe{EQN}
  \IgNoRe{STM Assertion }
  \IgNoRe{EQN}
  \IgNoRe{EQN}
  \IgNoRe{EQN}
  \IgNoRe{EQN}
  \IgNoRe{EQN}
  \IgNoRe{STM Assertion }
  \IgNoRe{STM Assertion }
  \IgNoRe{STM Assertion }
  \IgNoRe{STM Assertion }
  \IgNoRe{STM Assertion }
  \IgNoRe{EQN}
  \IgNoRe{EQN}
  \IgNoRe{STM Assertion }
  \IgNoRe{STM Assertion }
  \IgNoRe{EQN}
  \IgNoRe{EQN}
  \IgNoRe{STM Assertion }
  \IgNoRe{EQN}
  \IgNoRe{STM Assertion }
  \IgNoRe{STM Assertion }
  \IgNoRe{EQN}
  \IgNoRe{STM Assertion }
  \IgNoRe{EQN}
  \IgNoRe{EQN}
  \IgNoRe{EQN}
  \IgNoRe{EQN}
  \IgNoRe{STM Assertion }
  \IgNoRe{PG}
  \IgNoRe{STM Assertion }
  \IgNoRe{STM Assertion }
  \IgNoRe{PG}
  \IgNoRe{STM Assertion }

  \IgNoRe{PG}
  \IgNoRe{STM Assertion }
  \IgNoRe{EQN}
  \IgNoRe{EQN}
  \IgNoRe{EQN}
  \IgNoRe{STM Assertion }
  \IgNoRe{STM Assertion }
  \IgNoRe{STM Assertion }
  \IgNoRe{STM Assertion }
  \IgNoRe{EQN}
  \IgNoRe{STM Assertion }
  \IgNoRe{EQN}
  \IgNoRe{EQN}
  \IgNoRe{EQN}
  \IgNoRe{STM Assertion }
  \IgNoRe{STM Assertion }
  \IgNoRe{STM Assertion }
  \IgNoRe{EQN}

  \IgNoRe{STM Assertion }
  \IgNoRe{PG}
  \IgNoRe{STM Assertion }
  \IgNoRe{EQN}
  \IgNoRe{EQN}
  \IgNoRe{STM Assertion }
  \IgNoRe{EQN}

  \IgNoRe{STM Assertion }
  \IgNoRe{EQN}
  \IgNoRe{EQN}
  \IgNoRe{PG}
  \IgNoRe{EQN}
  \IgNoRe{EQN}
  \IgNoRe{EQN}
  \IgNoRe{EQN}
  \IgNoRe{EQN}
  \IgNoRe{STM Assertion }
  \IgNoRe{EQN}
  \IgNoRe{EQN}
  \IgNoRe{EQN}
  \IgNoRe{EQN}
  \IgNoRe{EQN}
  \IgNoRe{STM Assertion }
  \IgNoRe{EQN}
  \IgNoRe{EQN}
  \IgNoRe{EQN}
  \IgNoRe{EQN}
  \IgNoRe{EQN}
  \IgNoRe{EQN}
  \IgNoRe{EQN}
  \IgNoRe{EQN}
  \IgNoRe{STM Assertion }

  \IgNoRe{STM Assertion }
  \IgNoRe{PG}
  \IgNoRe{EQN}
  \IgNoRe{EQN}
  \IgNoRe{STM Assertion }
  \IgNoRe{EQN}
  \IgNoRe{EQN}
  \IgNoRe{STM Assertion }
  \IgNoRe{EQN}
  \IgNoRe{STM Assertion }
  \IgNoRe{EQN}
  \IgNoRe{EQN}
  \IgNoRe{PG}
  \IgNoRe{PG}


  \IgNoRe{PG}
  \IgNoRe{EQN}
  \IgNoRe{STM Assertion }
  \IgNoRe{PG}
  \IgNoRe{STM Assertion }
  \IgNoRe{STM Assertion }
  \IgNoRe{STM Assertion }
  \IgNoRe{STM Assertion }
  \IgNoRe{STM Assertion }
  \IgNoRe{STM Assertion }
  \IgNoRe{EQN}
  \IgNoRe{STM Assertion }
  \IgNoRe{STM Assertion }
  \IgNoRe{STM Assertion }
  \IgNoRe{STM Assertion }
  \IgNoRe{STM Assertion }
  \IgNoRe{STM Assertion }
  \IgNoRe{STM Assertion }
  \IgNoRe{STM Assertion }
  \IgNoRe{PG}
  \IgNoRe{STM Assertion }
  \IgNoRe{STM Assertion }
  \IgNoRe{STM Assertion }
  \IgNoRe{STM Assertion }
  \IgNoRe{STM Assertion }
  \IgNoRe{STM Assertion }
  \IgNoRe{STM Assertion }
  \IgNoRe{STM Assertion }
  \IgNoRe{EQN}
  \IgNoRe{PG}
  \IgNoRe{PG}
  \IgNoRe{STM Assertion }
  \IgNoRe{STM Assertion }
  \IgNoRe{STM Assertion }
  \IgNoRe{STM Assertion }
  \IgNoRe{STM Assertion }
  \IgNoRe{STM Assertion }
  \IgNoRe{PG}
  \IgNoRe{EQN}
  \IgNoRe{EQN}
  \IgNoRe{EQN}
  \IgNoRe{EQN}
  \IgNoRe{EQN}
  \IgNoRe{EQN}
  \IgNoRe{STM Assertion }
  \IgNoRe{STM Assertion }
  \IgNoRe{STM Assertion }
  \IgNoRe{STM Assertion }
  \IgNoRe{PG}
  \IgNoRe{STM Assertion }
  \IgNoRe{EQN}
  \IgNoRe{STM Assertion }
  \IgNoRe{STM Assertion }
  \IgNoRe{STM Assertion }
 \def\exOSappMonoidI{\frefwarning A.3} \IgNoRe{STM Assertion }
  \IgNoRe{PG}
 \def\lemOSappMonoidIV{\frefwarning A.4} \IgNoRe{STM Assertion }
 \def\corOSappMonoidIV{\frefwarning A.5} \IgNoRe{STM Assertion }
  \IgNoRe{STM Assertion }
  \IgNoRe{STM Assertion }
  \IgNoRe{PG}
  \IgNoRe{EQN}
  \IgNoRe{EQN}
  \IgNoRe{PG}
  \IgNoRe{STM Assertion }
  \IgNoRe{STM Assertion }
  \IgNoRe{STM Assertion }
  \IgNoRe{EQN}
  \IgNoRe{PG}
  \IgNoRe{STM Assertion }
  \IgNoRe{STM Assertion }
  \IgNoRe{EQN}
  \IgNoRe{STM Assertion }
  \IgNoRe{STM Assertion }
  \IgNoRe{STM Assertion }
 \def\defOSscales{\frefwarning VIII.1} \IgNoRe{STM Assertion }
  \IgNoRe{PG}
  \IgNoRe{STM Assertion }
  \IgNoRe{STM Assertion }
  \IgNoRe{STM Assertion }
  \IgNoRe{STM Assertion }
 \def\thmOSfirststep{\frefwarning VIII.6} \IgNoRe{STM Assertion }
 \def\remOSthmV{\frefwarning VIII.7} \IgNoRe{STM Assertion }
  \IgNoRe{STM Assertion }
  \IgNoRe{STM Assertion }
  \IgNoRe{PG}
  \IgNoRe{STM Assertion }
 \def\defOSftcov{\frefwarning IX.3} \IgNoRe{STM Assertion }
  \IgNoRe{STM Assertion }
  \IgNoRe{STM Assertion }
  \IgNoRe{STM Assertion }
  \IgNoRe{EQN}
  \IgNoRe{STM Assertion }
  \IgNoRe{STM Assertion }
  \IgNoRe{PG}
  \IgNoRe{STM Assertion }
  \IgNoRe{STM Assertion }
  \IgNoRe{STM Assertion }
  \IgNoRe{STM Assertion }
  \IgNoRe{STM Assertion }
  \IgNoRe{STM Assertion }
  \IgNoRe{STM Assertion }
  \IgNoRe{STM Assertion }
  \IgNoRe{STM Assertion }
  \IgNoRe{EQN}
  \IgNoRe{STM Assertion }
 \def\defOSsymmetries{\frefwarning B.1} \IgNoRe{STM Assertion }
  \IgNoRe{PG}
  \IgNoRe{STM Assertion }
  \IgNoRe{STM Assertion }
  \IgNoRe{STM Assertion }
 \def\remOSrengrppreserves{\frefwarning B.5} \IgNoRe{STM Assertion }
  \IgNoRe{STM Assertion }
  \IgNoRe{STM Assertion }
  \IgNoRe{STM Assertion }
  \IgNoRe{PG}
  \IgNoRe{STM Assertion }
  \IgNoRe{PG}
  \IgNoRe{PG}
  \IgNoRe{STM Assertion }
  \IgNoRe{STM Assertion }
  \IgNoRe{PG}
 \def\lemOSsectpartunit{\frefwarning XII.3} \IgNoRe{STM Assertion }
 \def\defOSsectrepr{\frefwarning XII.4} \IgNoRe{STM Assertion }
  \IgNoRe{STM Assertion }
 \def\defOStens{\frefwarning XII.6} \IgNoRe{STM Assertion }
  \IgNoRe{STM Assertion }
 \def\propOSfunctorialitySect{\frefwarning XII.8} \IgNoRe{STM Assertion }
  \IgNoRe{STM Assertion }
  \IgNoRe{STM Assertion }
 \def\remOSdiffnorm{\frefwarning XII.11} \IgNoRe{STM Assertion }
 \def\lemOSNormMom{\frefwarning XII.12} \IgNoRe{STM Assertion }
  \IgNoRe{STM Assertion }
  \IgNoRe{STM Assertion }
  \IgNoRe{STM Assertion }
 \def\propOScontrintboundsectors{\frefwarning XII.16} \IgNoRe{STM Assertion }
  \IgNoRe{STM Assertion }
  \IgNoRe{STM Assertion }
  \IgNoRe{EQN}
  \IgNoRe{STM Assertion }
  \IgNoRe{STM Assertion }
  \IgNoRe{PG}
  \IgNoRe{STM Assertion }
  \IgNoRe{EQN}
  \IgNoRe{EQN}
  \IgNoRe{STM Assertion }
  \IgNoRe{EQN}
  \IgNoRe{EQN}
  \IgNoRe{STM Assertion }
  \IgNoRe{EQN}
  \IgNoRe{STM Assertion }
  \IgNoRe{EQN}
 \def\lemOSdiffpropbound{\frefwarning XIII.6} \IgNoRe{STM Assertion }
 \def\lemOSumu{\frefwarning XIII.7} \IgNoRe{STM Assertion }
  \IgNoRe{STM Assertion }
  \IgNoRe{STM Assertion }
  \IgNoRe{PG}
  \IgNoRe{STM Assertion }
  \IgNoRe{STM Assertion }
  \IgNoRe{STM Assertion }
  \IgNoRe{STM Assertion }
  \IgNoRe{EQN}
 \def\eqnOSrhomn{\frefwarning XV.1} \IgNoRe{EQN}
 \def\defOSscalednorms{\frefwarning XV.1} \IgNoRe{STM Assertion }
 \def\remOSscalednorms{\frefwarning XV.2} \IgNoRe{STM Assertion }
  \IgNoRe{PG}
 \def\thOSrengroupestimate{\frefwarning XV.3} \IgNoRe{STM Assertion }
  \IgNoRe{STM Assertion }
 \def\lemOSconcreteintconst{\frefwarning XV.5} \IgNoRe{STM Assertion }
  \IgNoRe{EQN}
  \IgNoRe{STM Assertion }
  \IgNoRe{EQN}
 \def\thOSrengroupdiffestimate{\frefwarning XV.7} \IgNoRe{STM Assertion }
  \IgNoRe{STM Assertion }
  \IgNoRe{EQN}
  \IgNoRe{STM Assertion }
  \IgNoRe{EQN}
  \IgNoRe{EQN}
  \IgNoRe{EQN}
  \IgNoRe{EQN}
 \def\propOSresidualrengroupest{\frefwarning XV.10} \IgNoRe{STM Assertion }
 \def\remOSchoiceofrep{\frefwarning XV.11} \IgNoRe{STM Assertion }
  \IgNoRe{STM Assertion }
  \IgNoRe{STM Assertion }
  \IgNoRe{PG}
  \IgNoRe{STM Assertion }
  \IgNoRe{STM Assertion }
  \IgNoRe{STM Assertion }
  \IgNoRe{STM Assertion }
  \IgNoRe{STM Assertion }
  \IgNoRe{STM Assertion }
  \IgNoRe{STM Assertion }
  \IgNoRe{STM Assertion }
  \IgNoRe{STM Assertion }
  \IgNoRe{STM Assertion }
  \IgNoRe{EQN}
  \IgNoRe{EQN}
  \IgNoRe{STM Assertion }
  \IgNoRe{PG}
  \IgNoRe{STM Assertion }
  \IgNoRe{STM Assertion }
  \IgNoRe{EQN}
  \IgNoRe{STM Assertion }
  \IgNoRe{STM Assertion }
  \IgNoRe{EQN}
  \IgNoRe{EQN}
  \IgNoRe{EQN}
  \IgNoRe{EQN}
  \IgNoRe{EQN}
  \IgNoRe{STM Assertion }
  \IgNoRe{STM Assertion }
  \IgNoRe{STM Assertion }
  \IgNoRe{EQN}
  \IgNoRe{EQN}
  \IgNoRe{EQN}
  \IgNoRe{EQN}
  \IgNoRe{EQN}
  \IgNoRe{EQN}
  \IgNoRe{STM Assertion }
  \IgNoRe{STM Assertion }
  \IgNoRe{STM Assertion }
  \IgNoRe{PG}
  \IgNoRe{STM Assertion }
  \IgNoRe{STM Assertion }
  \IgNoRe{EQN}
  \IgNoRe{EQN}
  \IgNoRe{STM Assertion }
  \IgNoRe{EQN}
  \IgNoRe{EQN}
  \IgNoRe{STM Assertion }
  \IgNoRe{EQN}
  \IgNoRe{EQN}
  \IgNoRe{STM Assertion }
  \IgNoRe{STM Assertion }
  \IgNoRe{PG}
  \IgNoRe{STM Assertion }
  \IgNoRe{STM Assertion }
  \IgNoRe{STM Assertion }
  \IgNoRe{PG}
  \IgNoRe{STM Assertion }
 \def\propOSthreetoonenorm{\frefwarning XIX.1} \IgNoRe{STM Assertion }
  \IgNoRe{STM Assertion }
 \def\remOSresector{\frefwarning XIX.3} \IgNoRe{STM Assertion }
  \IgNoRe{PG}
 \def\propOSresectorI{\frefwarning XIX.4} \IgNoRe{STM Assertion }
 \def\remOStoresectorI{\frefwarning XIX.5} \IgNoRe{STM Assertion }
 \def\defOSresectorII{\frefwarning XIX.6} \IgNoRe{STM Assertion }
  \IgNoRe{STM Assertion }
 \def\corOSirrelevantresect{\frefwarning XIX.8} \IgNoRe{STM Assertion }
  \IgNoRe{STM Assertion }
  \IgNoRe{STM Assertion }
  \IgNoRe{STM Assertion }
 \def\corOSIntUp{\frefwarning XIX.12} \IgNoRe{STM Assertion }
 \def\corOSresectorvanishkzero{\frefwarning XIX.13} \IgNoRe{STM Assertion }
 \def\defOScreateSectoriz{\frefwarning XIX.14} \IgNoRe{STM Assertion }
 \def\propOScreateSectoriz{\frefwarning XIX.15} \IgNoRe{STM Assertion }
  \IgNoRe{STM Assertion }
  \IgNoRe{STM Assertion }
  \IgNoRe{PG}
  \IgNoRe{STM Assertion }
  \IgNoRe{PG}
  \IgNoRe{STM Assertion }
  \IgNoRe{STM Assertion }
  \IgNoRe{PG}
  \IgNoRe{STM Assertion }
  \IgNoRe{STM Assertion }
  \IgNoRe{EQN}
  \IgNoRe{PG}
  \IgNoRe{EQN}
  \IgNoRe{STM Assertion }
  \IgNoRe{STM Assertion }
  \IgNoRe{PG}
  \IgNoRe{EQN}
  \IgNoRe{EQN}
  \IgNoRe{EQN}
  \IgNoRe{EQN}
  \IgNoRe{EQN}
  \IgNoRe{STM Assertion }
  \IgNoRe{STM Assertion }
  \IgNoRe{EQN}
  \IgNoRe{STM Assertion }
  \IgNoRe{PG}
  \IgNoRe{PG}
  \IgNoRe{PG}
  \IgNoRe{STM Assertion }
  \IgNoRe{EQN}
  \IgNoRe{STM Assertion }
  \IgNoRe{PG}
  \IgNoRe{STM Assertion }
  \IgNoRe{EQN}
  \IgNoRe{STM Assertion }
  \IgNoRe{STM Assertion }
  \IgNoRe{PG}
  \IgNoRe{EQN}
  \IgNoRe{EQN}
  \IgNoRe{EQN}
  \IgNoRe{STM Assertion }
  \IgNoRe{STM Assertion }
  \IgNoRe{STM Assertion }
  \IgNoRe{EQN}
  \IgNoRe{STM Assertion }
  \IgNoRe{STM Assertion }
 \def\theoremOSLadA{\frefwarning XXII.8} \IgNoRe{STM Assertion }
 \def\defOSzeroext{\frefwarning E.1} \IgNoRe{STM Assertion }
  \IgNoRe{STM Assertion }
 \def\defOSzerosectorext{\frefwarning E.3} \IgNoRe{STM Assertion }
  \IgNoRe{PG}
 \def\remOSzerosectorext{\frefwarning E.4} \IgNoRe{STM Assertion }
 \def\lemOSsectorext{\frefwarning E.5} \IgNoRe{STM Assertion }
  \IgNoRe{STM Assertion }
 \def\defOSindresector{\frefwarning E.7} \IgNoRe{STM Assertion }
  \IgNoRe{STM Assertion }
  \IgNoRe{STM Assertion }
 \def\propOSindresectorI{\frefwarning E.10} \IgNoRe{STM Assertion }
  \IgNoRe{PG}
  \IgNoRe{PG}


\newcount\CHAPNO
\newcount\APPNO
\CHAPNO=0
\APPNO=1
\def\advCHAPNO{\advance\CHAPNO by 1}
\def\advAPPNO{\advance\APPNO by 1}

\def\caproman#1{\ifcase#1\or I\or II\or III\or IV\or V\or VI\or VII\or
VIII\or IX\or X\or XI\or XII\or XIII\or XIV\or XV\or XVI\or XVII\or XVIII\or
XIX\or XX\or XXI\or XXII\or XXIII\or XXIV\or XXV\or XXVI\or XXVII\or XXVIII\or XXIX\or XXX\or XXXI\or XXXII\or XXXIII\or XXXIV\or XXXV\or XXXVI\or XXXVII\or XXXVIII\or XXXIX\fi}%

\def\capletter#1{\ifcase#1\or A\or B\or C\or D\or E\or F\or G\or
H\or I\or J\or K\or L\or M\or N\or O\or P\or Q\or R\or
S\or T\or U\or V\or W\or X\or Y\or Z\fi}%

\newcount\cHintroI \cHintroI=\CHAPNO \advCHAPNO 
                              \edef\CHintroI{\caproman\CHAPNO}
\newcount\cHnorms  \cHnorms=\CHAPNO \advCHAPNO 
                              \edef\CHnorms{\caproman\CHAPNO}
\newcount\cHproprengrp \cHproprengrp=\CHAPNO \advCHAPNO 
                              
\newcount\cHcovbounds  \cHcovbounds=\CHAPNO \advCHAPNO 
                              
\newcount\cHinsulator \cHinsulator=\CHAPNO \advCHAPNO

 \advAPPNO

\newcount\cHintroII \cHintroII=\CHAPNO \advCHAPNO 
                              \edef\CHintroII{\caproman\CHAPNO}
\newcount\cHamputate \cHamputate=\CHAPNO \advCHAPNO
                              
\newcount\cHscales \cHscales=\CHAPNO \advCHAPNO
                              
\newcount\cHfourier \cHfourier=\CHAPNO \advCHAPNO
                              
\newcount\cHmomentum \cHmomentum=\CHAPNO \advCHAPNO

 \advAPPNO
 \advAPPNO

\newcount\cHintroIII \cHintroIII=\CHAPNO \advCHAPNO
                              \edef\CHintroIII{\caproman\CHAPNO}
\newcount\cHsectors \cHsectors=\CHAPNO \advCHAPNO
                              \edef\CHsectors{\caproman\CHAPNO}
\newcount\cHsecpropbounds \cHsecpropbounds=\CHAPNO \advCHAPNO
                              
\newcount\cHladdersNotn  \cHladdersNotn=\CHAPNO \advCHAPNO
                              \edef\CHladdersNotn{\caproman\CHAPNO}
\newcount\cHestren  \cHestren=\CHAPNO \advCHAPNO
                              
\newcount\cHsecmomnorm \cHsecmomnorm=\CHAPNO \advCHAPNO
                              
\newcount\cHmomestren \cHmomestren=\CHAPNO \advCHAPNO

 \advAPPNO

\newcount\cHintroIV  \cHintroIV=\CHAPNO \advCHAPNO
                              
\newcount\cHcomparison   \cHcomparison=\CHAPNO \advCHAPNO
                              
\newcount\cHsumsmom  \cHsumsmom=\CHAPNO \advCHAPNO
                              
\newcount\cHsectorsmom   \cHsectorsmom=\CHAPNO \advCHAPNO
                              
\newcount\cHppladsect    \cHppladsect=\CHAPNO \advCHAPNO

 \advAPPNO


 \def\defNPCTMSpace{\frefwarning I.1} \IgNoRe{STM Assertion }
  \IgNoRe{PG}
 \def\eqnNPreal{\frefwarning I.1} \IgNoRe{EQN}
  \IgNoRe{EQN}
 \def\eqnNPinteraction{\frefwarning I.3} \IgNoRe{EQN}
  \IgNoRe{EQN}
  \IgNoRe{EQN}
 \def\defNPscales{\frefwarning I.2} \IgNoRe{STM Assertion }
  \IgNoRe{STM Assertion }
 \def\theoremNPmainthI{\frefwarning I.4} \IgNoRe{STM Assertion }
  \IgNoRe{STM Assertion }
  \IgNoRe{STM Assertion }
  \IgNoRe{STM Assertion }
  \IgNoRe{STM Assertion }
  \IgNoRe{STM Assertion }
 \def\defNPstrongasymm{\frefwarning I.10} \IgNoRe{STM Assertion }
  \IgNoRe{STM Assertion }
 \def\hypNPdisprel{\frefwarning I.12} \IgNoRe{STM Assertion }
  \IgNoRe{PG}
  \IgNoRe{PG}
  \IgNoRe{PG}
  \IgNoRe{PG}
  \IgNoRe{EQN}
  \IgNoRe{EQN}
  \IgNoRe{EQN}
  \IgNoRe{EQN}
  \IgNoRe{PG}
  \IgNoRe{PG}
  \IgNoRe{PG}
  \IgNoRe{EQN}
  \IgNoRe{PG}
  \IgNoRe{PG}
 \def\eqnOVtriponesi{\frefwarning II.6} \IgNoRe{EQN}
  \IgNoRe{EQN}
  \IgNoRe{EQN}
  \IgNoRe{EQN}
  \IgNoRe{EQN}
 \def\eqnOVsectorvertexreq{\frefwarning II.11} \IgNoRe{EQN}
  \IgNoRe{EQN}
  \IgNoRe{EQN}
 \def\eqnOVtripthreesi{\frefwarning II.14} \IgNoRe{EQN}
  \IgNoRe{EQN}
  \IgNoRe{PG}
  \IgNoRe{PG}
 \def\eqnNPcovFT{\frefwarning III.1} \IgNoRe{EQN}
 \def\eqnNPantisymmCov{\frefwarning III.2} \IgNoRe{EQN}
 \def\eqnNPjdef{\frefwarning III.3} \IgNoRe{EQN}
 \def\defNPrengroupmap{\frefwarning III.1} \IgNoRe{STM Assertion }
  \IgNoRe{PG}
 \def\eqnNPsemigrp{\frefwarning III.4} \IgNoRe{EQN}
  \IgNoRe{EQN}
  \IgNoRe{EQN}
  \IgNoRe{EQN}
  \IgNoRe{STM Assertion }
  \IgNoRe{STM Assertion }
 \def\defNPeffinttriple{\frefwarning III.4} \IgNoRe{STM Assertion }
 \def\defNPformalCovariances{\frefwarning III.5} \IgNoRe{STM Assertion }
 \def\defNPinoutmap{\frefwarning III.6} \IgNoRe{STM Assertion }
 \def\remNPinoutmap{\frefwarning III.7} \IgNoRe{STM Assertion }
 \def\defNPinputData{\frefwarning III.8} \IgNoRe{STM Assertion }
 \def\defNPoutputData{\frefwarning III.9} \IgNoRe{STM Assertion }
  \IgNoRe{STM Assertion }
  \IgNoRe{EQN}
 \def\eqnNPinouttriple{\frefwarning III.9} \IgNoRe{EQN}
 \def\eqnNPgenfnrengrp{\frefwarning III.10} \IgNoRe{EQN}
 \def\eqnNPKuprime{\frefwarning III.11} \IgNoRe{EQN}
 \def\lemNPformalselfconsistent{\frefwarning III.11} \IgNoRe{STM Assertion }
  \IgNoRe{EQN}
  \IgNoRe{STM Assertion }
  \IgNoRe{EQN}
  \IgNoRe{STM Assertion }
  \IgNoRe{PG}
  \IgNoRe{EQN}
  \IgNoRe{STM Assertion }
  \IgNoRe{STM Assertion }
  \IgNoRe{EQN}
  \IgNoRe{PG}
 \def\pgNPIV{\frefwarning 1} \IgNoRe{PG}
 \def\defNPdecayop{\frefwarning V.1} \IgNoRe{STM Assertion }
 \def\defNPFancynormdomain{\frefwarning V.2} \IgNoRe{STM Assertion }
 \def\pgNPV{\frefwarning 2} \IgNoRe{PG}
 \def\defNPSymmNorm{\frefwarning V.3} \IgNoRe{STM Assertion }
 \def\notNPfourierTI{\frefwarning V.4} \IgNoRe{STM Assertion }
 \def\remNPnorminteraction{\frefwarning V.5} \IgNoRe{STM Assertion }
 \def\defNPrhomn{\frefwarning V.6} \IgNoRe{STM Assertion }
 \def\remNPrhomn{\frefwarning V.7} \IgNoRe{STM Assertion }
 \def\thmNPfirststep{\frefwarning V.8} \IgNoRe{STM Assertion }
 \def\lemNPfirststep{\frefwarning V.9} \IgNoRe{STM Assertion }
 \def\defNPfourtrans{\frefwarning VI.1} \IgNoRe{STM Assertion }
 \def\pgNPVI{\frefwarning 8} \IgNoRe{PG}
 \def\defNPsectors{\frefwarning VI.2} \IgNoRe{STM Assertion }
 \def\defNPsectrepr{\frefwarning VI.3} \IgNoRe{STM Assertion }
 \def\defNPresector{\frefwarning VI.4} \IgNoRe{STM Assertion }
 \def\defNPtens{\frefwarning VI.5} \IgNoRe{STM Assertion }
 \def\eqnNPsectGrf{\frefwarning VI.1} \IgNoRe{EQN}
 \def\defNPsectnorm{\frefwarning VI.6} \IgNoRe{STM Assertion }
 \def\defNPsectGrnorm{\frefwarning VI.7} \IgNoRe{STM Assertion }
 \def\remNPnvsnj{\frefwarning VI.8} \IgNoRe{STM Assertion }
 \def\eqnNPnormmotivate{\frefwarning VI.2} \IgNoRe{EQN}
 \def\defNPCTmSpace{\frefwarning VI.9} \IgNoRe{STM Assertion }
 \def\remNPInclusionK{\frefwarning VI.10} \IgNoRe{STM Assertion }
 \def\defNPCovariances{\frefwarning VI.11} \IgNoRe{STM Assertion }
 \def\thmNPsetupinduction{\frefwarning VI.12} \IgNoRe{STM Assertion }
 \def\pgNPVII{\frefwarning 16} \IgNoRe{PG}
 \def\defbubbleprop{\frefwarning VII.1} \IgNoRe{STM Assertion }
 \def\deNPdefsectbubbleprop{\frefwarning VII.2} \IgNoRe{STM Assertion }
 \def\NPsomespaces{\frefwarning VII.3} \IgNoRe{STM Assertion }
 \def\defParicleHoleDecomp{\frefwarning VII.4} \IgNoRe{STM Assertion }
 \def\lemunordord{\frefwarning VII.5} \IgNoRe{STM Assertion }
 \def\propparticleparticleladder{\frefwarning VII.6} \IgNoRe{STM Assertion }
 \def\defcompLadder{\frefwarning VII.7} \IgNoRe{STM Assertion }
 \def\theoremcompLadder{\frefwarning VII.8} \IgNoRe{STM Assertion }
 \def\defNPintquad{\frefwarning VIII.1} \IgNoRe{STM Assertion }
 \def\eqnNPpbound{\frefwarning VIII.1} \IgNoRe{EQN}
 \def\eqnNPudecomp{\frefwarning VIII.2} \IgNoRe{EQN}
 \def\eqnNPdeubound{\frefwarning VIII.3} \IgNoRe{EQN}
 \def\remNPintquad{\frefwarning VIII.2} \IgNoRe{STM Assertion }
 \def\pgNPVIII{\frefwarning 22} \IgNoRe{PG}
 \def\defNPprojectivesyst{\frefwarning VIII.3} \IgNoRe{STM Assertion }
 \def\remNPprojectivesyst{\frefwarning VIII.4} \IgNoRe{STM Assertion }
 \def\theoremNPinduction{\frefwarning VIII.5} \IgNoRe{STM Assertion }
 \def\eqnNPgdiff{\frefwarning VIII.4} \IgNoRe{EQN}
 \def\eqnNPnbndA{\frefwarning VIII.5} \IgNoRe{EQN}
 \def\eqnNPnbndB{\frefwarning VIII.6} \IgNoRe{EQN}
 \def\defNPconvol{\frefwarning VIII.6} \IgNoRe{STM Assertion }
 \def\lemNPpptyu{\frefwarning VIII.7} \IgNoRe{STM Assertion }
 \def\eqnNPdenomA{\frefwarning VIII.7} \IgNoRe{EQN}
 \def\eqnNPdenomB{\frefwarning VIII.8} \IgNoRe{EQN}
 \def\eqnNPdenomC{\frefwarning VIII.9} \IgNoRe{EQN}
 \def\stepInputData{\frefwarning IX.1} \IgNoRe{STM Assertion }
 \def\pgNPIX{\frefwarning 31} \IgNoRe{PG}
 \def\pgNPIXa{\frefwarning 31} \IgNoRe{PG}
 \def\stepOutputData{\frefwarning IX.2} \IgNoRe{STM Assertion }
 \def\remNPOandRconditions{\frefwarning IX.3} \IgNoRe{STM Assertion }
 \def\pgNPIXb{\frefwarning 33} \IgNoRe{PG}
 \def\defNPomegaj{\frefwarning IX.4} \IgNoRe{STM Assertion }
 \def\thmIntoOut{\frefwarning IX.5} \IgNoRe{STM Assertion }
 \def\eqnNPwpwppgp{\frefwarning IX.1} \IgNoRe{EQN}
 \def\eqnNPodI{\frefwarning IX.2} \IgNoRe{EQN}
 \def\eqnNPodIb{\frefwarning IX.3} \IgNoRe{EQN}
 \def\eqnNPodII{\frefwarning IX.4} \IgNoRe{EQN}
 \def\eqnNPodIII{\frefwarning IX.5} \IgNoRe{EQN}
 \def\eqnNPodIIb{\frefwarning IX.6} \IgNoRe{EQN}
 \def\eqnNPodFI{\frefwarning IX.7} \IgNoRe{EQN}
 \def\eqnNPodFIa{\frefwarning IX.8} \IgNoRe{EQN}
 \def\eqnNPodFII{\frefwarning IX.9} \IgNoRe{EQN}
 \def\eqnNPodFIII{\frefwarning IX.10} \IgNoRe{EQN}
 \def\eqnNPFjplusI{\frefwarning IX.11} \IgNoRe{EQN}
 \def\eqnNPgdiffinout{\frefwarning IX.12} \IgNoRe{EQN}
 \def\eqnNPderivgprimeI{\frefwarning IX.13} \IgNoRe{EQN}
 \def\eqnNPderivgprimeIV{\frefwarning IX.14} \IgNoRe{EQN}
 \def\remNPFnochangeI{\frefwarning IX.6} \IgNoRe{STM Assertion }
 \def\eqnNPreWickDefs{\frefwarning IX.15} \IgNoRe{EQN}
 \def\pgNPIXc{\frefwarning 40} \IgNoRe{PG}
 \def\eqnNPtwrewick{\frefwarning IX.16} \IgNoRe{EQN}
 \def\lemreWick{\frefwarning IX.7} \IgNoRe{STM Assertion }
 \def\eqndefkkprime{\frefwarning IX.17} \IgNoRe{EQN}
 \def\lemNPdeKbnd{\frefwarning IX.8} \IgNoRe{STM Assertion }
 \def\propNPctmmap{\frefwarning IX.9} \IgNoRe{STM Assertion }
 \def\eqnNPtwwpp{\frefwarning IX.18} \IgNoRe{EQN}
 \def\eqnNPgup{\frefwarning IX.19} \IgNoRe{EQN}
 \def\eqnNPcheckup{\frefwarning IX.20} \IgNoRe{EQN}
 \def\thmOuttoIn{\frefwarning IX.10} \IgNoRe{STM Assertion }
 \def\eqnNPpboundal{\frefwarning IX.21} \IgNoRe{EQN}
 \def\eqnNPdelicateA{\frefwarning IX.22} \IgNoRe{EQN}
 \def\eqnNPdelicateI{\frefwarning IX.23} \IgNoRe{EQN}
 \def\eqnNPdelicateII{\frefwarning IX.24} \IgNoRe{EQN}
 \def\eqnNPdelicateB{\frefwarning IX.25} \IgNoRe{EQN}
 \def\eqnNPdelicateIII{\frefwarning IX.26} \IgNoRe{EQN}
 \def\eqnNPdelicateIV{\frefwarning IX.27} \IgNoRe{EQN}
 \def\eqnNPwpwpp{\frefwarning IX.28} \IgNoRe{EQN}
 \def\eqnNPwickwtw{\frefwarning IX.29} \IgNoRe{EQN}
 \def\eqnNPreWickBnd{\frefwarning IX.30} \IgNoRe{EQN}
 \def\eqnNPreWickBndbis{\frefwarning IX.31} \IgNoRe{EQN}
 \def\eqnNPderivwickwtw{\frefwarning IX.32} \IgNoRe{EQN}
 \def\eqnNPderivwprimeI{\frefwarning IX.33} \IgNoRe{EQN}
 \def\eqnNPderivwprimeII{\frefwarning IX.34} \IgNoRe{EQN}
 \def\eqnNPderivwtildeest{\frefwarning IX.35} \IgNoRe{EQN}
 \def\eqnNPgdiffouttoin{\frefwarning IX.36} \IgNoRe{EQN}
 \def\eqnNPderivgprimeI{\frefwarning IX.37} \IgNoRe{EQN}
 \def\eqnNPdifffourlegged{\frefwarning IX.38} \IgNoRe{EQN}
 \def\eqnNPgoodFprimeest{\frefwarning IX.39} \IgNoRe{EQN}
 \def\eqnNPfourlegest{\frefwarning IX.40} \IgNoRe{EQN}
 \def\eqnNPnonfourlegest{\frefwarning IX.41} \IgNoRe{EQN}
 \def\remNPFnochangeII{\frefwarning IX.11} \IgNoRe{STM Assertion }
 \def\pgNPX{\frefwarning 54} \IgNoRe{PG}
 \def\pgNPXa{\frefwarning 54} \IgNoRe{PG}
 \def\lemNPprojSysBnds{\frefwarning X.1} \IgNoRe{STM Assertion }
 \def\pgNPXb{\frefwarning 56} \IgNoRe{PG}
 \def\eqnNPrenijInda{\frefwarning X.1} \IgNoRe{EQN}
 \def\eqnNPrenijIndb{\frefwarning X.2} \IgNoRe{EQN}
 \def\eqnNPrenijbnd{\frefwarning X.3} \IgNoRe{EQN}
 \def\eqnNPrenijsupbnd{\frefwarning X.4} \IgNoRe{EQN}
 \def\lemRWintbnd{\frefwarning B.1} \IgNoRe{STM Assertion }
 \def\eqnNPwkKofKp{\frefwarning B.1} \IgNoRe{EQN}
 \def\pgNPB{\frefwarning 61} \IgNoRe{PG}
 \def\eqnNPwkce{\frefwarning B.2} \IgNoRe{EQN}
 \def\eqnNPmomspE{\frefwarning B.3} \IgNoRe{EQN}
 \def\eqnNPwkdecomp{\frefwarning B.4} \IgNoRe{EQN}
 \def\eqnNPwka{\frefwarning B.5} \IgNoRe{EQN}
 \def\eqnNPwkb{\frefwarning B.6} \IgNoRe{EQN}
 \def\eqnNPwkc{\frefwarning B.7} \IgNoRe{EQN}
 \def\eqnNPwkd{\frefwarning B.8} \IgNoRe{EQN}
 \def\eqnNPwke{\frefwarning B.9} \IgNoRe{EQN}
 \def\eqnNPwkf{\frefwarning B.10} \IgNoRe{EQN}
 \def\eqnNPwki{\frefwarning B.11} \IgNoRe{EQN}
 \def\eqnNPwkg{\frefwarning B.12} \IgNoRe{EQN}
 \def\eqnNPwkh{\frefwarning B.13} \IgNoRe{EQN}
 \def\lemRWrhsbounds{\frefwarning B.2} \IgNoRe{STM Assertion }
 \def\lemNPcontrmap{\frefwarning B.3} \IgNoRe{STM Assertion }
 \def\eqnNPQ{\frefwarning B.14} \IgNoRe{EQN}
 \def\eqnNPQprime{\frefwarning B.15} \IgNoRe{EQN}
 \def\pgNPIIref{\frefwarning 72} \IgNoRe{PG}
  \IgNoRe{PG}
  \IgNoRe{STM Assertion }
  \IgNoRe{EQN}
  \IgNoRe{STM Assertion }
  \IgNoRe{PG}
  \IgNoRe{EQN}
  \IgNoRe{EQN}
  \IgNoRe{EQN}
  \IgNoRe{STM Assertion }
  \IgNoRe{STM Assertion }
  \IgNoRe{EQN}
  \IgNoRe{EQN}
  \IgNoRe{EQN}
  \IgNoRe{EQN}
  \IgNoRe{EQN}
  \IgNoRe{STM Assertion }
  \IgNoRe{EQN}
  \IgNoRe{EQN}
  \IgNoRe{EQN}
  \IgNoRe{EQN}
  \IgNoRe{STM Assertion }
  \IgNoRe{STM Assertion }
  \IgNoRe{EQN}
  \IgNoRe{STM Assertion }
  \IgNoRe{STM Assertion }
  \IgNoRe{STM Assertion }
  \IgNoRe{STM Assertion }
  \IgNoRe{PG}
  \IgNoRe{STM Assertion }
  \IgNoRe{STM Assertion }
  \IgNoRe{STM Assertion }
  \IgNoRe{STM Assertion }
  \IgNoRe{STM Assertion }
  \IgNoRe{STM Assertion }
  \IgNoRe{STM Assertion }
  \IgNoRe{STM Assertion }
  \IgNoRe{STM Assertion }
  \IgNoRe{STM Assertion }
  \IgNoRe{STM Assertion }
  \IgNoRe{STM Assertion }
  \IgNoRe{STM Assertion }
  \IgNoRe{STM Assertion }
  \IgNoRe{STM Assertion }
  \IgNoRe{PG}
  \IgNoRe{STM Assertion }
  \IgNoRe{STM Assertion }
  \IgNoRe{STM Assertion }
  \IgNoRe{STM Assertion }
  \IgNoRe{STM Assertion }
  \IgNoRe{STM Assertion }
 \def\tildepropparticleparticleladder{\frefwarning XIV.9} \IgNoRe{STM Assertion }
 \def\remtildepropparticleparticleladder{\frefwarning XIV.10} \IgNoRe{STM Assertion }
  \IgNoRe{STM Assertion }
 \def\theoremtildecompLadder{\frefwarning XIV.12} \IgNoRe{STM Assertion }
 \def\remtildecompLadder{\frefwarning XIV.13} \IgNoRe{STM Assertion }
  \IgNoRe{STM Assertion }
  \IgNoRe{EQN}
  \IgNoRe{STM Assertion }
  \IgNoRe{STM Assertion }
  \IgNoRe{STM Assertion }
  \IgNoRe{STM Assertion }
  \IgNoRe{STM Assertion }
  \IgNoRe{STM Assertion }
  \IgNoRe{EQN}
  \IgNoRe{STM Assertion }
  \IgNoRe{PG}
  \IgNoRe{PG}
  \IgNoRe{STM Assertion }
  \IgNoRe{STM Assertion }
  \IgNoRe{STM Assertion }
  \IgNoRe{EQN}
  \IgNoRe{STM Assertion }
  \IgNoRe{PG}
  \IgNoRe{EQN}
  \IgNoRe{STM Assertion }
  \IgNoRe{STM Assertion }
  \IgNoRe{EQN}
  \IgNoRe{EQN}
  \IgNoRe{EQN}
  \IgNoRe{EQN}
  \IgNoRe{EQN}
  \IgNoRe{EQN}
  \IgNoRe{EQN}
  \IgNoRe{EQN}
  \IgNoRe{EQN}
  \IgNoRe{EQN}
  \IgNoRe{EQN}
  \IgNoRe{EQN}
  \IgNoRe{EQN}
  \IgNoRe{EQN}
  \IgNoRe{EQN}
  \IgNoRe{EQN}
  \IgNoRe{EQN}
  \IgNoRe{EQN}
  \IgNoRe{EQN}
  \IgNoRe{EQN}
  \IgNoRe{STM Assertion }
  \IgNoRe{STM Assertion }
  \IgNoRe{PG}
  \IgNoRe{EQN}
  \IgNoRe{EQN}
  \IgNoRe{EQN}
  \IgNoRe{EQN}
  \IgNoRe{EQN}
  \IgNoRe{EQN}
  \IgNoRe{EQN}
  \IgNoRe{EQN}
  \IgNoRe{EQN}
  \IgNoRe{EQN}
  \IgNoRe{EQN}
  \IgNoRe{EQN}
  \IgNoRe{EQN}
  \IgNoRe{EQN}
  \IgNoRe{EQN}
  \IgNoRe{EQN}
  \IgNoRe{EQN}
  \IgNoRe{EQN}
  \IgNoRe{EQN}
  \IgNoRe{EQN}
  \IgNoRe{EQN}
  \IgNoRe{STM Assertion }
  \IgNoRe{PG}
  \IgNoRe{STM Assertion }
  \IgNoRe{STM Assertion }
  \IgNoRe{EQN}
  \IgNoRe{EQN}
  \IgNoRe{PG}
  \IgNoRe{EQN}
  \IgNoRe{EQN}
  \IgNoRe{EQN}
  \IgNoRe{EQN}
  \IgNoRe{EQN}
  \IgNoRe{EQN}
  \IgNoRe{EQN}
  \IgNoRe{EQN}
  \IgNoRe{STM Assertion }
  \IgNoRe{STM Assertion }
  \IgNoRe{STM Assertion }
  \IgNoRe{STM Assertion }
  \IgNoRe{STM Assertion }
  \IgNoRe{PG}
  \IgNoRe{PG}
 \def\pgNPIInot{\frefwarning 73} \IgNoRe{PG}
  \IgNoRe{PG}


\newcount\CHAPNO
\newcount\APPNO
\CHAPNO=0
\APPNO=1
\def\advCHAPNO{\advance\CHAPNO by 1}
\def\advAPPNO{\advance\APPNO by 1}

\def\caproman#1{\ifcase#1\or I\or II\or III\or IV\or V\or VI\or VII\or
VIII\or IX\or X\or XI\or XII\or XIII\or XIV\or XV\or XVI\or XVII\or XVIII\or
XIX\or XX\or XXI\or XXII\or XXIII\or XXIV\or XXV\or XXVI\or XXVII\or XXVIII\or XXIX\or XXX\or XXXI\or XXXII\or XXXIII\or XXXIV\or XXXV\or XXXVI\or XXXVII\or XXXVIII\or XXXIX\fi}%

\def\capletter#1{\ifcase#1\or A\or B\or C\or D\or E\or F\or G\or
H\or I\or J\or K\or L\or M\or N\or O\or P\or Q\or R\or
S\or T\or U\or V\or W\or X\or Y\or Z\fi}%

\newcount\cHintroI \cHintroI=\CHAPNO \advCHAPNO 
                              \edef\CHintroI{\caproman\CHAPNO}         
\newcount\cHintroOverview  \cHintroOverview=\CHAPNO \advCHAPNO 
                              \edef\CHintroOverview{\caproman\CHAPNO}  
\newcount\cHrenmap \cHrenmap=\CHAPNO \advCHAPNO 
                              \edef\CHrenmap{\caproman\CHAPNO}         

\edef\APappModelComp{\capletter\APPNO} \advAPPNO

\newcount\cHintroII \cHintroII=\CHAPNO \advCHAPNO 
                              \edef\CHintroII{\caproman\CHAPNO}
\newcount\cHfirstscale \cHfirstscale=\CHAPNO \advCHAPNO
                              
\newcount\cHnewsectors \cHnewsectors=\CHAPNO \advCHAPNO
                              \edef\CHnewsectors{\caproman\CHAPNO}
\newcount\cHphladders \cHphladders=\CHAPNO \advCHAPNO
                              
\newcount\cHfinitescale \cHfinitescale=\CHAPNO \advCHAPNO
                              
\newcount\cHstep \cHstep=\CHAPNO \advCHAPNO
                              
\newcount\cHrecurs \cHrecurs=\CHAPNO \advCHAPNO
                              \edef\CHrecurs{\caproman\CHAPNO}
\edef\APappRewick{\capletter\APPNO} \advAPPNO

\newcount\cHintroIII \cHintroIII=\CHAPNO \advCHAPNO
                              \edef\CHintroIII{\caproman\CHAPNO}
\newcount\cHtildefinitescale \cHtildefinitescale=\CHAPNO \advCHAPNO
                              
\newcount\cHtildenewsectors \cHtildenewsectors=\CHAPNO \advCHAPNO
                              
\newcount\cHtildephladders \cHtildephladders=\CHAPNO \advCHAPNO
                              
\newcount\cHtildestep  \cHtildestep=\CHAPNO \advCHAPNO
                              \edef\CHtildestep{\caproman\CHAPNO}

\edef\APappHoelder{\capletter\APPNO} \advAPPNO
\edef\APappPhladders{\capletter\APPNO} \advAPPNO

\chapno=\cHintroII

{\nopagenumbers
\multiply\baselineskip by \spacingDenominator\divide \baselineskip by\spacingNumerator

\null\vskip3truecm

%
%
\centerline{\tafontt A Two Dimensional Fermi Liquid }
\vskip0.1in
\centerline{\tbfontt Part 2: Convergence}

\vskip0.75in
\centerline{Joel Feldman{\parindent=.15in\footnote{$^{*}$}{Research supported 
in part by the
 Natural Sciences and Engineering Research Council of Canada and the Forschungsinstitut f\"ur Mathematik, ETH Z\"urich}}}
\centerline{Department of Mathematics}
\centerline{University of British Columbia}
\centerline{Vancouver, B.C. }
\centerline{CANADA\ \   V6T 1Z2}
\centerline{feldman@math.ubc.ca}
\centerline{http:/\hskip-3pt/www.math.ubc.ca/\squiggle
feldman/}
\vskip0.3in
\centerline{Horst Kn\"orrer, Eugene Trubowitz}
\centerline{Mathematik}
\centerline{ETH-Zentrum}
\centerline{CH-8092 Z\"urich}
\centerline{SWITZERLAND}
\centerline{knoerrer@math.ethz.ch, trub@math.ethz.ch}
\centerline{http:/\hskip-3pt/www.math.ethz.ch/\squiggle
knoerrer/}

\vskip0.75in
\noindent
%
{\bf Abstract.\ \ \ } 
Using results established in other papers in our series, 
we prove the existence of the infinite volume, temperature zero, 
thermodynamic Green's functions of a two dimensional,  weakly 
coupled fermion gas with an asymmetric Fermi curve and short range interactions. This is done by showing that our sequence of 
renormalization group maps converges.

\vfill
\eject


\titleb{Table of Contents}
\halign{\hfill#\ &\hfill#\ &#\hfill&\ p\ \hfil#&\ p\ \hfil#\cr
\noalign{\vskip0.05in}
\S IV&\omit Introduction                          \span&\:\pgNPIV&\omit\cr
\noalign{\vskip0.05in}
\S V&\omit The First Scales                        \span&\:\pgNPV\cr
\noalign{\vskip0.05in}
\S VI&\omit Sectors and Sectorized Norms           \span&\:\pgNPVI\cr
\noalign{\vskip0.05in}
\S VII&\omit Ladders                                 \span&\:\pgNPVII\cr
\noalign{\vskip0.05in}
\S VIII&\omit Infrared Limit of Finite Scale Green's Functions
                                                    \span&\:\pgNPVIII\cr
\noalign{\vskip0.05in}
\S IX&\omit One Recursion Step                      \span&\:\pgNPIX\cr
&&Input and Output Data                               &\omit&\:\pgNPIXa\cr
&&Integrating Out a Scale                             &\omit&\:\pgNPIXb\cr
&&Sector Refinement, ReWick ordering and Renormalization  &\omit&\:\pgNPIXc\cr
\noalign{\vskip0.05in}
\S X&\omit The Recursive Construction of the Green's Functions\span&\:\pgNPX\cr
&&Initialization at $j=j_0$                               &\omit&\:\pgNPXa\cr
&&Recursive step $j\rightarrow j+1$                      &\omit&\:\pgNPXb\cr
\noalign{\vskip0.05in}
{\bf Appendices}\span\cr
\noalign{\vskip0.05in}
\S B&\omit Self--consistent ReWick Ordering             \span&\:\pgNPB\cr
\noalign{\vskip0.05in}
 &\omit References                                    \span&\:\pgNPIIref \cr
\noalign{\vskip0.05in}
 &\omit Notation                                      \span&\:\pgNPIInot \cr
}
\vfill\eject
\multiply\baselineskip by \spacingNumerator\divide \baselineskip by\spacingDenominator}
\pageno=1


\chap{Introduction}\PG\pgNPIV

This paper, together with [FKTf1] and [FKTf3] provides a construction of a
two dimensional Fermi liquid at zero temperature. It contains
Sections \CHintroII\ through \CHrecurs\ and  Appendix \APappRewick.  
Sections \CHintroI\ through \CHrenmap\ and 
Appendix \APappModelComp\ are in [FKTf1] and 
Sections \CHintroIII\ through \CHtildestep\ and 
Appendices \APappHoelder\  and \APappPhladders\ are in [FKTf3]. 
Cumulative notation tables are provided at the end of each part.
The main goal
of this part is the proof of convergence of the Green's functions
stated in Theorem  \theoremNPmainthI. In the proof of this theorem, which
follows the statement of Theorem \:\theoremNPinduction, we compare 
$\cG_\il(\phi,\bar\phi)$ with a generating functional $\cG^\rg_j(\phi,\bar\phi)$
constructed by iterating a renormalization group map $j$ times for some 
$\il-2<j<\il$. See also \S\CHrenmap. To aid in the derivation of bounds on the 
renormalization group map, we fix a scale parameter $M$ that is 
sufficiently big (depending on the dispersion relation $e(\k)$ and 
the ultraviolet cutoff $U(\k)$). This $M$ is used throughout the rest of this paper, with the exception of the proof of Theorem  \theoremNPmainthI\ from 
Theorem \:\theoremNPinduction, where we also explain that fixing $M$
gives no loss of generality.

\vfill\eject

\chap{The First Scales}\PG\pgNPV

In \S\CHrenmap, we outlined the algebraic aspects of our
 strategy for proving Theorem \theoremNPmainthI.
To state and prove the convergence of $\de e_j(0)$ and $\tilde\cG_j(\phi,0)$,
we clearly have to introduce norms for these and various related objects. 
There are at least two places where control over derivatives will be needed. 
The analog, Lemma \lemreWick, of the formal power series Lemma 
\lemNPformalselfconsistent\ will involve an application of the 
implicit function theorem and will require control of derivatives with respect 
to $K$.
Secondly, we need to control the size of 
$\check u\big((k_0,\k);K\big)$ in a neighbourhood of the Fermi surface 
when $k_0\ne 0$, using the fact that this quantity is
small when $k_0=0$. This is done using the $k_0$--derivatives. 
For this reason we shall also control momentum
space derivatives, through position space decay, of quantities appearing in the
strategy outlined in the last section. The following notation is convenient to keep track of
the effect of the chain rule and Leibniz's rule in the estimates of derivatives
(see [FKTo1, \S\CHnorms]).

\definition{\STM\defNPdecayop (Decay operators)}{
\Item i) Recall that, for a  multiindex $\de$,
$x=(x_0,\x,\si),\,x'=(x'_0,\x',\si') \in \bbbr\times\bbbr^d
\times \{ \uparrow,\,\downarrow\}$,
$$
(x-x')^\de = (x_0-x_0')^{\de_0}\,(\x_1-\x'_1)^{\de_1}\cdots
(\x_d-\x_d')^{\de_d}
$$
If  $\xi=(x,a),\,\xi'=(x',a') \in \cB$, we define
$\ (\xi -\xi')^\de = (x-x')^\de$. 
\Item ii)
Let $n$ be a positive integer. 
For a function $f(\xi_1,\cdots,\xi_n)$ on $\cB^n$, a multiindex $\de$, and
$1\le i,j\le n;\,i\ne j$ set
$$
\cD_{i,j}^\de f\,(\xi_1,\cdots,\xi_n) 
= (\xi_i-\xi_j)^\de f(\xi_1,\cdots,\xi_n)
$$
A decay operator $\cD$ on the set of 
functions on $\cB^n$ is an operator of the form
$$
\cD = \cD^{\de^{(1)}}_{u_1,v_1} \cdots 
 \cD^{\de^{(k)}}_{u_k,v_k}
$$
with multiindices $\de^{(1)},\cdots,\de^{(k)}$ and 
$1\le u_j,v_j\le n,\ u_j\ne v_j$. The indices $u_j,v_j$ are called variable
indices.  The total order of $\cD$ is 
$$
\de(\cD) = \de^{(1)} + \cdots + \de^{(k)}
$$
In a similar way, we define the action of a decay operator on the set of 
functions on $\big(\bbbr\times\bbbr^d \big)^n$ or on
$\big(\bbbr\times\bbbr^d \times \{\uparrow,\downarrow\} \big)^n$.
}

\definition{\STM\defNPFancynormdomain}{
\Item{i)}
On $\bbbr_+\cup\{\infty\} = \set{x\in\bbbr}{x\ge0}\cup\{+\infty\}$, addition
and the total ordering $\le$ are defined in the standard way. With the 
convention that $0\cdot\infty=\infty$, multiplication is also defined in the
standard way.
\Item{ii)}
Let $d\ge 0$. The $(d+1)$--dimensional norm domain $\fN_{d+1}$ is the 
set of all formal power series
$$
X = \sum_{\de\in\bbbn_0\times\bbbn_0^d} X_\de \
t_0^{\de_0}t_1^{\de_1}\cdots t_d^{\de_d}
$$
in the variables $t_0,t_1,\cdots,t_d$ with coefficients $X_\de \in \bbbr_+\cup\{\infty\}$. To shorten notation, we set
 $t^\de=t_0^{\de_0}t_1^{\de_1}\cdots t_d^{\de_d} $.
 Addition and partial ordering on $\fN_{d+1}$ are defined componentwise. Multiplication is defined by
$$
(X\cdot X')_\de = \smsum_{\be+\ga=\de} X_\be X'_\ga
$$
The max and min of two elements of $\fN_{d+1}$ are again defined componentwise.

\noindent
We identify $\bbbr_+\cup\{\infty\}$ with the set of all $X\in \fN_{d+1}$ having
$X_\de=0$ for all $\de\ne \0=(0,\cdots,0)$.

\noindent
If $a>0$, $X_\0\ne\infty$ and $a-X_\0 >0$ then $(a-X)^{-1}$ is defined as
$$
(a-X)^{-1} = \sfrac{1}{a-X_\0} 
\smsum_{n=0}^\infty \big(\sfrac{X-X_\0}{a-X_\0}\big)^n
$$

\noindent
For an element 
$\ 
X = \sum_{\de\in\bbbn_0\times\bbbn_0^d} X_\de \ t^{\de}
\ $
of $\fN_{d+1}$ and $0\le j\le d$ the formal 
derivative $\sfrac{\partial\hfill}{\partial t_j}X$ is defined as
$$
\sfrac{\partial\hfill}{\partial t_j}X 
= \sum_{\de\in\bbbn_0\times\bbbn_0^d} (\de_j+1)X_{\de+\ep_j} \ t^{\de}
$$
where $\ep_j$ is the $j^{\rm th}$ unit vector.
\Item{iii)}
For $j\ge 0$ we set
$$
\cb_j=\sum_{|\bde|\le r\atop |\de_0|\le r_0}  M^{j|\de|}\,t^\de
+\sum_{|\bde|> r\atop {\rm or\ }|\de_0|> r_0}\infty\, t^\de
\in\fN_{d+1}
$$ 
and for $X \in \fN_{d+1}$ with $X_\0<\sfrac{1}{M^j}$ 
$$
\fe_j(X) = \sfrac{\cb_j}{1-M^j X}
$$
}

\definition{\STM\defNPSymmNorm}{
For a function $f$ on $\cB^m \times \cB^n$ we 
define the (scalar valued) $L_1$--$L_\infty$--norm as
$$
\tn f\tn_{1,\infty} 
= \cases{ 
\max\limits_{1\le j_0 \le n}\ 
\sup\limits_{\xi_{j_0} \in \cB}\  
\int \prod\limits_{j=1,\cdots, n \atop j\ne j_0} d\xi_j\, 
| f( \xi_1,\cdots,\xi_n) | & if \ $m=0$  \cr
\sup\limits_{\eta_1,\cdots,\eta_m \in \cB}
\int \prod\limits_{j=1,\cdots, n}  d\xi_j\  
| f( \eta_1,\cdots,\eta_m;\,\xi_1,\cdots,\xi_n) | & if \ $m\ne0$ 
}$$
and the $(d+1)$--dimensional $L_1$--$L_\infty$ seminorm
$$
\| f\|_{1,\infty}=\cases{
{\dst \sum_{\de\in\bbbn_0\times\bbbn_0^d}}\ \frac{1}{\de!}\ \Big(
\max\limits_{\cD\ {\rm decay\ operator} \atop {\rm with\ } \de(\cD) =\de} 
 \tn\cD\,f\tn_{1,\infty}\Big)\  t^{\de} & if $m=0$\cr
\noalign{\vskip.1in}
\tn f\tn_{1,\infty}& if $m\ne 0$\cr
}
$$
Here $\tn f\tn_{1,\infty}$ stands for the formal power series with constant 
coefficient $\tn f\tn_{1,\infty}$ and all other coefficients zero
and 
$\int d\xi\ g(\xi)=\smsum_{a\in\{0,1\}}\smsum_{\si\in\{\uparrow,\downarrow\}}
\int dx_0 d\x\ g\big((x_0,\x,\si,a)\big)$.

}
 
\noindent
Given a function on momentum space, we apply the above norms using the
\notation{\STM\notNPfourierTI}{
 If $\chi(k)$ is a function on $\bbbr\times \bbbr^d$, we define the Fourier transform $\hat\chi$ by 
$$
\hat\chi(\xi,\xi') 
= \de_{\si,\si'} \de_{a,a'}\int e^{(-1)^a\imath<k,x-x'>_-}\,\chi(k)\,
\sfrac{d^{d+1}k}{(2\pi)^{d+1}} 
$$
for $\xi=(x,a)=(x_0,\x,\si,a),\,\xi'=(x',a') =(x'_0,\x',\si',a')\in \cB$.

}

\remark{\STM\remNPnorminteraction}{
\Item{i)}
Let $V(x_1,x_2,x_3,x_4)$ be an interaction kernel as in Theorem 
\theoremNPmainthI\  and define, by abuse of notation,
 the function $V$ on $\cB^4$ by
$$
V({\sst (x_1,b_1),(x_2,b_2),(x_3,b_3),(x_4,b_4)}) 
= \de_{b_1,1}\de_{b_2,0}\de_{b_3,1}\de_{b_4,0} \,V(x_1,x_2,x_3,x_4)
$$
Then the hypothesis of Theorem \theoremNPmainthI\ is equivalent to
$\ \| V\|_{1,\infty} \le \veps\cb_0\ $ for some sufficiently small $\veps$.

\Item{ii)}
The constants $\cb_j$ will be used to describe the behaviour of momentum space derivatives of the covariance $C^{(j)}$.
The quantities $\fe_j(X)$ are used in bounding the differentiability 
properties of various kernels depending on a counterterm whose norm is bounded 
by $X$. This allows us to take into account the fact that the
 characteristics, as regards both size and smoothness, 
of counterterms are very different from the characteristics of kernels built
purely from $C^{(j)}$ and various smooth functions. The characteristics
of the counterterms are a consequence of their construction from 
various $C^{(j')}$'s, including those with $j'\gg j$. As $j'$ increases,
the contribution to the counterterm from $C^{(j')}$ becomes smaller and
smaller, and more and more concentrated near the Fermi surface,
 but less and less smooth.

}

We also wish to use our norms to control the coupling constant dependence
of various kernels. This is done using 

\definition{\STM\defNPrhomn}{ Fix $0<\upsilon<\sfrac{1}{4}$. 
 Set, for a coupling constant $0<\la<1$,
$$
\rho_{m;n}(\la)
=\frac{1}{\la^{(1-\upsilon)\max\{m+n-2,2\}/2}}
=\cases{ 
                      \la^{-(1-\upsilon)(m+n-2)/2}& if $m+n\ge 4$\cr
                      \noalign{\vskip.05in}
                      \la^{-(1-\upsilon)}& if $m+n=2$\cr}
$$

}

\remark{\STM\remNPrhomn}{
The exponent of Definition \defNPrhomn\ is motivated by the 
following considerations.  For this discussion, introduce 
a coupling constant $\la$ and replace $\cV(\psi)$ by $\la\cV(\psi)$.

The exponent of the initial generating functional contains, 
aside from the counterterm, two vertices with $\psi$ fields.
One, $\la\cV(\psi)$, has four $\psi$ fields and is proportional to the 
coupling constant $\la$.  The other, $\psi\phi$, has one $\psi$ field, one
$\phi$ field and is independent of $\la$. Consider any connected graph $G$ 
with $m$ external $\phi$ legs, $n$ external $\psi$ legs, $\rv\ge 1$ of 
the $\la \cV(\psi)$ vertices and $m$ of the $\psi\phi$ vertices. 
Since the $\phi$ field is always external, $G$ must have precisely 
$m$ $\psi\phi$ vertices to have $m$ external $\phi$ legs. 
The graph has $\sfrac{4\rv+2m-(m+n)}{2}$ internal
lines. To be connected, $G$ must have at least $\rv+m-1$ internal lines,
so that
$$
\sfrac{4\rv+2m-(m+n)}{2}\ge \rv+m-1\qquad\Longrightarrow\qquad \rv\ge \sfrac{m+n-2}{2}
$$
Thus $G$ is proportional to $\la^\rv$ with
$$
\rv\ge \max\big\{\sfrac{m+n-2}{2},1\big\}
$$
We set aside $\la^{\upsilon\max\{m+n-2,2\}/2}$, which we bound by 
$\la^{\upsilon n/10}$ 
 to achieve good inductive behaviour, i.e. to control various
constants that arise in the course of the expansion. 
Ultimately, we choose a maximum allowed coupling constant $\la_0$, rename
$\sfrac{1}{\la_0^{\upsilon/10}}=\al_0$ and consider
$|\la|<\la_0$ and $\al\ge\al_0$. Then,
our bound on the $m$ $\phi$--legged, $n$ $\psi$--legged part of the effective
interaction will be proportional to
$$
\sfrac{1}{\al^n}\la_0^{(1-\upsilon)\max\{m+n-2,2\}/2}
$$

We now further explain the phrase ``good inductive behaviour'' used in the last
paragraph. Consider, more generally, a connected graph $G$ with 
$m$ external $\phi$ legs, $n$ external $\psi$ legs, $\tilde m$ of the 
$\psi\phi$ vertices and $\rv\ge 1$ other vertices. Suppose that the 
$i^{\rm th}$ other vertex has $m_i$ $\phi$--legs and $n_i$ $\psi$--legs.
The number $\sfrac{\Si_i(m_i+n_i)+2\tilde m-(m+n)}{2}$ of internal lines
must be at least $\rv+\tilde m-1$ so
$$
\sfrac{\Si_i(m_i+n_i)+2\tilde m-(m+n)}{2}\ge \rv+\tilde m-1
\quad\Rightarrow\quad 
\sum_{i=1}^\rv\sfrac{m_i+n_i-2}{2}\ge \sfrac{m+n-2}{2}
$$
As $\rv\ge 1$
$$
\sum_{i=1}^\rv\max\big\{\sfrac{m_i+n_i-2}{2},1\big\}
\ge \max\big\{\sfrac{m+n-2}{2},1\big\}
$$
We thus have
$$
\sfrac{\al^n }{\la_0^{(1-\upsilon)\max\{m+n-2,2\}/2}}
\le \sfrac{1}{\al^{\Si n_i-n}}
\prod_{i=1}^\rv\sfrac{\al^{n_i} }{\la_0^{(1-\upsilon)\max\{m_i+n_i-2,2\}/2}}
$$
The small factors $\sfrac{1}{\al^{\Si n_i-n}}$
are available for controlling various
constants that arise in the course of the expansion. Observe that, as 
$m=\tilde m+\Si m_i$ and $\tilde m\le  \Si n_i-n$, the number of internal
lines of $G$, 
$\sfrac{\Si_i(m_i+n_i)+2\tilde m-(m+n)}{2}$ is bounded by $\Si n_i-n$.
}

\vskip .3cm 
We choose  an arbitrary but fixed scale, $j_0\ge 2$, and integrate the first scales, between 1 and $j_0$, in one fell swoop. 

\theorem{\STM\thmNPfirststep}{ 
There are ($M$ and $j_0$--dependent) constants
$\mu$, $\bar\la$ and $\be_0$ such that, for all 
$\la<\bar\la$ and $\be_0\le\be\le\sfrac{1}{\la^{\up/5}}$, 
the following holds:

\noindent
Let $X \in \fN_{d+1}$ with $X_\0<\mu$, $\de e\in\cE$ with
$ \|\de\hat e\|_{1,\infty}\le X$ and
$$
\cV(\psi) = \int_{\cB^4} {\sst d\xi_1 \cdots d\xi_4} \,V({\sst \xi_1,\cdots,\xi_4})
\,\psi({\sst \xi_1})\cdots \psi({\sst \xi_1})
$$
with an antisymmetric function $V$ fulfilling 
$$
\|V\|_{1,\infty} \le \la \,\fe_0(X)
$$
Write
$$\eqalign{
&\tilde\Om_{C_{-\de e}^{(\le j_0)}}\big(\cV{\sst(\psi)}\big) (\phi,\psi) 
= \cV(\psi) +\half\phi JC^{(\le j_0)}_{-\de e}J\phi\cr 
& \hskip 1.3cm+ \smsum_{m,n\ge 0\atop m+n\ {\rm even}} 
\int_{\cB^{m+n}}\hskip-15pt {\sst d\et_1\cdots d\et_m\ d\xi_1\cdots d\xi_n}\ 
W_{m,n}({\sst\et_1,\cdots ,\et_m,\xi_1,\cdots,\xi_n};\de e)\,
\phi{\sst(\et_1)}\cdots\phi{\sst(\et_m)}\,\psi{\sst(\xi_1)}\cdots{\sst\psi(\xi_n)}
}$$
with kernels $W_{m,n}$ that are separately antisymmetric under 
permutations of their $\et$ and $\xi$ arguments. Then
$$
\smsum_{m+n \ge 2 \atop m+n\ {\rm even}} \be^n \,\rho_{m;n}(\la)\,
\| W_{m,n}(\de e)\|_{1,\infty} \le  \const\be^3\la^\upsilon\,\fe_0(X) 
$$
and
$$
\smsum_{m+n \ge 2 \atop m+n\ {\rm even}} \be^n \,\rho_{m;n}(\la)\,
\big\|\sfrac{d\hfill}{ds} W_{m,n}(\de e+s\de e')\big|_{s=0}\big\|_{1,\infty} 
\le  \const\be^3\la^\upsilon\,\fe_0(X)\ \|\de\hat e'\|_{1,\infty}
$$
Furthermore, each $W_{m,n}$ is jointly analytic\footnote{$^{(1)}$}
{As in the discussion leading up to Theorem \theoremNPmainthI,
the $W_{m,n}$'s are initially defined as formal Taylor series in $V$.
The conclusions of the Theorem implicitly include the convergence of 
the formal Taylor series for $V$ obeying $\|V\|_{1,\infty} \le \la \,\fe_0(X)$ and $\de e$ obeying $ \|\de\hat e\|_{1,\infty}\le X$.} in $V$ and $\de  e$.
If $V$ fulfills the reality condition of (\eqnNPreal) and 
$\de e(\k)$ is real valued, then 
$\tilde\Om_{C_{-\de e}^{(\le j_0)}}\big(\cV{\sst(\psi)}\big) (\phi,\psi)$ is
$k_0$--reversal real, in the sense of Definition \defOSsymmetries.R of [FKTo2].

}

\prf 
Apply Theorem \thmOSfirststep\ of [FKTo2] with $\rho_{m,n}=\rho_{m;n}(\la)$  and $\veps=\abcst\,\be^4\la^\upsilon$.
Observe that, by Remark \remOSthmV.iii of [FKTo2], the hypotheses on 
$\rho_{m;n}$  are fulfilled. If $\bar\la$ is chosen small enough, then 
the hypothesis $\veps<\veps_0$ is also fulfilled.The reality statement is 
a consequence of Remark \remOSrengrppreserves\ of [FKTo2].

\endproof

In Theorem \thmNPfirststep, we integrated out the part of the field $\psi$ with covariance $C^{(\le j_0)}$. To recover the full, infrared cutoff covariance
$C^{\IR(\il)}$ of Theorem \theoremNPmainthI, we must also integrate out the
part of the field with covariance
$$
C^{(i,j)}_u(k) 
= \frac{\nu^{(> i)}(k)-\nu^{(\ge j)}(k)}{\imath k_0 - e(\k)
                -\check u(k)[1-\nu^{(\ge\il)}(k)]}
$$

\lemma{\STM\lemNPfirststep}{
Let $\il\ge j_0+2$ be an infrared cutoff. For  $\tn V\tn_{1,\infty}$
and $\tn \de e\tn_{1,\infty}$ sufficiently small
$$
\tilde\Om_{C^{(j_0,\il)}_{-\de e}}\Big( \tilde\Om_{C^{(\le j_0)}_{-\de e}}
\big(\cV{\sst(\psi)}\big)   \Big)
=\tilde\Om_{C^{\IR(\il)}(\de e)}\big(\cV{\sst(\psi)}\big) 
$$

}
\prf We just apply the semi--group property (\eqnNPsemigrp) using
$$
C^{\IR(\il)}(\de e)
=C^{(\le j_0)}_{-\de e}+C^{(j_0,\il)}_{-\de e}
$$
\endproof

\vfill\eject

\chap{ Sectors and Sectorized Norms}\PG\pgNPVI

From now on we consider only $d=2$, so that the Fermi ``surface'' is a curve
in $\bbbr\times\bbbr^2$. We choose a projection $\pi_F$ from the first extended
neighbourhood onto the Fermi surface. 

\vskip .3cm \noindent
{\bf Convention:} Generic constants that depend only on the dispersion 
relation $e(\k)$ and the ultraviolet cutoff $U(\k)$ will be denoted by 
``$\abcst$". Generic constant that may also depend on the scale parameter $M$,
but still not on the scale $j$, will be denoted ``$\const$".

To systematically deal with Fourier transforms, we call
$$
\check \cB=\bbbr\times\bbbr^d\times\{\uparrow,\downarrow\}\times\{0,1\}
$$
``momentum space''. For 
$ \check\xi = (k,\si',a') = (k_0,\k,\si',a') \in \check\cB$ and
$\xi = (x,a) = (x_0,\x,\si,a) \in \cB$ we define the inner product
$$
\<\check\xi,\xi\>\ =\ \de_{\si',\si} \de_{a',a}\,(-1)^a\,\<k,x\>_-\  
=\ \de_{\si',\si} \de_{a',a}\,(-1)^a\, \big(- k_0 x_0+\k_1\x_1+\cdots+\k_d\x_d \big)
$$
 ``characters''
$$\deqalign{
E_+(\check\xi,\xi)\ &=\ \de_{\si',\si} \de_{a',a}e^{\imath\<\check\xi,\xi\>}
&=\ \de_{\si',\si}\de_{a',a}e^{\imath(-1)^a
                          \big(-k_0 x_0+\k_1\x_1+\cdots+\k_d\x_d \big)}\cr
E_-(\check\xi,\xi)\ &=\ \de_{\si',\si} \de_{a',a}e^{-\imath\<\check\xi,\xi\>}
&=\ \de_{\si',\si}\de_{a',a}e^{-\imath(-1)^a
                            \big(-k_0 x_0+\k_1\x_1+\cdots+\k_d\x_d \big)}\cr
}$$
and integrals
$$
\int d\xi\ \cdot\ = \sum_{a\in\{0,1\}\atop\si\in\{\uparrow,\downarrow\}}
\int_{\bbbr\times\bbbr^d} dx_0\, d^d\x\ \cdot\ \qquad\qquad
\int d\check\xi\ \cdot\ = \sum_{a\in\{0,1\}\atop\si\in\{\uparrow,\downarrow\}}
\int_{\bbbr\times\bbbr^d} dk_0\, d^d\k\ \cdot\ 
$$
For $\check\xi = (k,\si,a), \,\check\xi' = (k',\si',a') \in \check \cB$ we set
$$
\check\xi+\check\xi' = (-1)^a\,k + (-1)^{a'}\,k'\ \in\bbbr \times \bbbr^d
$$

\definition{\STM\defNPfourtrans (Fourier transforms)}{
Let $f({\sst \eta_1,\cdots,\eta_m;\,\xi_1,\cdots,\xi_n})$
be a translation invariant function on $\cB^m \times \cB^n$.
The total Fourier transform $\check f$ of $f$ is defined by 
$$\eqalign{
&\check f({\sst
\check\eta_1,\cdots,\check\eta_m;\,\check\xi_1,\cdots,\check\xi_n }) 
\ (2\pi)^{d+1}\de({\sst \check\eta_1+\cdots+\check\eta_m+\check\xi_1+\cdots+\check\xi_n}) \cr
&\hskip2in= \int \smprod_{i=1}^m E_+(\check\eta_i,\eta_i)\,d\et_i\,
\smprod_{j=1}^n E_+(\check\xi_j,\xi_j)\,d\xi_j\,
f({\sst \eta_1,\cdots,\eta_m;\,\xi_1,\cdots,\xi_n }) 
\cr 
}$$ 
or, equivalently, by 
$$\eqalign{
&f({\sst \eta_1,\cdots,\eta_m;\,\xi_1,\cdots,\xi_n }) \cr
&= \int \smprod_{i=1}^m
\sfrac{ E_-(\check\eta_i,\eta_i)\,d\check\et_i}{(2\pi)^{d+1}}\,
\smprod_{j=1}^n \sfrac{E_-(\check\xi_j,\xi_j)\,d\check\xi_j }{(2\pi)^{d+1}}\,
\check f({\sst
\check\eta_1,\cdots,\check\eta_m;\,\check\xi_1,\cdots,\check\xi_n }) 
\ (2\pi)^{d+1}\de({\sst \check\eta_1+\cdots+\check\eta_m+\check\xi_1+\cdots+\check\xi_n}) \cr 
}$$ 
$\check f$ is defined on the set
$ \set{ ({\sst
\check\eta_1,\cdots,\check\eta_m;\,\check\xi_1,\cdots,\check\xi_n }) \in
\check \cB^m \times \cB^n }
{ {\sst \check\eta_1+\cdots+\check\eta_m+\check\xi_1+\cdots+\check\xi_n=0}}$.

If $m=0$, $n=2$ and $f({\sst \xi_1,\xi_2 })$ conserves particle number and is spin independent and antisymmetric, we define $\check f(k)$ by
$$
\check f({\sst (k,\si,1),(k,\si',0)}) =\de_{\si,\si'}\check f(k)
$$ 

}

\vskip.3cm
We now introduce sectors. 
\definition{\STM\defNPsectors (Sectors and sectorizations)}{ 
\Item i)
Let $I$ be an interval on the Fermi surface $F$ and $j\ge 2$. Then
$$
s=\set{k\ {\rm in\ the\ } j^{\rm th}\ {\rm neighbourhood}}{\pi_F(k)\in I}
$$
is called a sector of length $|I|$ at scale $j$. 
Two different sectors $s$ and $s'$ are called neighbours if  
$s'\cap s\ne \emptyset$.
\Item ii)
If $s$ is a sector at scale $j$, its extension is 
$$
\tilde s=\set{k\ {\rm in\ the\ } j^{\rm th}\ {\rm extended\ neighbourhood}}
{\pi_F(k)\in s}
$$
\Item iii)
A sectorization of length $\fl$ at scale $j$ is a set $\Si$
of sectors of length $\fl$ at scale $j$ that obeys 
\item{-} the set $\Si$ of sectors covers the Fermi surface
\item{-} each sector in $\Si$ has precisely two neighbours in $\Si$, 
one to its left and one to its right
\item{-} if $s,\ s'\in\Si$ are neighbours then 
$\sfrac{1}{16}\fl\le |s\cap s'\cap F|\le\sfrac{1}{8}\fl$

\noindent Observe that there are at most 
$2\,{\rm length}(F)/\fl$ sectors in $\Si$.

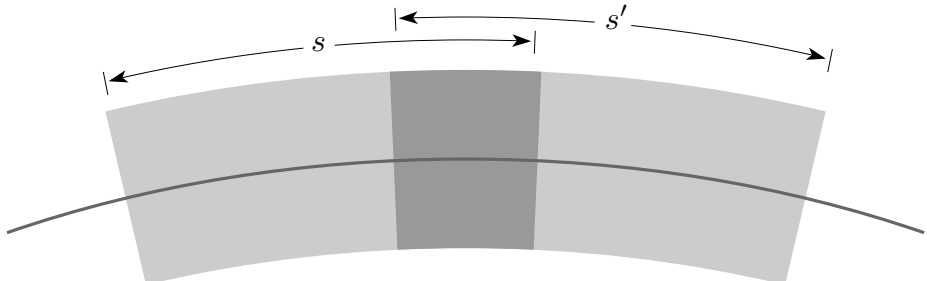
\centerline{\figput{sector2}}
}

\definition{\STM\defNPsectrepr (Sectorized representatives)}{
Let $\Si$ be a  sectorization at scale $j$, and let $m,n \ge 0$. 
\Item{i)}
The antisymmetrization of a function $\varphi$ on 
$\cB^m\times \big(\cB \times \Si \big)^n$ is
$$
{\rm Ant}\, \varphi ({\sst \eta_1,\cdots,\eta_m;\,(\xi_1,s_1),\cdots,(\xi_n,s_n)})
=\sfrac{1}{m!\,n!} \smsum_{\pi \in S_m \atop \pi'\in S_n} 
\varphi({\sst
\eta_{\pi(1)},\cdots,\eta_{\pi(m)};
\,(\xi_{\pi'(1)},s_{\pi'(1)}),\cdots,(\xi_{\pi'(n)},s_{\pi'(n)})})
$$
\Item{ii)} 
Denote by  $\cF_m(n;\Si)$ the space
of all translation invariant, complex valued functions 
$$
\varphi({\sst\eta_1,\cdots,\eta_m;\,(\xi_1,s_1),\cdots,(\xi_n,s_n)} )
$$
on $\cB^m \times  \big( \cB \times\Si \big)^n$ that are antisymmetric in their 
external ($=\eta$) variables and whose Fourier transform
$\check\varphi({\sst\check\eta_1,\cdots,\check\eta_m;
\,(\check\xi_1,s_1),\cdots,(\check\xi_n,s_n)} )$
vanishes unless 
$ k_i\in \tilde s_i$ for all $1\le j\le n$. 
Here, $\check\xi_i=(k_i,\si_i,a_i)$.
\Item{iii)}
Let $f$ be a translation invariant, complex valued function
on $\cB^m \times  \cB^n$ that is antisymmetric in its first $m$ variables.
A $\Si$--sectorized representative for $f$ is a function
$\varphi\in\cF_m(n;\Si)$ obeying
$$
\check f({\sst\check\eta_1,\cdots,\check\eta_m;\,\check\xi_1,\cdots,\check\xi_n }) 
=\smsum_{s_i\in\Si \atop i=1,\cdots,n} \check \varphi
({\sst\check\eta_1,\cdots,\check\eta_m;
\,(\check\xi_1,s_1),\cdots,(\check\xi_n,s_n) } )
$$
for all $\check\xi_i=(k_i,\si_i,a_i)$ with $k_i$ in the $j^{\rm th}$ neighbourhood. 
\Item{iv)} Let $u({\sst(\xi,s)},{\sst(\xi',s')})$ be a translation invariant, spin independent, particle number conserving function on
$(\cB\times\Si)^2$. We define $\check u(k)$ by
$$
\de_{\si,\si'}\check u(k)=
\sum_{s,s'\in\Si}\check u({\sst(k,\si,1,s)},{\sst(k,\si',0,s')})
$$

}

We now fix a constant $\sfrac{1}{2} < \aleph < \sfrac{2}{3}$, and for each scale
$j\ge 2$, a sectorization $\Si_j$ of length $\fl_j=\sfrac{1}{M^{\aleph j}}$.
Also, we fix for each $j\ge 2$, a system $\chi_s(k),\ s\in \Si_j$ of functions
that take values in $[0,1]$ such that
\Item{i)} $\chi_s$ is supported in the extended sector $\tilde s$ and
$$
\smsum_{s\in \Si} \chi_s(k) =1 \qquad {\rm for\ } k\ {\rm in\ the\ } j^{\rm th}\
{\rm neighbourhood}
$$
\Item{ii)}
$$
\|\hat\chi_s\|_{1,\infty},\ 
\le\  \abcst\,\cb_{j-1}
$$
with a constant $\abcst$ that does not depend $M$, $j$,  or $s$. The existence of
such a ``partition of unity'' is shown in Lemma \lemOSsectpartunit\ of [FKTo3]. They are used to construct sectorized representatives.

\definition{\STM\defNPresector}{ Let $j,i\ge 2$. If $i\ne j$, define, for
functions $\varphi$ on $\cB^m\times\big(\cB\times\Si_i\big)^n$ and $f$ on
$\check\cB^m\times\big(\cB\times\Si_i\big)^n$,
$$\eqalign{
\varphi_{\Si_j}({\sst\eta_1,\cdots,\eta_m;\,(\xi_1,s_1),\cdots,(\xi_n,s_n)} )
&=\smsum_{s'_1,\cdots,s'_{n} \in \Si_i} \int {\sst d\xi'_1\cdots d\xi'_{n}}\,
\varphi({\sst\eta_1,\cdots,\eta_m;\,(\xi'_1,s'_1),\cdots,(\xi'_n,s'_n)} )
\smprod_{\ell=1}^n\hat\chi_{s_\ell}(\xi'_\ell,\xi_\ell) \cr
f_{\Si_j}({\sst\check\eta_1,\cdots,\check\eta_m;\,(\xi_1,s_1),\cdots,(\xi_n,s_n)} )
&=\smsum_{s'_1,\cdots,s'_{n} \in \Si_i} \int {\sst d\xi'_1\cdots d\xi'_{n}}\,
f({\sst\check\eta_1,\cdots,\check\eta_m;\,(\xi'_1,s'_1),\cdots,(\xi'_n,s'_n)} )
\smprod_{\ell=1}^n\hat\chi_{s_\ell}(\xi'_\ell,\xi_\ell) \cr
}$$
If $\varphi$ is antisymmetric
under permutation of its $\et$ arguments, then $\varphi_{\Si_j}\in\cF_m(n,\Si_j)$. 
For $i= j$ define $\varphi_{\Si_j}=\varphi$ and $f_{\Si_j}=f$.

\noindent 
Similarly, define, for
functions $\varphi$ on $\cB^m\times \cB^n$ and $f$ on
$\check\cB^m\times \cB^n$,
$$\eqalign{
\varphi_{\Si_j}({\sst\eta_1,\cdots,\eta_m;\,(\xi_1,s_1),\cdots,(\xi_n,s_n)} )
&= \int {\sst d\xi'_1\cdots d\xi'_{n}}\,
\varphi({\sst\eta_1,\cdots,\eta_m;\,\xi'_1,\cdots,\xi'_n} )
\smprod_{\ell=1}^n\hat\chi_{s_\ell}(\xi'_\ell,\xi_\ell) \cr
f_{\Si_j}({\sst\check\eta_1,\cdots,\check\eta_m;\,(\xi_1,s_1),\cdots,(\xi_n,s_n)} )
&= \int {\sst d\xi'_1\cdots d\xi'_{n}}\,
f({\sst\check\eta_1,\cdots,\check\eta_m;\,\xi'_1,\cdots,\xi'_n} )
\smprod_{\ell=1}^n\hat\chi_{s_\ell}(\xi'_\ell,\xi_\ell) \cr
}$$
They are $\Si_j$--sectorized representatives for $\varphi$ resp. $f$.
}

\definition{\STM\defNPtens}{
Let $j\ge 2$ be a scale.
We consider fermionic fields $\phi(\eta),\,\eta\in\cB$ and $\psi({\sst \xi,s})$, 
$\xi \in\cB,\,s\in\Si_j$.
\Item{i)}
A $\Si_j$--sectorized Grassmann function is of the form
$$\eqalign{
w &=\smsum_{m,n \ge0} \smsum_{s_1,\cdots,s_n \in \Si_j}
 \int \smprod_{i=1}^m {\sst d\eta_i}\, \smprod_{j=1}^n {\sst d\xi_j}\ 
w_{m,n}({\sst\eta_1,\cdots,\eta_m;\,(\xi_1,s_1),\cdots,(\xi_n,s_n)})\cr
\noalign{\vskip-.1in}
 &\hskip3in\phi{\sst( \eta_1)}\cdots\phi{\sst(\eta_m)}\ 
\psi{\sst(\xi_1,s_1)} \cdots  \psi{\sst (\xi_n,s_n)}
}\EQN\eqnNPsectGrf$$
\Item{ii)}
Let 
$$
\cW =\smsum_{m,n \ge0} \int \smprod_{i=1}^m d\eta_i\, \smprod_{j=1}^n d\xi_j\ 
W_{m,n}({\sst\eta_1,\cdots,\eta_m;\, \xi_1,\cdots,\xi_n })\,
 \phi({\sst \eta_1})\cdots\phi({\sst\eta_m})\ 
\psi({\sst\xi_1}) \cdots  \psi({\sst \xi_n})
$$
be a Grassmann function with each $W_{m,n}$ a function on $\cB^m\times\cB^n$
that is separately antisymmetric in its external ($=\phi$) variables and in its internal ($=\psi$) variables. A $\Si_j$--sectorized representative for 
$\cW$ is a $\Si_j$--sectorized Grassmann function of the form (\eqnNPsectGrf),
where, for each $m,n$, $\ w_{m,n}$ is a $\Si_j$--sectorized representative for
$W_{m,n}$ that is also antisymmetric in the variables
$(\xi_1,s_1),\cdots,(\xi_n,s_n)$. }

\definition{\STM\defNPsectnorm(Norms for sectorized functions)}{
Let $j\ge 2$ and $m,\,n \ge 0$.

\Item{i)}
For a function $\varphi$ on $\cB^m\times (\cB\times\Si_j)^n$ and an integer $p>0$ we
define the seminorm
$\v \varphi\v_{p,\Si_j}$ to be zero 
if $m\ge 1$, $p\ge 2$ or if $m=0,\ p>n$.

\noindent
In the case $m \ge 1$, $p=1$ we set
$$
\v \varphi\v_{p,\Si_j} 
= \smsum_{s_i \in \Si_j}\, 
\|\varphi({\sst\eta_1,\cdots,\eta_m;\,(\xi_1,s_1),\cdots,(\xi_n,s_n)})
\|_{1,\infty} 
$$
 
\noindent
In the case $m =0$, $p\le n$ we set
$$
\v \varphi\v_{p,\Si_j}\ 
=\max_{1\le i_1<\cdots<i_p\le n}\ \ 
 \max_{s_{i_1},\cdots,s_{i_p} \in \Si_j} \ \ 
 \sum_{s_i \in \Si_j \ {\rm for} \atop i \ne i_1,\cdots,i_p} 
 \|\varphi({\sst (\xi_1,s_1),\cdots,(\xi_n,s_n)} ) \|_{1,\infty}
$$

\noindent
In both cases, the $\|\cdot\|_{1,\infty}$ norm applies to all the
position space variables.

\Item{ii)}  
We shall fix a $\la_0>0$, sufficiently small depending on $j_0$ 
and the scale parameter $M$.
For $\varphi\in \cF_m(n,\Si_j)$ set
$$
\v \varphi \v_j = \rho_{m;n}\cases{
\v \varphi \v_{1,\Si_j} + \sfrac{1}{\fl_j}\,\v \varphi \v_{3,\Si_j}
+ \sfrac{1}{\fl_j^2}\,\v \varphi \v_{5,\Si_j} 
    & if $m=0$ \cr
\sfrac{\fl_j}{M^{2j}}\,\v \varphi \v_{1,\Si_j} & if $m\ne0$ \cr
}$$
where 
$$
\rho_{m;n}=\rho_{m;n}^{(j)}
=\frac{1}{\la_0^{(1-\upsilon)\max\{m+n-2,2\}/2}}\cases{1 & if $m=0$\cr
\noalign{\vskip.05in}
      \root{4}\of{\fl_jM^j}& if $m>0$\cr}
$$
and $\upsilon$ was fixed in Definition \defNPrhomn.
}

\definition{\STM\defNPsectGrnorm(Norms for sectorized Grassmann functions)}{
\Item{i)}
A $\Si_j$--sectorized Grassmann function $w$ can be uniquely written in the form
$$\eqalign{
w(\phi,\psi) = \smsum_{m,n}\ \smsum_{s_1,\cdots,s_n\in\Si_j}\
\int {\sst d\eta_1\cdots d\eta_m\,d\xi_1\cdots d\xi_n}\ & 
w_{m,n}({\sst \eta_1,\cdots, \eta_m\,(\xi_1,s_1),\cdots ,(\xi_n,s_n)})\cr
& \hskip 1cm \phi({\sst \eta_1})\cdots \phi({\sst \eta_m})\
\psi({\sst (\xi_1,s_1)})\cdots \psi({\sst (\xi_n,s_n)\,})\cr
}$$
with $w_{m,n}$ antisymmetric separately in the $\eta$ and in the $\xi$ variables.
Set, as in Definition \defOSscalednorms\  of [FKTo3], for $\al >0$
and $X\in \fN_{d+1}$,
$$
N_j(w,\al,\,X)
=\sfrac{M^{2j}}{\fl_j}\,\fe_j(X) 
\smsum_{m,n\ge 0}\,
\al^{n}\,\big(\sfrac{\fl_j\,\IB}{M^j}\big)^{n/2} \,\v w_{m,n}\v_j 
$$
The constant $\IB$ depends on $M$, but not $j$ and was specified in Definition
\defOSscalednorms\ of [FKTo3].

\Item{ii)} 
A Grassmann function $\cG(\phi)$ can be uniquely written in the form
$$
\cG(\phi) = \smsum_{m} 
\int {\sst d\eta_1\cdots d\eta_m}\ 
G_{m}({\sst \eta_1,\cdots, \eta_m})
\ \phi({\sst \eta_1})\cdots \phi({\sst \eta_m})
$$
with $G_{m}$ antisymmetric.
Set
$$
N(\cG)
=\smsum_{m> 0} \sfrac{1}{\la_0^{(1-\upsilon)\max\{m-2,2\}/2}}\ 
\tn G_{m}\tn_\infty
$$
where
$\ 
\tn G_{m}\tn_\infty=\sup_{\eta_1,\cdots, \eta_m}
\big| G_{m}({\sst \eta_1,\cdots, \eta_m})\big|
\ $.

}

\remark{\STM\remNPnvsnj}{
\Item i)
The system $\vec \rho = (\rho_{m;n})$ of Definition \defNPsectnorm.ii
fulfill the inequalities (\eqnOSrhomn) of [FKTo3].

\Item ii) 
If $w(\phi,\psi)$ is a $\Si_j$--sectorized Grassmann function, 
then
$$
N\big(w(\phi,0)\big)\le \sfrac{1}{\root{4}\of{\fl_jM^j}}N_j(w,\al,\,X)
$$
for all $\al$ and $X\in\fN_{d+1}$.

\Item iii) 
The $j$ independent part of the coefficient of 
$\v w_{m,n}\v_{1,\Si_j} $
in $N_j(w,\al,X)$ is, up to a factor of $\IB^{n/2}$, equal to
  $\sfrac{\al^n }{\la_0^{(1-\upsilon)\max\{m+n-2,2\}/2}}$. This choice was 
motivated in Remark \remNPrhomn. 

\Item iv) 
If $N_j(w,\al,\,X)_\0\le 1$, then, up to $\smsum_{\de\ne \0}\infty\, t^\de$,
$$
\v w_{m,n}\v_{1,\Si_j}\le\cases{
\sfrac{1}{\fl^{{n\over2}-1}_j}M^{j({n\over 2}-2)}
&if $m=0$\cr
\sfrac{1}{\root 4\of {M^j\fl_j}}\big(\sfrac{M^j}{\fl_j}\big)^{n/2}
&if $m\ne 0$ }
\EQN\eqnNPnormmotivate$$
The case $m=0$ was motivated in (\eqnOVsectorvertexreq). 
Next consider the case that $m+n=4$, $m,n\ge 1$ and $w_{m,n}$ is the coefficient of $\phi^m\psi^n$ in $\cV(\phi+\psi)$.
Then, allowing a full sector sum for each $\psi$ leg, $\tn V\tn_{1,\infty}<\infty$ 
implies that $\v w_{m,n}\v_{1,\Si_j}=O\big(\sfrac{1}{\fl_j^n}\big)$, which is a tighter bound than (\eqnNPnormmotivate).
An argument similar to that in subsection 8 of \S\CHintroOverview\ may also be used to show that if $w'_{m,n}$ is a graph with vertices obeying (\eqnNPnormmotivate), then
$w'_{m,n}$  obeys a bound of the same order as (\eqnNPnormmotivate).

}

We now define the space of functions from which the various counterterm
kernels will be chosen, using the concepts in
Definition \defOSzerosectorext\ of [FKTo4].

\definition{\STM\defNPCTmSpace}{ 
Let $\fK_j$ be the space of all 
translation invariant, sectorized functions $K\big((\x,s),(\x',s')\big)$
on $\big(\bbbr^2\times\Si_{j}\big)^2$ for which
\item{i)}
$\ 
\| K\|_{1,\Si_{j}} < \la_0^{1-\upsilon} \sfrac{\fl_{j+1}}{M^{j+1}}
+\smsum_{\de\ne \0}\infty\, t^\de
\ $
\item{ii)} the Fourier transform $\check K(\k)$ 
is supported on ${\rm supp}\,\nu^{(\ge j+1)}\big((0,\k)\big)$

\noindent The counterterm $K$ is said to be {\bf real} if,
for each $s,s'\in\Si_j$, the Fourier transform 
$\check K\big((\k,s),(\k',s')\big)$ is real valued.

}

\remark{\STM\remNPInclusionK}{ 
If $K\in \fK_{j}$, then $K_{\Si_{j-1}}\in\fK_{j-1}$.
To see this, observe that, by Lemma \propOSresectorI\ of [FKTo4],
$$
\| K_{\Si_{j-1}}\|_{1,\Si_{j-1}}
\le \abcst \sfrac{\fl_{j-1}}{\fl_{j}}\cb_{j-2}\| K\|_{1,\Si_{j}}
< \abcst \sfrac{\fl_{j-1}}{\fl_{j}}\la_0^{1-\upsilon} \sfrac{\fl_{j+1}}{M^{j+1}}
+\smsum_{\de\ne \0}\infty\, t^\de
< \la_0^{1-\upsilon} \sfrac{\fl_{j}}{M^{j}}
+\smsum_{\de\ne \0}\infty\, t^\de
$$
if $M$ is large enough.
}

\remark{}{ 
The final counterterm $\de e(\k)$ will be constructed in Theorem 
\theoremNPinduction\ using bounds proven in Lemma \lemNPprojSysBnds.

}

\noindent As in Definition \defNPformalCovariances, Remark \remNPinoutmap\ 
and Lemma \lemNPfirststep, we have the following covariances.

\definition{\STM\defNPCovariances}{ 
\item{(i)} 
Let $u({\sst(\xi,s)},{\sst(\xi',s')})$ be a translation invariant, spin independent, particle number conserving function on
$(\cB\times\Si_\ell)^2$.  Then
$$\eqalign{
C_u^{(j)}(k)&=\frac{\nu^{(j)}(k)}{ik_0-e(\k)-\check u(k)}\cr
C_u^{(\ge j)}(k)&=\frac{\nu^{(\ge j)}(k)}{ik_0-e(\k)-\check u(k)}\cr
C^{[i,j)}_u(k) 
&= \frac{\nu^{(\ge i)}(k)-\nu^{(\ge j)}(k)}{\imath k_0 - e(\k)
                -\check u(k)[1-\nu^{(\ge\il)}(k)]}
}$$

\item{(ii)} Let $u$ be a function from $\fK_{j}$ to the space 
of antisymmetric, 
translation invariant, spin independent, particle number conserving 
functions on $(\cB\times\Si_j)^2$. Then, for $K\in\fK_{j}$,
$$\eqalign{
C_j(u;K)(k) &= 
\sfrac{\nu^{(\ge j)}(k)}
{\imath k_0 -e(\k) -\check u(k;K) - \check  K(\k)\nu^{(\ge j+2)}(k)}\cr
D_j(u;K)(k) &= \sfrac{\nu^{(\ge j+1)}(k)}
{\imath k_0 -e(\k) -\check u(k;K)-\check  K(\k)\nu^{(\ge j+2)}(k)}\cr
}$$

\item{(iii)}
Let $C^{(j)}_u(\xi,\xi')$, $C^{(\ge j)}_u(\xi,\xi')$, $C^{[i,j)}_u(\xi,\xi')$,
$D_j(u;K)(\xi,\xi')$ and $C_j(u;K)(\xi,\xi')$ be their Fourier transforms as 
in (\eqnNPcovFT) and (\eqnNPantisymmCov).
}

\noindent To start the recursive construction of the Green's functions,
we reformulate Theorem \thmNPfirststep\ in terms of sectorized objects.

\theorem{\STM\thmNPsetupinduction }{
For $K\in \fK_{j_0}$, set 
$$
u(K)= - \big[K_{\rm ext}\big]_{\Si_{j_0}} \in \cF_0(2,\Si_{j_0})
$$
where $K_{\rm ext}$ was defined in Definition \defOSzerosectorext\ of [FKTo4]\footnote{$^{(1)}$}%
{By Remark \remOSzerosectorext.i of [FKTo4], under this definition,
$\check K_{\rm ext}\left((k_0,\k)\right)=\check K(\k)$.}. 
Then there exist constants $\bar\la,\bar\al>0$ such that for all 
$0\le \la_0 <\bar\la$, $\bar\al<\al<\sfrac{1}{\la_0^{\upsilon/10}}$ and all
$$
K\in\fK_{j_0}
 \qquad
\|V\|_{1,\infty}\le \la_0\fe_{j_0}\big(\|K\|_{1,\Si_{j_0}}\big)
$$
the Grassmann function
$$
\tilde\Om_{C_{u(K)}^{(\le j_0)}}\big(\cV{\sst(\psi)}\big) (\phi,\psi) 
 -\half\phi JC^{(\le j_0)}_{u(K)}J\phi
$$
has a $\Si_{j_0}$--sectorized representative 
$$\eqalign{
w(\phi,\psi;K) &= \smsum_{m,n}\ \smsum_{s_1,\cdots,s_n\in\Si_{j_0}}\
\int {\sst d\eta_1\cdots d\eta_m\,d\xi_1\cdots d\xi_n}\ 
w_{m,n}({\sst \eta_1,\cdots, \eta_m\,(\xi_1,s_1),\cdots ,(\xi_n,s_n)};K)\cr
& \hskip 5cm \phi({\sst \eta_1})\cdots \phi({\sst \eta_m})\
\psi({\sst (\xi_1,s_1)})\cdots \psi({\sst (\xi_n,s_n)\,})\cr
}$$
with $w_{m,n}$ antisymmetric separately in the $\eta$ and in the $\xi$ variables, 
$w_{0,0}=0$ and 
$$\eqalign{
N_{j_0}\big(w(K),\al,\|K\|_{1,\Si_{j_0}}\big) 
&\le \const\,\al^4\la_0^\upsilon\,\fe_{j_0}\big(\|K\|_{1,\Si_{j_0}}\big) \cr
N_{j_0}\big(\sfrac{d\hfill}{ds}w(K+sK')\big|_{s=0}
,\al,\,\| K\|_{1,\Si_{j_0}}\big) 
   &\le M^{j_0} \,\fe_{j_0}\big(\|K\|_{1,\Si_{j_0}}\big) \,  \|K'\|_{1,\Si_{j_0}}
\cr
}$$
for all $ K'$. $w$ is analytic in $V$ and $K$.
If $V$ fulfills the reality condition of (\eqnNPreal) and 
$K$ is real, then $w(\phi,\psi;K)$ is
$k_0$--reversal real, in the sense of Definition \defOSsymmetries.R of [FKTo2].
}

\prf
Write
$$\eqalign{
&\tilde\Om_{C_{u(K)}^{(\le j_0)}}\big(\cV{\sst(\psi)}\big) (\phi,\psi) 
= \cV(\psi) +\half\phi JC^{(\le j_0)}_{u(K)}J\phi\cr 
& \hskip 1.5cm+ \smsum_{m,n\ge 0\atop m+n\ {\rm even}} 
\int_{\cB^{m+n}}\hskip-15pt {\sst d\et_1\cdots d\et_m\ d\xi_1\cdots d\xi_n}\ 
W_{m,n}({\sst\et_1\cdots \et_m,\xi_1,\cdots,\xi_n})\,
\phi{\sst(\et_1)}\cdots\phi{\sst(\et_m)}\,\psi{\sst(\xi_1)}\cdots{\sst\psi(\xi_n)}
}$$
and set
$$
w_{m,n}=\cases{\big(W_{m,n}\big)_{\Si_{j_0}}& if $(m,n)\ne(0,4)$\cr
\noalign{\vskip0.05in}
                \big(W_{0,4}+V\big)_{\Si_{j_0}}& if $(m,n)=(0,4)$\cr}
$$
using the sectorization $f_\Si$ of Definition \defOScreateSectoriz\ of [FKTo4]. 
By Proposition \propOScreateSectoriz\ of [FKTo4]
$$\eqalign{
N_{j_0}\big(w(K),&\al,\|K\|_{1,\Si_{j_0}}\big) 
=\sfrac{M^{2{j_0}}}{\fl_{j_0}}\,\fe_{j_0}\big(\|K\|_{1,\Si_{j_0}}\big) 
\smsum_{m,n\ge 0}\,
\al^{n}\,\big(\sfrac{\fl_{j_0}\,\IB}{M^{j_0}}\big)^{n/2} \,\v w_{m,n}\v_{j_0} 
\cr
&\le\const\,\cb_{j_0}\,\fe_{j_0}\big(\|K\|_{1,\Si_{j_0}}\big) \Big[
\sfrac{\al^4}{\la_0^{1-\upsilon}}\,\| V\|_{1,\infty}+
\smsum_{m,n\ge 0}\,(\const \al)^{n}\,\rho_{m;n}(\la_0)\,\| W_{m,n}\|_{1,\infty} \Big]
}$$
since $\rho_{m;n}^{(j_0)}\le\const \rho_{m;n}(\la_0)$.
By hypothesis
$$
\sfrac{\al^4}{\la_0^{1-\upsilon}}\,\| V\|_{1,\infty}
\le \al^4\la_0^{\upsilon}\fe_{j_0}\big(\|K\|_{1,\Si_{j_0}}\big)
$$
and by Theorem \thmNPfirststep, with  $\de e=-\check u$, $X=\const \|K\|_{1,\Si_{j_0}}$ and $\be=\const\al$,
$$
\smsum_{m,n\ge 0}\,(\const \al)^{n}\,\rho_{m;n}(\la_0)\,\| W_{m,n}\|_{1,\infty}
\le\const \be^3\la^{\upsilon}\fe_0(X)
\le\const\al^3\la_0^\upsilon\, \fe_{j_0}\big(\|K\|_{1,\Si_{j_0}}\big)
$$
Therefore, by Corollary \corOSappMonoidIV.ii of [FKTo1],
$$\eqalign{
N_{j_0}\big(w(K),\al,\|K\|_{1,\Si_{j_0}}\big) 
&\le\const\,\al^4\la_0^\upsilon\,\cb_{j_0}\,\fe_{j_0}\big(\|K\|_{1,\Si_{j_0}}\big)^2\cr
&\le\const\,\al^4\la_0^\upsilon\,\fe_{j_0}\big(\|K\|_{1,\Si_{j_0}}\big)\cr
}$$
The proof of the bound on $N_{j_0}\big(\sfrac{d\hfill}{ds}w(K+sK')\big|_{s=0}
,\al,\,\| K\|_{1,\Si_{j_0}}\big)$ is similar.
\endproof

\vfill\eject

\chap{Ladders}\PG\pgNPVII

In naive power counting for our model, four-legged vertices are neutral. 
So there is a danger that the size of four legged kernels after $j$ steps of the
renormalization group flow is of order $j$. We shall show that this logarithmic
divergence does not occur. More precisely, let $(\cW,u,\cG)\in D^{(j)}_{\rm in}$
be an input datum before integrating out the $j$-th scale (see Definition
\stepInputData) and let $(\cW',\cG,u,\vec p)=\Om_j(\cW,\cG,u,\vec p)$ be the result
of integrating out scale $j$. Assume that $(\cW,\cG,u,\vec p)$ is bounded -- the
precise  hypothesis is given in Definition \stepInputData.
We shall show that the norm of the four point part of $\cW'$ does not exceed 
the norm of the four point part of $\cW$ by more than 
$\sfrac{\const}{M^{\ep j}}$ with constants $\ep,\ \const$ independent of $j$.
Most contributions to the four point part of $\cW'-\cW$ are controlled using 
``overlapping loops'', see [FKTr2]. The only exceptions are ladders. A ladder consists
of a sequence of four legged kernel ``rungs'' connected by pairs of propagators.

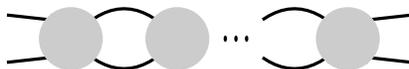
\centerline{\figplace{CmpLadder4}{-.1 in}{0 in} }

\noindent
For a formal definition, see \S\CHladdersNotn\  of [FKTo3].
Taking creation and annihilation indices into account, such a ladder is either a 
``particle--particle ladder''

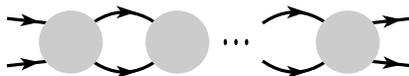
\centerline{\figplace{CmpLadder3}{-.1 in}{0 in} }

\noindent
or a ``particle--hole ladder''

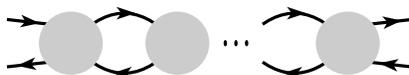
\centerline{\figplace{CmpLadder2}{-.1 in}{0 in} }

The strong asymmetry of the Fermi curve (see Definition \defNPstrongasymm) 
allows us to bound particle--particle ladders of scale $j$ 
by $\sfrac{\const}{M^{\ep j}}$. 
This estimate is stated precisely in Proposition \:\propparticleparticleladder\
below and proven in Theorem \:\theoremOSLadA\ of [FKTo4]. This is in 
contrast to the case of a symmetric Fermi curve, where the  
particle--particle ladders generate the Cooper instability (see 
[FW, Chapter 10], [FMRT, \S4]). 
The main estimates on particle--hole ladders are stated in 
Theorem \:\theoremcompLadder\ below. They are proven in 
[FKTl] for arbitrary strictly convex Fermi curves. 

Before we formulate the estimates on particle--hole and particle--particle 
ladders we give a precise definition of ladders. To treat ``undirected 
ladders'', particle--particle and particle--hole ladders  with or without spin
and with or without external momenta simultaneously, we first work over 
arbitrary measure spaces, like, for example, $\bbbr\times \bbbr^2$ or
$\bbbr\times \bbbr^2\times \{\uparrow,\downarrow\}$ or
$\cB=\bbbr\times \bbbr^2\times \{\uparrow,\downarrow\}\times \{0,1\}$.
See also \S\CHladdersNotn\ of [FKTo3].

\definition{\STM\defbubbleprop}{ Let $\fX$ be a measure space.
\Item{i)} A complex valued function on $\fX\times \fX$ is called a propagator 
over $\fX$.
\Item{ii)} A four legged kernel over $\fX$ is a complex valued function 
on $\fX^2\times \fX^2$. We sometimes consider it as a bubble propagator over $\fX$,  graphically depicted by 

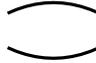
\centerline{\figplace{bubProp}{0 in}{0 in} }

\noindent
or as a rung over $\fX$, graphically depicted by 

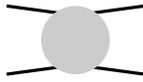
\centerline{\figplace{rung}{0 in}{0 in} }

\Item{ iii)} If $A$ and $B$ are propagators over $\fX$ then the tensor product
$$
A\otimes B(x_1,x_2,x_3,x_4)=A(x_1,x_3)B(x_2,x_4)
$$
is a bubble propagator over $\fX$. We set
$$
\cC(A,B) = A\otimes A + A\otimes B + B\otimes A 
$$

\Item iv)
Let $F$ be a four legged kernel over $\fX$. 
The antisymmetrization of $F$ is the four legged kernel
$$\eqalign{
\big({\rm Ant\,} F\big) (x_1,x_2,x_3,x_4)
& = \sfrac{1}{4!} \sum_{\pi\in S_4} {\rm sign}(\pi)\,
F(x_{\pi(1)},x_{\pi(2)},x_{\pi(3)},x_{\pi(4)})
}$$
$F$ is called antisymmetric if $F={\rm Ant\,} F$.

}

We will consider ladders with rungs taking values in the measure space $\cB\times\Si$, where $\Si$ is a sectorization. As propagators, we will use
the unsectorized propagators  $A~=~C^{(j)}_{u(K)}$ and $B=D_j(u(K);K)$
of Definition \defNPinoutmap.

\definition{\STM\deNPdefsectbubbleprop}{
Let $\fX$ be a measure space and let $S$ be a finite set\footnote{$^{(1)}$}{In practice, $S$ will be a set of sectors and $\fX$ will be $\cB$ 
or $\bbbr\times \bbbr^2\times \{\uparrow,\downarrow\}$ or
$\bbbr\times \bbbr^2$.}. It is endowed with the counting measure. Then $\fX\times S$ is also a measure space.

\Item{ i)} Let $P$ be a propagator over $\fX$, $f$ a four legged kernel over $\fX\times S$ and $F$ a function on $(\fX\times S)^2 \times \fX^2$. We define
$$\eqalign{
(f\bullet P)({\sst (x_1,s_1),(x_2,s_2);x_3,x_4}) 
&=  \smsum_{s'_1,s'_2\in S} \int {\sst dx'_1 dx'_2}\ 
f({\sst (x_1,s_1),(x_2,s_2),(x'_1,s'_1),(x'_2,s'_2)})\ 
P({\sst x'_1,x'_2;x_3,x_4}) \cr
(F\bullet f)({\sst(x_1,s_1),\cdots,(x_4,s_4)}) 
&=  \hskip-4pt\smsum_{s'_1,s'_2\in S}\int\hskip-4.2pt {\sst dx'_1 dx'_2 }\ 
F({\sst (x_1,s_1),(x_2,s_2);x'_1,x_2'})\ 
f({\sst (x_1',s_1'),(x_2',s_2'),(x_3,s_3),(x_4,s_4)}) \cr
}$$
whenever the integrals are well--defined. Observe that $(f\bullet P)$ is a function on $(\fX\times S)^2 \times \fX^2$ and $F\bullet f$ is a four legged kernel over $\fX\times S$.

\Item {ii)} Let
$\ell \ge 1\,$, $r_1,\cdots,r_{\ell+1}$ rungs over $\fX\times S$  and 
$P_1,\cdots,P_\ell$  bubble propagators over $\fX$.
The ladder with rungs $r_1,\cdots,r_{\ell+1}$ and 
bubble propagators $P_1,\cdots,P_\ell$ is defined to be
$$
r_1\bullet P_1 \bullet r_2 \bullet P_2\bullet \cdots \bullet
 r_{\ell}\bullet P_\ell\bullet r_{\ell+1}
$$
If $r$ is a  rung over $\fX\times S$
and $A,B$ are propagators over $\fX$, we define 
$L_\ell(r;A,B)$ as the ladder with   
$\ell+1$ rungs $r$ and  $\ell$ bubble propagators $\cC(A,B)$.

}

When we integrate out scale $j$ in our model, the contributions to the four 
legged kernel that are not controlled by ``overlapping loops'' are
antisymmetrizations of ladders that are defined over 
$\cB\times\Si$, where $\Si$ is a sectorization.
Such ladders decompose into particle--particle ladders and particle--hole 
ladders that are defined over smaller spaces that do not have 
creation/annihilation components.

\definition{\STM\NPsomespaces}{
Set
$\ \cB^\updownarrow=\set{(x_0,\x,\si)}
            {x_0\in\bbbr,\ \x\in\bbbr^2,\ \si\in\{\uparrow,\downarrow\} }\ $.

\noindent
If  $\Si$ is a sectorization and $z=(x,\si,b,s)\in\cB\times\Si$, we define its undirected part $u(z)\in\cB^\updownarrow\times\Si$ and its creation/annihilation index $b(z)\in\{0,1\}$ by
$u(z)=(x,\si,s)$ and $b(z)=b$, respectively.

\noindent
If $z'=(x,\si,s)\in\cB^\updownarrow\times\Si$ and $b\in\{0,1\}$, we define 
$\iota_b(z')=(x,\si,b,s)\in\cB\times\Si$.
}

\goodbreak
\definition{\STM\defParicleHoleDecomp}{
\Item{ i)} 
Let $f$ be a four legged kernel over $\cB\times\Si$. When $f$ is a rung,
its particle--particle reduction is the four legged kernel over $\cB^\updownarrow\times\Si$ given by
$$
f^{\rm pp}({\sst z'_1, z'_2, z'_3, z'_4}) = 
f({\sst \iota_0(z'_1),\iota_0(z'_2),\iota_1(z'_3),\iota_1(z'_4)}) = 
\figplace{PPvertex}{-.2in}{-.05in}
$$
and its particle--hole reduction is
$$
f^{\rm ph}({\sst z'_1, z'_2, z'_3, z'_4}) = 
f({\sst \iota_0(z'_1),\iota_1(z'_2),\iota_1(z'_3),\iota_0(z'_4)}) = 
\figplace{PHvertex}{-.2in}{-.05in}
$$
When $f$ is a bubble propagator, the corresponding reductions are
$$\eqalign{
{^{\rm pp}\!}f({\sst z'_1, z'_2, z'_3, z'_4}) &= 
f({\sst \iota_1(z'_1),\iota_1(z'_2),\iota_0(z'_3),\iota_0(z'_4)})\cr 
{^{\rm ph}\!}f({\sst z'_1, z'_2, z'_3, z'_4}) &= 
f({\sst \iota_1(z'_1),\iota_0(z'_2),\iota_0(z'_3),\iota_1(z'_4)})\cr 
}$$
\Item{ ii)} 
Let $f'$ be a four legged kernel over $\cB^\updownarrow\times\Si$. Its particle--particle value
is the four legged kernel over $\cB\times\Si$ given by
$$\eqalign{
V_{\rm pp}(f')({\sst z_1, z_2, z_3, z_4}) 
&=\de_{b(z_1) , 0}
  \de_{b(z_2) , 0}
  \de_{b(z_3) , 1}
  \de_{b(z_4) , 1}
\, f'({\sst u(z_1), u(z_2), u(z_3), u(z_4)}) \cr
&+\de_{b(z_1) , 1}
  \de_{b(z_2) , 1}
  \de_{b(z_3) , 0}
  \de_{b(z_4) , 0}
\, f'({\sst u(z_3), u(z_4), u(z_1), u(z_2)}) \cr  
}$$
and its particle--hole value is
$$\eqalign{
V_{\rm ph}(f')({\sst z_1, z_2, z_3, z_4}) 
&=\de_{b(z_1) , 0}
  \de_{b(z_2) , 1}
  \de_{b(z_3) , 1}
  \de_{b(z_4) , 0}
\, f'({\sst u(z_1), u(z_2), u(z_3), u(z_4)}) \cr
&+\de_{b(z_1) , 1}
  \de_{b(z_2) , 0}
  \de_{b(z_3) , 0}
  \de_{b(z_4) , 1}
\, f'({\sst u(z_2), u(z_1), u(z_4), u(z_3)}) \cr  
&-\de_{b(z_1) , 1}
  \de_{b(z_2) , 0}
  \de_{b(z_3) , 1}
  \de_{b(z_4) , 0}
\, f'({\sst u(z_2), u(z_1), u(z_3), u(z_4)}) \cr  
&-\de_{b(z_1) , 0}
  \de_{b(z_2) , 1}
  \de_{b(z_3) , 0}
  \de_{b(z_4) , 1}
\, f'({\sst u(z_1), u(z_2), u(z_4), u(z_3)}) \cr  
}$$
}

The decomposition of ladders over $\cB$ into particle--particle and 
particle--hole ladders is given by the following Lemma, whose proof is trivial.

\lemma{\STM\lemunordord}{
\Item{i)}
Let $f_1,\cdots,f_{\ell+1}$ be particle number preserving four legged 
kernels over $\cB\times\Si$ that are separately antisymmetric in their first two 
and their last two arguments. Let $P_1,\cdots,P_\ell$ be particle number preserving bubble
propagators over $\cB$ that satisfy 
$P_i(\xi_1,\xi_2,\xi_3,\xi_4) =P_i(\xi_2,\xi_1,\xi_4,\xi_3)$ for
$i=1,\cdots,\ell$.  Then
$$\eqalign{
\big( f_1\bullet P_1\bullet \cdots \bullet P_\ell\bullet f_{\ell+1} 
\big)^{\rm pp}  
& =\ \ \ \ f_1^{\rm pp}\bullet {^{\rm pp}\!}P_1\bullet \cdots 
\bullet {^{\rm pp}\!}P_\ell\bullet f_{\ell+1}^{\rm pp} \cr
\big( f_1\bullet P_1\bullet \cdots \bullet P_\ell\bullet f_{\ell+1} 
\big)^{\rm ph} 
 & =2^\ell\ f_1^{\rm ph}\bullet {^{\rm ph}\!}P_1\bullet \cdots 
\bullet {^{\rm ph}\!}P_\ell\bullet f_{\ell+1}^{\rm ph} \cr
}$$
\Item{ii)} Let $f$ be an antisymmetric, particle number preserving, four
legged  kernel over $\cB\times\Si$. Then
$$
f = V_{\rm pp}(f^{\rm pp}) + V_{\rm ph}(f^{\rm ph})
$$
}

\vskip .3cm

We now state the ladder estimates used in the rest of the paper.

\proposition{\STM\propparticleparticleladder}{
Let $0<\La<\sfrac{\tau_2}{2M^j}$, where $\tau_2$ is the constant of 
Lemma \lemOSdiffpropbound\ of [FKTo3].
Let $u({\sst(\xi,s)},{\sst(\xi',s')})\in\cF_0(2,\Si_j)$ 
be an antisymmetric, spin independent, particle number conserving function whose
Fourier transforms obeys
 $|\check u(k)| \le \half |\imath k_0-e(k)|$ 
and such that 
$\ \v  u \v_{1,\Si_j} \le \La\cb_j\ $. 
Furthermore let $f \in\cF_0(4,\Si_j)$. Then for all $\ell \ge 1$
$$\eqalign{
\V L_\ell \big(f;C^{(j)}_u,C^{(\ge j+1)}_u \big) \V_{3,\Si_j}
& \le \big( \const\,\cb_j \big)^\ell\  \v f\v_{3,\Si_j}^{\ell+1} \cr
\V V_{\rm pp} \Big( L_\ell \big(f;C^{(j)}_u,C^{(\ge j+1)}_u \big)^{\rm pp}
\Big) \V_{3,\Si_j}
& \le \big( \const \,\fl_j^{\raise2pt\hbox{$\scriptscriptstyle{1/n_0}$}}\,
\cb_j \big)^\ell\   \v f\v_{3,\Si_j}^{\ell+1} \cr
}$$
if the Fermi curve $F$ is strongly asymmetric in the sense of Definition
\defNPstrongasymm.
Here, $n_0$ is the constant of Definition
\defNPstrongasymm.
}

Proposition \propparticleparticleladder\ is a special case of Proposition \:\tildepropparticleparticleladder. See Remark \:\remtildepropparticleparticleladder.
\vskip .3cm

The first inequality of Proposition \propparticleparticleladder\ is not good enough
for the control of the four point function, since replacing
$\cC^{(j)}_u$ by $\cC^{(\le j)}_u$ would give an additional factor of $j^n$.
The second inequality of Proposition \propparticleparticleladder\ gives estimates
for particle--particle ladders at each individual scale $j$ that are good enough
to be summable over $j$. Particle--hole ladders do
not, at least in general, obey such estimates.  If they did, they would
be continuous in momentum space, even after all cutoffs are
removed. Therefore, we  bound the sum of all particle--hole ladders of scales
up to $j$ together, making use of cancellations between neighbouring scales.
Building up such sums  of ladders leads to ``compound particle--hole ladders''.

\definition{\STM\defcompLadder}{
Let $\vec F=\set{F^{(i)}}{i=2,3,\cdots}$ be a family of 
antisymmetric functions in $\cF_0(4,\Si_i)$. Let
$\vec p=\big(p^{(2)},p^{(3)},\cdots \big)$ be a sequence of 
antisymmetric, spin independent, particle number conserving functions
$p^{(i)}({\sst(\xi,s)}, {\sst(\xi',s')}) \in \cF_0(2,\Si_i)$.
We define, recursively on 
$0\le j <\infty$, the iterated particle hole (or wrong way) ladders up
to scale $j$, denoted by $\,\cL^{(j)}(\vec p,\vec F)\,$, as
$$\eqalign{
\cL^{(0)}(\vec p,\vec F)&=0 \cr
\cL^{(j+1)}(\vec p,\vec F)&= \cL^{(j)}(\vec p,\vec F)_{\Si_j}
+ 2 \smsum_{\ell=1}^\infty {\sst (-1)^\ell(12)^{\ell+1}}
  L_\ell\big(w_j;C^{(j)}_{u_j}, C^{(\ge j+1)}_{u_j}\big)^{\rm ph}
\cr }$$
where $u_j=\smsum_{i=2}^{j-1}p^{(i)}_{\Si_j}$ and 
$w_j = \sum_{i=2}^{j}{F}_{\Si_j}^{(i)}
+\sfrac{1}{8} {\rm Ant\,}\Big( V_{\rm ph} 
 \big( \cL^{(j)}(\vec p,\vec F)\big)\Big)_{\Si_j}$. The resectorization
$\cL^{(j)}(\vec p,\vec F)_{\Si_j}$ is defined by the natural analog 
of Definition \defNPresector. For the details,
see Definition \defOSresectorII\ of [FKTo4].

\noindent
Observe that $\,\cL^{(j)}(\vec p,\vec F)\,$ is a four legged kernel over $\cB^\updownarrow\times\Si_{j-1}$ and depends only on the
components 
$\,F^{(2)},\cdots,F^{(j-1)}\,$ of $\,\vec F\,$ and 
$\,p^{(2)},\cdots,p^{(j-2)}\,$ of $\,\vec p\,$.  Also observe that
$w_0,\ \cL^{(1)}(\vec p,\vec F), w_1$ and $\cL^{(2)}(\vec p,\vec F)$
all vanish.

}

When we apply Definition \defcompLadder, $F^{(i)}$ will be the volume improved 
part of the contribution to the four--point function generated by
integrating out scale $i$. Furthermore, $p^{(i)}$ will be, roughly speaking,
the contribution to the renormalized two--point function at $K=0$ that is moved into the covariance at scale $i$. In particular, $F^{(2)}$ through $F^{(j_0)}$
and $p^{(2)}$ through $p^{(j_0-1)}$ will be zero.

The main estimate on iterated particle hole ladders is

\theorem{\STM\theoremcompLadder}{
For every $\veps>0$ there are constants $\rho_0,\,\const$ such that the following
holds.  
Let $\vec F=\big(F^{(2)},F^{(3)},\cdots \big)$ be a sequence of 
antisymmetric, spin independent, particle number conserving functions
$F^{(i)}\in \cF_0(4,\Si_i)$
and $\vec p=\big(p^{(2)},p^{(3)},\cdots \big)$ be a sequence of 
antisymmetric, spin independent, particle number conserving functions
$p^{(i)} \in \cF_0(2,\Si_i)$.
Assume that there is $\rho \le \rho_0$ such that for $i\ge 2$
$$
\v F^{(i)}\v_{3,\Sigma_i} \le \sfrac{\rho}{M^{\veps i}}\cb_i\qquad\quad 
\v p^{(i)}\v_{1,\Sigma_i} \le \sfrac{\rho\,\fl_i}{M^i}\cb_i\qquad
\check p^{(i)}(0,\k)=0
$$
Then for all $j\ge 2$
$$
\V V_{\rm ph}\big( \cL^{(j)}(\vec p,\vec F)_{\Si_{j}} \big)\V_{3,\Si_{j}} 
\le \const \rho^2 \,\cb_{j}
$$
}

\noindent
Theorem \theoremcompLadder\ is a special case of Theorem \theoremtildecompLadder\ in part 3. See Remark \:\remtildecompLadder. 
Both Theorems
are consequences of the estimates on iterated particle hole ladders derived in [FKTl].

\vfill\eject

\chap{ Infrared Limit of Finite Scale Green's Functions}\PG\pgNPVIII

The nonperturbative construction of the infrared limit will be similar to the 
formal construction outlined in \S\CHrenmap. We first adapt the notion of 
a formal interaction triple $(\cW,\cG,u)$ at scale $j$ of Definition 
\defNPeffinttriple\ to the needs of the 
nonperturbative construction.
The function $u$ modifying the covariance at scale $j$ is built up of contributions created at scales up to $j-1$. To bound $u$, we keep track of all of these individual
contributions. They are encoded in the additional datum $\vec p$ of 
\definition{\STM\defNPintquad (Interaction Quadruple)}{
An interaction quadruple  at scale $j$ is a quadruple $(\cW,\cG,u,\vec p)$ that obeys the following conditions. 
\item{$\circ$}
$\cW$ is a map from the space $\fK_{j}$ of counterterms to the space of 
even, translation invariant, spin independent, particle number 
conserving Grassmann functions in $\phi$ and $\psi$, that obeys 
$\cW(\phi,0,K)=0$. 
\item{$\circ$}
$\cG$ is a map from $\fK_{j}$ to the space of even, translation invariant, 
spin independent, particle number conserving Grassmann functions in $\phi$,
that obeys $\cG(0,K)=0$.
\item{$\circ$}
$\vec p=\big(p^{(2)},\cdots,p^{(j-1)}\big)$ where each 
$p^{(i)}({\sst(\xi,s)}, {\sst(\xi',s')})$ is
an antisymmetric, spin independent, particle number conserving function 
in $\cF_0(2,\Si_i)$ that obeys
$$
\V p^{(i)}\V_{1,\Si_i}\le \la_0^{1-\upsilon} \sfrac{\fl_i}{M^i}\cb_i 
\EQN\eqnNPpbound$$
The Fourier transform $\check p^{(i)}(k)$ of $p^{(i)}$
is supported in the $i^{\rm th}$ neighbourhood and 
vanishes at $k_0=0$.
\item{$\circ$}
$u$ is a map from $\fK_{j}$ to the space of antisymmetric, 
spin independent, particle number conserving functions in $\cF_0(2,\Si_j)$.
The function $u(K)$ has a decomposition
$$
u(K)=  \smsum_{i=2}^{j-1}p^{(i)}_{\Si_j}
+\big[\de u(K)- K_{\rm ext}\big]_{\Si_j}
\EQN\eqnNPudecomp$$
with $\de u({\sst(\xi,s)}, {\sst(\xi',s')};K)$
 an antisymmetric function  in $\cF_0(2,\Si_{j-1})$ that vanishes at $k_0=0$ and when $K=0$ and obeys
$$\eqalign{
\V \sfrac{d\hfill}{ds}\de u(K+sK')\big|_{s=0}\,  \V_{1,\Si_{j-1}}
&\le\la_0^{1-\upsilon}\,\fe_j\big(\|K\|_{1,\Si_j}\big) \,  \|K'\|_{1,\Si_j}\cr
\V  \cD^{(1,0,0)}_{1,2}\sfrac{d\hfill}{ds} \de u(K+sK')\big|_{s=0}\,  \V_{1,\Si_{j-1}}
&\le\la_0^{1-\upsilon}M^{j-3}
\,\fe_j\big(\|K\|_{1,\Si_j}\big) \,  \|K'\|_{1,\Si_j}
+\smsum_{\de_0=r_0}\infty\, t^\de\cr
}\EQN\eqnNPdeubound$$
for all $ K\in\fK_{j}$ and all $ K'$.

\noindent The interaction quadruple $(\cW,\cG,u,\vec p)$ is said to be 
{\bf real}, if $\cW(\phi,\psi,K)$, $\cG(\phi,K)$, $u(K)$ and $p^{(1)},\cdots,
p^{(j-1)}$ are $k_0$--reversal real, in the sense of Definition \defOSsymmetries.R of [FKTo2], for all real $K\in\fK_j$. In particular
$\check p^{(i)}(-k_0,\k)=\overline{\check p^{(i)}(k_0,\k)}$.

}

\remark{\STM\remNPintquad}{ 
\Item (i) We remind the reader that 
\item{} The space $\fK_j$ of counterterms was defined in Definition \defNPCTmSpace. The extension $K_{\rm ext}\big((\xi,s),(\xi',s')\big)$ of 
$K\big((\x,s),(\x',s')\big)$ was defined in Definition \defOSzeroext\ of [FKTo4].
\item{} The sector length $\fl_j$ was fixed after Definition \defNPsectrepr,
the resectorization $p_{\Si_j}$ was defined in Definition \defNPresector,
the space $\cF_0(2,\Si_{j})$ was defined in Definition \defNPsectrepr.ii
and the seminorm $\v\ \cdot\ \v_{1,\Si_j}$ was defined in Definition \defNPsectnorm.i.
\item{} The decay operator $\cD_{i,j}^\de$ was defined in Definition
\defNPdecayop.ii and the elements $\cb_j$ and $\fe_j(X)$ of the norm domain were defined in Definition \defNPFancynormdomain.iii.

\Item (ii) 
Observe that $\vec p$ is independent of $K$, so that, 
in (\eqnNPudecomp), the only $K$ dependence of $u$ is through 
$\de u(K)-K_{\rm ext}$.
\Item (iii)
The representation (\eqnNPudecomp) of $u$ 
implies that
$$
\check u(k; K)=  \smsum_{i=2}^{j-1} \check p^{(i)}(k) \ 
+\big[\de\check u(k;K) 
  -\check K(\k)\big]
$$
for $k$ in the $j^{\rm th}$ neighbourhood. Bounds on $u(K)$ will be provided in Lemma \:\lemNPpptyu.
}

\noindent
The relation between counterterms at different scales is formalized in
\definition{\STM\defNPprojectivesyst}{
A projective system of counterterms consists of analytic maps
$$\meqalign{
 \ren_{i,j}: \cK_{j+1} &\longrightarrow \cK_{i+1}
\qquad &&{\rm  for\ } j_0\le i\le j \cr
\de e_j: \cK_{j+1} &\longrightarrow \cE
\qquad &&{\rm  for\ } j_0 \le j \cr
}$$
such that 
$$\eqalign{
&\ren_{j,j}\ \  {\rm is \  the \ identity\ map\ of \ } \cK_{j+1} \cr
&\ren_{i,i'} \circ \ren_{i',j} = \ren_{i,j}
\qquad {\rm  for\ } j_0\le i \le i' \le j \cr
&\de e_i \circ \ren_{i,j} = \de e_j \qquad {\rm  for\ } j_0\le i  \le j \cr
}$$
and
$$\eqalign{
\sup_{K\in\fK_{j}}\tn\de \hat e_j(K)\tn_{1,\infty}
&\le \la_0^{1-\upsilon}\cr
\big\| \ren_{i,j}(0)-\ren_{i,j'}(0)\big\|_{1,\Si_i}
&\le \la_0^{1-\upsilon}\sfrac{1}{2^j}+\sum_{\de\ne 0}\infty t^\de\cr
\tn\de \hat e_j(0)-\de\hat e_{j'}(0)\tn_{1,\infty}
&\le  \la_0^{1-\upsilon}\sfrac{1}{2^j}\cr
}$$
for all $j_0\le i\le j\le j'$.

\noindent A projective system is said to be {\bf real} if $\ren_{i,j}(K)$ is real and $\de e_j(\k;K)$ is real--valued for all $i$, $j$
and all real $K$.

}

\remark{\STM\remNPprojectivesyst}{
For any projective system of counterterms, the sequence $\de e_j(K)\big|_{K=0}$ 
of infrared cutoff counterterms converges in the topology of $\cE$.

}

\noindent
We shall prove, in \S\CHrecurs, the following bounds on the analogs 
 of the formal interaction triple $(\cW^\out_j,\cG^\out_{j},u_j)$
of (\eqnNPinouttriple).

\theorem{\STM\theoremNPinduction}{
Assume that $d=2$, that $e(\k)$ fulfills the Hypotheses  \hypNPdisprel\  and that the scale parameter $M$ has been chosen big enough.
Then there exist constants $\bar\al,\ \bar\la>0$ such that for all 
$0\le \la_0 <\bar\la$, $\bar\al<\al<\sfrac{1}{\la_0^{\upsilon/10}}$ the following holds. For each translation invariant, spin independent 
interaction kernel $V$ obeying 
$$
\|V\|_{1,\infty}\le \la_0\cb_0
$$
 there exist 
\item{$\circ$}
a projective system of counterterms $\big( \ren_{i,j},\ \de e_j \big)$
\item{$\circ$}
a family $\vec p=\big(p^{(2)},p^{(3)}\cdots\big)$ of  functions 
$p^{(i)}\in\cF_0(2,\Si_i)$ 
\item{$\circ$}
a family $\vec F=\big(F^{(2)},F^{(3)}\cdots\big)$ of antisymmetric kernels
$$
F^{(i)}({\sst(\xi_1,s_1)},\cdots,{\sst(\xi_4,s_4)})\in\cF_0(4,\Si_i)
$$ 
such that
$$
\V F^{(i)}\V_{3,\Si_i}\le \sfrac{\la_0^{1-\upsilon}}{\al^7}\,\vi_i\,\cb_i
$$

\noindent
All of this data depends analytically on $V$.
Also, for each scale $j \ge j_0$ there exist $\cW_j$, $\cG^\rg_j$ and $u_j$,
depending analytically on $V$ and $K$,
such that $\big(\cW_j,\cG^\rg_j,u_j,({\sst p^{(2)},\cdots,p^{(j-1)}})\big)$
is an interaction quadruple at scale $j$. Furthermore
\item{(R1)}  
$\cW_j(K)$  has a $\Si_j$--sectorized representative, 
$$\eqalign{
w(\phi,\psi;K) &= \smsum_{m,n}\ \smsum_{s_1,\cdots,s_n\in\Si_j}\
\int {\sst d\eta_1\cdots d\eta_m\,d\xi_1\cdots d\xi_n}\ 
w_{m,n}({\sst \eta_1,\cdots, \eta_m\,(\xi_1,s_1),\cdots ,(\xi_n,s_n)};K)\cr
& \hskip 5cm \phi({\sst \eta_1})\cdots \phi({\sst \eta_m})\
\psi({\sst (\xi_1,s_1)})\cdots \psi({\sst (\xi_n,s_n)\,})\cr
}$$
with $w_{m,n}$ antisymmetric separately in the $\eta$ and in the $(\xi,s)$ variables,  $w_{m,0}=0$ for all $m\ge 0$ and 
$$\eqalign{
\V w_{0,2}(K)\V_{1,\Si_j} 
& \le\sfrac{\la_0^{1-\upsilon}}{\al^{7}} \,\sfrac{\fl_j}{M^j}\,
 \fe_j\big(\|K\|_{1,\Si_j}\big)\cr
N_j\big(w(K),\al,\|K\|_{1,\Si_j}\big) 
&\le \fe_j\big(\|K\|_{1,\Si_j}\big) \cr
N_j\big( \sfrac{d\hfill}{ds} w(K+sK')\big|_{s=0},\al,\|K\|_{1,\Si_j}\big) 
   &\le M^j \,\fe_j\big(\|K\|_{1,\Si_j}\big) \,  \|K'\|_{1,\Si_j}
\cr
}$$
for all $ K\in\fK_{j}$ and all $ K'$. $w$ depends analytically on $V$ and $K$.
\item{(R2)}
The function $w_{0,4}(K)$ has a decomposition  
$$
w_{0,4}(K)=\de F^{(j+1)}(K)+\smsum_{i=2}^{j}F^{(i)}_{\Si_j}+
\sfrac{1}{8}{\rm Ant\,}
\Big( V_{\rm ph}\big( \cL^{(j+1)}(\vec p,\vec F)\big) \Big)
$$
with an antisymmetric kernel 
$\de F^{(j+1)}({\sst(\xi_1,s_1)},\cdots,{\sst(\xi_4,s_4)};K)\in\cF_0(4,\Si_j)$ such that  
$$
\V \de F^{(j+1)}(K)\V_{3,\Si_j}\le 
\sfrac{\la_0^{1-\upsilon}}{\al^4}\Big\{
      \sfrac{\vi_{j+1}}{\al^4}
     +\sfrac{1}{\IB^2}  M^j\|K\|_{1,\Si_j}
\Big\}\fe_j({\sst \| K\|_{1,\Si_{j}}})
\qquad\hbox{for all }K\in\fK_{j}
$$
The particle--hole projector $V_{\rm ph}$ is defined in 
Definition \defParicleHoleDecomp. 
\item{(R3)}
For each $K\in\fK_{j}$, 
$$
N\big(\cG^\rg_j(K)-\half\phi JC^{(\le j)}J\phi\big)\le 4\smsum_{i=2}^{j} 
\sfrac{1}{\root{4}\of{\fl_iM^i}} \hbox{ for all }K\in\fK_{j}
$$
Let the part of $\cG^\rg_j(K)$ that is homogeneous of degree two
be 
$$
\cG^\rg_{j,2}(K)=\int d\eta_1 d\eta_2\ G^\rg_{j,2}(\eta_1, \eta_2,K)\ \phi(\eta_1)\phi(\eta_2)
$$ 
Then
$$\eqalign{
N\big(\sfrac{d\hfill}{ds}\big[\cG^\rg_j(K+sK')-\cG^\rg_{j,2}(K+sK')\big]_{s=0}\big)
&\le M^j \|K'\|_{1,\Si_j} \cr
\TN \sfrac{d\hfill}{ds}G^\rg_{j,2}(K+sK')\big|_{s=0}\TN_\infty
&\le M^j \|K'\|_{1,\Si_j} \cr
}$$
for all $K\in\fK_{j}$ and all $K'$.
\item{(R4)}
For $K\in \cK_{j+2}$ and $K'=\ren_{j,j+1}(K)$ 
$$
N\big(\cG^\rg_{j+1}(K)-\half\phi JC^{(j+1)}J \phi -\cG^\rg_{j}(K') \big)
\le  \sfrac{4}{\root{4}\of{\fl_jM^j}}
$$
\item{(R5)}
For infrared cutoffs $\jbar\ge j+2$, 
the generating function of the connected Green's 
functions at scale $\jbar$ of Theorem \theoremNPmainthI, considered as a
formal Taylor series in $V$, fulfills
$$\eqalign{
\cG_\jbar(\phi,\bar\phi;\de  e_j(K)) 
& = \cG^\rg_{j}(\phi;K) +
\log\frac{\int e^{\phi J\psi}\, e^{\lw \cW_j(\phi,\psi;K)\rw_{\psi,D_{j}(u_j;K)}}\,
 d\mu_{C^{[j+1,\jbar)}_{u_j(K)}}(\psi)}{
\int e^{\lw \cW_j(0,\psi;K)}\rw_{\psi,D_{j}(u_j;K)}\, d\mu_{C^{[j+1,\jbar)}_{u_j(K)}}(\psi)}
}$$
for $K\in\fK_{j}$.

\noindent If, in addition, $V$  satisfies the reality condition of
 (\eqnNPreal) then
\item{$\circ$} the projective system $\big( \ren_{i,j},\ \de e_j \big)$ is real
\item{$\circ$} each $F^{(i)}$ is $k_0$--reversal real, in the sense 
of Definition \defOSsymmetries.R of [FKTo2]
\item{$\circ$} each interaction quadruple $\big(\cW_j,\cG^\rg_j,u_j,({\sst p^{(2)},\cdots,p^{(j-1)}})\big)$ is real
\item{$\circ$} for real $K$, the $\Si_j$--sectorized representative
$w(\phi,\psi;K)$ of $\cW_j(K)$ is $k_0$--reversal real.

}
\goodbreak
\proof{ of Theorem \theoremNPmainthI\ from Theorem \theoremNPinduction}
Observe that $\cG_{\il}(\phi;\de e)$ depends on $M$ and $\il$ only through
the combination ${M}^{\raise1pt\hbox{$\sst\il$}}$. Hence, for constructing 
$\lim\limits_{\il\rightarrow+\infty \atop  \il\in\bbbr}\cG_{\il}(\phi;\de e)$,
we may, without loss of generality, use any $M>1$ we wish.

Choose $M$, $\al$ and $\la_0$ fulfilling the hypotheses of Theorem 
\theoremNPinduction.
By Remark \remNPnorminteraction, the conditions on the interaction kernel 
in Theorems \theoremNPmainthI\ and \theoremNPinduction\ agree. By
Remark \remNPprojectivesyst,
$$
\de e=\lim_{j\rightarrow \infty} \de e_j(0)
$$
exists. If $V$ is $k_0$--reversal real, as in (\eqnNPreal), 
then $\de e(\k)$ is real for all $\k$.
By Definition \defNPprojectivesyst,
$$
\tn\de \hat e\tn_{1,\infty}
\le \la_0^{1-\upsilon}\qquad
\tn\de \hat e_j(0)-\de\hat e\tn_{1,\infty}
\le  \la_0^{1-\upsilon}\sfrac{1}{2^j}
$$
for all $j\ge j_0$. By Weierstrass' Theorem, $\de e$ is analytic in $V$.

We now show that the sequence $\cG^\rg_j(0)-\half \phi JC^{(\le j)}J\phi$ 
is a Cauchy sequence. Let $j\ge j_0$ and $K'=\ren_{j,j+1}(0)$. By
Definition \defNPprojectivesyst, 
$\big\| K'\big\|_{1,\Si_j}
\le \la_0^{1-\upsilon}\sfrac{\fl_j}{M^j}+\sum_{\de\ne 0}\infty t^\de$.
Hence, by (R3), (R4) and the Definition \defNPsectGrnorm.ii of the norm
$N(\cG)$,
$$\eqalign{
&N\big(\cG^\rg_{j+1}(0)-\half\phi JC^{(\le j+1)}J \phi 
         -\cG^\rg_{j}(0)+\half\phi JC^{(\le j)}J \phi \big)\cr
&\hskip0.3in
\le N\big(\cG^\rg_{j+1}(0)-\half\phi JC^{(j+1)}J \phi -\cG^\rg_{j}(K') \big)
+N\big(\cG^\rg_{j}(K')-\cG^\rg_{j,2}(K') -\cG^\rg_{j}(0)+\cG^\rg_{j,2}(0) \big)
\cr
&\hskip1in
+N\big(\cG^\rg_{j,2}(K') -\cG^\rg_{j,2}(0) \big)\cr
&\hskip0.3in
\le \sfrac{4}{\root{4}\of{\fl_jM^j}}
+ M^j \|K'\|_{1,\Si_j}+ \sfrac{1}{\la_0^{1-\upsilon}}M^j \|K'\|_{1,\Si_j}\cr
&\hskip0.3in\le \sfrac{4}{\root{4}\of{\fl_jM^j}}
+ 2\fl_j\cr
}$$
Let 
$$
\cG(\phi) = 
\sum_{n=1}^\infty \sfrac{1}{n!} \int\smprod_{i=1}^{n} d\xi_i\ 
G_{n}(\xi_1,\cdots,\xi_{n}) 
\smprod_{i=1}^{n} \phi(\xi_i) 
$$
be the limit of the $\cG^\rg_j(0)$'s. It is analytic in $V$.

We now show that the generating functionals $\cG_\il(\phi)$ of the  connected Green's functions at scale $\il$ also converge to $\cG(\phi)$. Let 
$\il\ge j_0+3$ and let $j=j(\il)$ be the integer with $j+1<\il\le j+2$.
By Definition \defNPprojectivesyst,
$$
K=\lim_{j'\rightarrow \infty} \ren_{j,j'}(0)
$$
exists in $\fK_{j}$ and obeys $\ \de e=\de e_j(K)$. Observe that,
by (R5), 
$$\eqalign{
\cG_{\il}(\phi;\de e)-\cG^{\rg}_j(\phi;0)
&= \cG_{\il}(\phi;\de e_j(K))-\cG^{\rg}_j(\phi;K)
+\cG^{\rg}_j(\phi;K)-\cG^{\rg}_j(\phi;0)\cr
&= \log\frac{\int e^{\phi J\psi}\, e^{\lw \cW_j(\phi,\psi;K)\rw_{\psi,D_{j}(u_j;K)}}\,
 d\mu_{S}(\psi)}{
\int e^{\lw \cW_j(0,\psi;K)}\rw_{\psi,D_{j}(u_j;K)}\, d\mu_{S}(\psi)}
+\cG^{\rg}_j(\phi;K)-\cG^{\rg}_j(\phi;0)\cr
&= \tilde \Om_S\big(\lw \cW_j(K)\rw_{\psi,D_{j}(u_j;K)}\big)(\phi,0)
+\cG^{\rg}_j(\phi;K)-\cG^{\rg}_j(\phi;0)\cr
}\EQN\eqnNPgdiff$$
where
$$
S= \frac{\nu^{(\ge j+1)}(k)-\nu^{(\ge\il)}(k)}
          {\imath k_0 - e(\k) -\check u_j(k;K)[1-\nu^{(\ge \il)}(k)]}
= \frac{\nu^{(\ge j+1)}(k)-\nu^{(\ge\il)}(k)}{\imath k_0 - e(\k)
                -\check u_j(k;K)[\nu^{(\ge j)}(k)-\nu^{(\ge \il)}(k)]}
$$
As in the previous paragraph
$$
N\big(\cG^{\rg}_j(K)-\cG^{\rg}_j(0)\big)\le 2\fl_j
\EQN\eqnNPnbndA$$
In Lemma \lemNPpptyu, below, we prove that the hypotheses of Proposition
\propOSresidualrengroupest\ of [FKTo3], with $\check u(k)=\check u_j(k;K)[\nu^{(\ge j)}(k)-\nu^{(\ge \il)}(k)]$, $\check v(k)=\check u_j(k;K)+ \check K(\k)\nu^{(\ge j+2)}(k)$ and $X=\|K\|_{1,\Si_j}$ are satisfied. 
Consequently,
$$
{\sst \root{4}\of{\fl_j M^j}}\ N\Big(\tilde \Om_S\big(\lw \cW_j(K)\rw_{\psi,D_{j}(u_j;K)}\big)(\phi,0)
-\half \phi JSJ\phi\Big)\le 10
$$
As
$$\eqalign{
 N\big(\half \phi JSJ\phi\big)&=\sfrac{1}{\la_0^{1-\upsilon}}\tn JSJ\tn_\infty
\le \sfrac{1}{\la_0^{1-\upsilon}}\| S(k)\|_{L^1}\cr
&\le \sfrac{1}{\la_0^{1-\upsilon}}\ 
{\rm Vol}\big({\rm support\,}\nu^{(\ge j+1)}-\nu^{(\ge\il)}\big)\ 
\sup_k |S(k)|\cr
&\le \sfrac{\const}{\la_0^{1-\upsilon} M^j}
}$$
we have that 
$$
N\Big(\tilde \Om_S\big(\lw \cW_j(K)\rw_{\psi,D_{j}(u_j;K)}\big)(\phi,0)\Big)
\le \sfrac{10}{\root{4}\of{\fl_j M^j}}+\sfrac{\const}{\la_0^{1-\upsilon} M^j}
\EQN\eqnNPnbndB$$
Combining (\eqnNPgdiff), (\eqnNPnbndA) and (\eqnNPnbndB),
$$
\lim_{\il\rightarrow\infty}N\big(\cG_{\il}(\phi;\de e)
-\cG^{\rg}_{j(\il)}(\phi;0)\big)=0
$$
so that 
$$
\lim_{\il\rightarrow\infty}\cG_\il=\cG
$$
in the $N(\ \cdot\ )$ norm. Consequently, for each $n$, $G_{n;\il}$ converges
uniformly to $G_{n}$.
\endproof 
\definition{\STM\defNPconvol}{
If $u({\sst(\xi,s)},{\sst(\xi',s')})$ is an antisymmetric,
translation invariant, spin independent, particle number conserving function 
on $(\cB\times\Si)^2$ and $\mu(k)$ is a  function on $\bbbr\times\bbbr^2$, set 
$$\eqalign{
(u*\hat\mu)({\sst(\xi,s)},{\sst(\xi',s')})
&=\int_\cB d\ze\ u({\sst(\xi,s)},{\sst(\ze,s')})\hat\mu(\ze,\xi')\cr
(\hat\mu*u)({\sst(\xi,s)},{\sst(\xi',s')})
&=\int_\cB d\ze\ u({\sst(\ze,s)},{\sst(\xi',s')})\hat\mu(\ze,\xi)\cr
}$$
where $\hat\mu$ was defined in Notation \notNPfourierTI. 

}

\noindent With this definition
$(u*\hat\mu)^{\check{}}(k)=(\hat\mu*u)^{\check{}}(k)=\check u(k)\mu(k)$.

\lemma{\STM\lemNPpptyu}{ Let $\big(\cW,\cG,u,\vec p\big)$ be an interaction quadruple at scale $j$. Then
\Item i)
$$\eqalign{
&\big|\check u(k;K) + \check K(\k)\nu^{(\ge j+2)}(k)\big|
\le\la_0^{1-\upsilon}\big|\imath k_0-e(\k)\big|
\le\half\big|\imath k_0-e(\k)\big|\cr
&\Big|\sfrac{d\hfill}{ds}\check u(k;K+sK')\big|_{s=0} 
+ \check K'(\k)\nu^{(\ge j+2)}(k)\Big|
\le 4M^{j+{3\over 2}}\|K'\|_{1,\Si_j}\, \big|\imath k_0-e(\k)\big|
}$$
for all $k$ in the $j^{\rm th}$ neighbourhood, all $K\in\fK_{j}$ and all $K'$.
\Item ii)
$$\eqalign{
&\v u(K)\v_{1,\Si_j}
\le\abcst\,\Big[
\sfrac{\la_0^{1-\upsilon}}{M^{j-1}}+\| K\|_{1,\Si_{j}}
\Big]\fe_j({\sst \| K\|_{1,\Si_{j}}})\cr
&\V \sfrac{d\hfill}{ds} u(K+sK')\big|_{s=0} \V_{1,\Si_j}
\le\abcst\,\fe_j({\sst \| K\|_{1,\Si_{j}}})\,\| K'\|_{1,\Si_{j}}\cr
}$$
\Item iii) Let $\il\in(j,j+2]$. Then
$$
\V u(K)*\big(\nu^{(\ge j)}-\nu^{(\ge \il)}\big)^{\widehat{}}\ \V_{1,\Si_j}
\le\const\,\Big[
\sfrac{\la_0^{1-\upsilon}}{M^{j-1}}+\| K\|_{1,\Si_{j}}
\Big]\fe_j({\sst \| K\|_{1,\Si_{j}}})
$$
}
\prf i)
By Remark \remNPintquad.iii, 
$$\eqalign{
\big|\check u(k;K) + \check K(\k)\nu^{(\ge j+2)}(k)\big|
&=\Big| \smsum_{i=2}^{j-1} \check p^{(i)}(k) + \de\check u(k;K)
  -\check K(\k)\big[ 1- \nu^{(\ge j+2)}(k)\big]\Big|\cr
&\le\smsum_{i=2}^{j-1} \big|\check p^{(i)}(k)\big| +\big|\de\check u(k;K)\big|
  +\big|\check K(\k)\big|\big[1- \nu^{(\ge j+2)}(k)\big]\cr
}$$
and
$$
\Big|\sfrac{d\hfill}{ds}\check u(k;K+sK')\big|_{s=0} 
+ \check K'(\k)\nu^{(\ge j+2)}(k)\Big|
\le\Big|\sfrac{d\hfill}{ds}\de\check u(k;K+sK')\big|_{s=0}\Big|
  +\big|\check K'(\k)\big|\big[1- \nu^{(\ge j+2)}(k)\big]
$$
for all $k$ in the $j^{\rm th}$ neighbourhood. Because $\check p^{(i)}$ 
vanishes at $k_0=0$, Lemma \lemOSNormMom\ of [FKTo3] and (\eqnNPpbound) imply
that
$$
\big|\check p^{(i)}(k)\big|
\le 2|k_0|\,\sfrac{\partial\hfill}{\partial t_{(1,0,0)}}
                \v p^{(i)}\v_{1,\Si_i}\Big|_{t=0}
\le 2|k_0|\,\la_0^{1-\upsilon}\sfrac{\fl_i}{M^i}M^i
\EQN\eqnNPdenomA$$
Similarly, by Lemma \lemOSNormMom\ of [FKTo3], (\eqnNPdeubound) and Definition
\defNPCTmSpace,
$$\eqalign{
\big|\de\check u(k;K)\big|
&\le 2|k_0|\, \v  \cD^{(1,0,0)}_{1,2}\de u(K)\v_{1,\Si_{j-1}}\Big|_{t=0}\cr
&\le 2|k_0|\,\la_0^{1-\upsilon}M^{j-3}
\frac{1}{1-M^j\la_0^{1-\upsilon}\sfrac{\fl_{j+1}}{M^{j+1}}}
\la_0^{1-\upsilon}\sfrac{\fl_{j+1}}{M^{j+1}}\cr
&\le 4|k_0|\,\la_0^{2-2\upsilon}\fl_j\cr
\Big|\sfrac{d\hfill}{ds}\de\check u(k;K+sK')\big|_{s=0}\Big|
&\le 2|k_0|\, \V  \cD^{(1,0,0)}_{1,2}\sfrac{d\hfill}{ds}
\de u(K+sK')\big|_{s=0}\V_{1,\Si_{j-1}}\Big|_{t=0}\cr
&\le 2|k_0|\,\la_0^{1-\upsilon}M^{j-3}
\frac{1}{1-M^j\la_0^{1-\upsilon}\sfrac{\fl_{j+1}}{M^{j+1}}}\|K'\|_{1,\Si_j}\cr
&\le 4|k_0|\,\la_0^{1-\upsilon}M^{j-3}\|K'\|_{1,\Si_j}\cr
}\EQN\eqnNPdenomB$$
and, by Lemma \lemOSNormMom\ of [FKTo3], Definition
\defNPCTmSpace\ and Definition \defOSscales.i of [FKTo2],
$$\eqalign{
\big|\check K(\k)\big|\big[1- \nu^{(\ge j+2)}(k)\big]
&\le 2\| K\|_{1,\Si_j}\big|_{t=0}\ \big[1- \nu^{(\ge j+2)}(k)\big]\cr
&\le 2\la_0^{1-\upsilon}\sfrac{\fl_{j+1}}{M^{j+1}} 
               \big[1- \nu^{(\ge j+2)}(k)\big]\cr
&\le 2\la_0^{1-\upsilon}\sfrac{\fl_{j+1}}{M^{j+1}} 
               \frac{|\imath k_0-e(\k)|}{\sqrt{1/M}\sfrac{1}{M^{j+1}}}\cr
&\le 2\sqrt{M}\la_0^{1-\upsilon}\fl_{j+1}|\imath k_0-e(\k)|\cr
&\le \sfrac{1}{10}\la_0^{1-\upsilon}\fl_j|\imath k_0-e(\k)|\cr
\big|\check K'(\k)\big|\big[1- \nu^{(\ge j+2)}(k)\big]
&\le 2\| K'\|_{1,\Si_j}\big|_{t=0}\ \big[1- \nu^{(\ge j+2)}(k)\big]\cr
&\le 2\| K'\|_{1,\Si_j}\  
               \frac{|\imath k_0-e(\k)|}{\sqrt{1/M}\sfrac{1}{M^{j+1}}}\cr
&\le 2M^{j+{3\over 2}}\| K'\|_{1,\Si_j}\ |\imath k_0-e(\k)|\cr
}\EQN\eqnNPdenomC$$
if $M$ is large enough. In the last inequality of the bound on 
$\big|\check K(\k)\big|\big[1- \nu^{(\ge j+2)}(k)\big]$, we used that $\sfrac{\fl_{j+1}}{\fl_j}=\sfrac{1}{M^\aleph}$
with $\aleph>\half$. Combining (\eqnNPdenomA), (\eqnNPdenomB) and (\eqnNPdenomC),
$$\eqalign{
\big|\check u(k;K) + \check K(\k)\nu^{(\ge j+2)}(k)\big|
&\le\Big\{\smsum_{i=2}^{j-1} 2\fl_i +4\la_0^{1-\upsilon}\fl_j+\sfrac{1}{10}\fl_j
 \Big\}
\la_0^{1-\upsilon}|\imath k_0-e(\k)|\cr
&\le \la_0^{1-\upsilon}|\imath k_0-e(\k)|\cr
\Big|\sfrac{d\hfill}{ds}\check u(k;K+sK')\big|_{s=0} 
+ \check K'(\k)\nu^{(\ge j+2)}(k)\Big|
&\le\Big\{ 4\la_0^{1-\upsilon}M^{-3}+   2M^{{3\over 2}}
 \Big\}
M^{j}\| K'\|_{1,\Si_j}\ |\imath k_0-e(\k)|\cr
&\le 4M^{j+{3\over 2}}\|K'\|_{1,\Si_j}\, \big|\imath k_0-e(\k)\big|\cr
}$$

\Item ii) By Corollary \corOSresectorvanishkzero, 
Remark \remOStoresectorI\ and Proposition \propOSindresectorI.i of [FKTo4]
$$\eqalign{
\v u(K)\v_{1,\Si_j}
&=\VV \smsum_{i=2}^{j-1}p^{(i)}_{\Si_j}
+\big[\de u(K)- K_{\rm ext}\big]_{\Si_j}\VV_{1,\Si_j}\cr
&\le\abcst\,\cb_{j-1}\Big[
\smsum_{i=2}^{j-1}\la_0^{1-\upsilon}M\sfrac{\fl_i}{M^j} +
\la_0^{1-\upsilon}\fe_j({\sst \| K\|_{1,\Si_{j}}})
\big\| K\big\|_{1,\Si_{j}} +\big\| K\big\|_{1,\Si_{j}}\Big]\cr
&\le\abcst\,
\Big[
\sfrac{\la_0^{1-\upsilon}}{M^{j-1}}\cb_j +
\la_0^{1-\upsilon}\fe_j({\sst \| K\|_{1,\Si_{j}}})
\cb_j\big\| K\big\|_{1,\Si_{j}} +\cb_j\big\| K\big\|_{1,\Si_{j}}\Big]\cr
&\le\abcst\,
\Big[
\sfrac{\la_0^{1-\upsilon}}{M^{j-1}}+
\big(1+\la_0^{1-\upsilon}\big)\big\| K\big\|_{1,\Si_{j}}
\Big]\fe_j({\sst \| K\|_{1,\Si_{j}}})\cr
\V \sfrac{d\hfill}{ds} u(K+sK')\big|_{s=0} \V_{1,\Si_j}
&\le\abcst\,\cb_{j-1}\Big[
\la_0^{1-\upsilon}\fe_j({\sst \| K\|_{1,\Si_{j}}})
\| K'\|_{1,\Si_{j}} +\| K'\|_{1,\Si_{j}}\Big]\cr
&\le\abcst\,
\fe_j({\sst \| K\|_{1,\Si_{j}}})\,\| K'\|_{1,\Si_{j}}\cr
}$$

\Item iii) By Lemma \lemOSumu\ of [FKTo3] with 
$\mu(t)=\varphi(t/M)-\varphi(M^{2(\il-j)}t/M)$, where $\varphi$ is the 
function used in Definition \defNPscales, and  $\La=M^j$
we have
$$\eqalign{
\V u(K)*\big(\nu^{(\ge j)}-\nu^{(\ge\il)}\big)^{\widehat{}}\ \V_{1,\Si_j}
&\le\const\, \cb_{j}\, \v u(K)\v_{1,\Si_j}\cr
&\le\const\, \cb_{j}\, \Big[
\sfrac{\la_0^{1-\upsilon}}{M^{j-1}}+\| K\|_{1,\Si_{j}}
\Big]\fe_j({\sst \| K\|_{1,\Si_{j}}})\cr
&\le\const\, \Big[
\sfrac{\la_0^{1-\upsilon}}{M^{j-1}}+\| K\|_{1,\Si_{j}}
\Big]\fe_j({\sst \| K\|_{1,\Si_{j}}})\cr
}$$

\endproof

\vfill\eject

\chap{One Recursion Step}\PG\pgNPIX

The data of Theorem \theoremNPinduction\ are constructed recursively. In this section, we implement one recursion step, analogous  to the map 
$\Om_j\circ\cO_j$ of \S\CHrenmap.

\titleb{Input and Output Data}\PG\pgNPIXa 
We now impose the actual conditions on the input and output data,
analogous to Definitions \defNPinputData\ and \defNPoutputData.
\definition{\STM\stepInputData (Input Data)}{
 The input data just before integrating 
out the $j^{\rm th}$ scale is the set $\cD^{(j)}_{\rm in}$ of 
interaction quadruples, in the sense of Definition \defNPintquad,
 $(\cW,\cG,u,\vec p)$  that fulfill
\item{(I1)}  
$\cW(K)$  has a $\Si_j$--sectorized representative 
$$\eqalign{
w(\phi,\psi;K) &= \smsum_{m,n}\ \smsum_{s_1,\cdots,s_n\in\Si_j}\
\int {\sst d\eta_1\cdots d\eta_m\,d\xi_1\cdots d\xi_n}\ 
w_{m,n}({\sst \eta_1,\cdots, \eta_m\,(\xi_1,s_1),\cdots ,(\xi_n,s_n)};K)\cr
& \hskip 5cm \phi({\sst \eta_1})\cdots \phi({\sst \eta_m})\
\psi({\sst (\xi_1,s_1)})\cdots \psi({\sst (\xi_n,s_n)\,})\cr
}$$
with $w_{m,n}$ antisymmetric separately in the $\eta$ and in the $\xi$ variables, $w_{0,2}=0$, $w_{m,0}=0$ for all $m\ge 0$ and 
$$\eqalign{
N_j\big(w(K),64\al,\|K\|_{1,\Si_j}\big) 
&\le \fe_j\big(\|K\|_{1,\Si_j}\big) \cr
N_j\big( \sfrac{d\hfill}{ds} w(K+sK')\big|_{s=0},64\al,\|K\|_{1,\Si_j}\big) 
   &\le M^j \,\fe_j\big(\|K\|_{1,\Si_j}\big) \,  \|K'\|_{1,\Si_j}
\cr
}$$
for all $ K\in\fK_{j}$ and all $ K'$.
\item{(I2)}  
There is a family $\vec F$ of  antisymmetric kernels
$$
F^{(i)}({\sst(\xi_1,s_1)},\cdots,{\sst(\xi_4,s_4)})\in\cF_0(4,\Si_i),\qquad
2\le i\le j-1
$$ 
(independent of $K$) and an antisymmetric kernel 
$\de F^{(j)}({\sst(\xi_1,s_1)},\cdots,{\sst(\xi_4,s_4)};K)\in\cF_0(4,\Si_j)$
 such that
$$
w_{0,4}(K)=\de F^{(j)}(K)+\smsum_{i=2}^{j-1}F^{(i)}_{\Si_j}+
\sfrac{1}{8}{\rm Ant\,}
\Big( V_{\rm ph}\big( \cL^{(j)}(\vec p,\vec F)\big) \Big)_{\Si_j}
$$
where the particle--hole value $V_{\rm ph}$ was defined in 
Definition \defParicleHoleDecomp. Furthermore, for all $2\le i\le j-1$,
$$\eqalign{
\V F^{(i)}\V_{3,\Si_i}&\le \sfrac{\la_0^{1-\upsilon}}{\al^7}\,\vi_i\,\cb_i\cr
}$$ 
and
$$
\V \de F^{(j)}(K)\V_{3,\Si_j}\le 
\sfrac{\la_0^{1-\upsilon}}{\al^4}\Big\{
      \sfrac{\vi_j}{\al^3}
     +\sfrac{1}{\IB^2}  M^j\|K\|_{1,\Si_j}
\Big\}\fe_j({\sst \| K\|_{1,\Si_{j}}})
\qquad\hbox{for all }K\in\fK_{j}
$$
\item{(I3)}
For each $K\in\fK_{j}$
$$\meqalign{
N\big(\cG(K)-\half\phi JC^{(<j)}J\phi\big) &\le 4\smsum_{i=2}^{j-1} 
\sfrac{1}{\root{4}\of{\fl_iM^i}} \cr
}$$
Let $\cG_2(K)=\int d\eta_1 d\eta_2\ G_{2}(\eta_1, \eta_2,K)\ \phi(\eta_1)\phi(\eta_2)$
be the part of $\cG(K)$ that is homogeneous of degree two. Then
$$\eqalign{
N\big(\sfrac{d\hfill}{ds}\big[\cG(K+sK')-\cG_2(K+sK')\big]_{s=0}\big)
&\le \half M^j \|K'\|_{1,\Si_j} \cr
\TN \sfrac{d\hfill}{ds}G_2(K+sK')\big|_{s=0}\TN_\infty
&\le \half M^j \|K'\|_{1,\Si_j} \cr
}$$
for all $K\in\fK_{j}$ and all $K'$.

\noindent The input data is said to be {\bf real} if
\item{$\circ$} the interaction quadruple $(\cW,\cG,u,\vec p)$ is real
\item{$\circ$} for real $K$, the $\Si_j$--sectorized representative
$w(\phi,\psi;K)$ of $\cW(K)$ is $k_0$--reversal real, in the sense 
of Definition \defOSsymmetries.R of [FKTo2] and
\item{$\circ$} each $F^{(i)}$ is $k_0$--reversal real

}

\definition{\STM\stepOutputData (Output Data)}{
The output data just after integrating 
out the $j^{\rm th}$ scale is the set $\cD^{(j)}_{\rm out}$ of interaction 
quadruples $(\cW,\cG,u,\vec p)$ that fulfill
\item{(O1)}  
$\cW(K)$  has a $\Si_j$--sectorized representative 
$$\eqalign{
w(\phi,\psi;K) &= \smsum_{m,n}\ \smsum_{s_1,\cdots,s_n\in\Si_j}\
\int {\sst d\eta_1\cdots d\eta_m\,d\xi_1\cdots d\xi_n}\ 
w_{m,n}({\sst \eta_1,\cdots, \eta_m\,(\xi_1,s_1),\cdots ,(\xi_n,s_n)};K)\cr
& \hskip 5cm \phi({\sst \eta_1})\cdots \phi({\sst \eta_m})\
\psi({\sst (\xi_1,s_1)})\cdots \psi({\sst (\xi_n,s_n)\,})\cr
}$$
with $w_{m,n}$ antisymmetric separately in the $\eta$ and in the $\xi$ variables, $w_{m,0}=0$ for all $m\ge 0$ and 
$$\eqalign{
\V w_{0,2}(K)\V_{1,\Si_j} 
& \le\sfrac{\la_0^{1-\upsilon}}{\al^{7}} \,\sfrac{\fl_j}{M^j}\,
 \fe_j\big(\|K\|_{1,\Si_j}\big)\cr
N_j\big(w(K),\al,\|K\|_{1,\Si_j}\big) 
&\le \fe_j\big(\|K\|_{1,\Si_j}\big) \cr
N_j\big(\sfrac{d\hfill}{ds} w(K+sK')\big|_{s=0}
,\al,\,\| K\|_{1,\Si_{j}}\big) 
   &\le M^j \,\fe_j\big(\|K\|_{1,\Si_j}\big) \,  \|K'\|_{1,\Si_j}
\cr
}$$
for all $ K\in\fK_{j}$ and all $ K'$.
\item{(O2)}  
There is a family $\vec F$ of  antisymmetric kernels
$$
F^{(i)}({\sst(\xi_1,s_1)},\cdots,{\sst(\xi_4,s_4)})\in\cF_0(4,\Si_i),\qquad
2\le i\le j
$$ 
(independent of $K$) and an antisymmetric kernel 
$\de F^{(j+1)}({\sst(\xi_1,s_1)},\cdots,{\sst(\xi_4,s_4)};K)\in\cF_0(4,\Si_j)$
 such that
$$
w_{0,4}(K)=\de F^{(j+1)}(K)+\smsum_{i=2}^{j}F^{(i)}_{\Si_j}+
\sfrac{1}{8}{\rm Ant\,}\Big( V_{\rm ph}\big( \cL^{(j+1)}(\vec p,\vec F)\big) \Big)
$$
 Furthermore, 
$$\meqalign{
\V F^{(i)}\V_{3,\Si_i}&\le \sfrac{\la_0^{1-\upsilon}}{\al^7}\,\vi_i\,\cb_i&&
&\hbox{for all $2\le i\le j$}\cr
}$$
and
$$
\V\de F^{(j+1)}(K)\V_{3,\Si_j}
\le \sfrac{\la_0^{1-\upsilon}}{\al^4}\Big\{
      \sfrac{\vi_{j+1}}{\al^4}
     +\sfrac{1}{\IB^2}  M^j\|K\|_{1,\Si_j}
\Big\}\fe_j({\sst \| K\|_{1,\Si_{j}}})
\qquad\hbox{for all }K\in\fK_{j}
$$
\item{(O3)}
For each $K\in\fK_{j}$, 
$$\meqalign{
N\big(\cG(K)-\half\phi JC^{(\le j)}J\phi\big) 
&\le 4\smsum_{i=2}^{j-1} \sfrac{1}{\root{4}\of{\fl_iM^i}} 
+ \sfrac{2}{\root{4}\of{\fl_jM^j}}&&&  \hbox{ for all }K\in\fK_{j} \cr
}$$
Let $\cG_2(K)=\int d\eta_1 d\eta_2\ G_{2}(\eta_1, \eta_2,K)\ \phi(\eta_1)\phi(\eta_2)$
be the part of $\cG(K)$ that is homogeneous of degree two. Then
$$\eqalign{
N\big(\sfrac{d\hfill}{ds}\big[\cG(K+sK')-\cG_2(K+sK')\big]_{s=0}\big)
&\le M^j \|K'\|_{1,\Si_j} \cr
\TN \sfrac{d\hfill}{ds}G_2(K+sK')\big|_{s=0}\TN_\infty
&\le M^j \|K'\|_{1,\Si_j} \cr
}$$
for all $K\in\fK_{j}$ and all $K'$.

\noindent The output data is said to be {\bf real} if
\item{$\circ$} the interaction quadruple $(\cW,\cG,u,\vec p)$ is real
\item{$\circ$} for real $K$, the $\Si_j$--sectorized representative
$w(\phi,\psi;K)$ of $\cW(K)$ is $k_0$--reversal real
\item{$\circ$} each $F^{(i)}$ is $k_0$--reversal real.

}
\remark{\STM\remNPOandRconditions}{ Conditions (O1) and (O2) 
coincide with conditions (R1) and (R2)  of Theorem \theoremNPinduction,
while condition (O3) implies condition (R3).

}
\goodbreak
\vskip.25in
\titleb{Integrating Out a Scale}\PG\pgNPIXb
In this subsection we implement the map $\Om_j:\cD^{(j)}_{\rm in}\rightarrow
\cD^{(j)}_{\rm out}$, analogous to that of Definition \defNPinoutmap, that integrates out fields of scale $j$. We use the covariances
$$\eqalignno{
C^{(j)}_u(k) &= \sfrac{\nu^{(j)}(k)}
{\imath k_0 -e(\k) -\check u(k;K) }\cr
C^{(\ge j)}_u(k) &= \sfrac{\nu^{(\ge j)}(k)}
{\imath k_0 -e(\k) -\check u(k;K) }\cr
D_j(u;K)(k) &= \sfrac{\nu^{(\ge j+1)}(k)}
{\imath k_0 -e(\k) -\check u(k;K) - \check K(\k)\nu^{(\ge j+2)}(k)}\cr
C_j(u;K)(k) &= C^{(j)}_u(k)+D_j(u;K)(k)\cr
}$$
from Definition \defNPformalCovariances\ and Remark \remNPinoutmap.

\definition{\STM\defNPomegaj}{
Integrating out the fields of scale $j$ is implemented
by the map $\Om_j$, 
which maps an interaction quadruple $(\cW,\cG,u,\vec p)\in\cD^{(j)}_{\rm in}$ 
 to the  quadruple  $(\cW',\cG',u,\vec p)$  determined by 
$$\eqalign{
\lw\cW'(\phi,\psi;K)\rw_{\psi,D_{j}(u;K)}
&=\log\sfrac{1}{Z(\phi)}\int e^{\phi J\ze}\, 
e^{\lw \cW(\phi,\psi+\ze;K)\rw_{\psi,C_{j}(u;K)}}
\, d\mu_{C^{(j)}_{u(K)}}(\ze)\cr
\cG'(\phi) &= \cG(\phi) 
+\log\sfrac{Z(\phi)}{Z(0)}\cr
}$$
where 
$$
\log Z(\phi)=\int\Big[\log\int e^{\phi J\ze}\, 
e^{\lw \cW(\phi,\psi+\ze;K)\rw_{\psi,C_{j}(u;K)}}\, 
d\mu_{C^{(j)}_{u(K)}}(\ze)\Big]d\mu_{D_j(u;K)}(\psi)
$$
}

\theorem{\STM\thmIntoOut}{
If $(\cW,\cG,u,\vec p)\in\cD^{(j)}_{\rm in}$ then $\Om_j(\cW,\cG,u,\vec p)\in\cD^{(j)}_{\rm out}$.
}
\prf
Let
$\ 
(\cW',\cG',u,\vec p)=\Om_j(\cW,\cG,u,\vec p)
\ $.
As in Definition \defNPinoutmap, one easily verifies that 
$(\cW',\cG',u,\vec p)$ is an interaction quadruple of scale $j$.

Define $\cW''$ by
$$\eqalign{
\lW \cW''(\phi,\psi;K)\rW_{\psi,D_j(u;K)}
&=\tilde\Om_{C^{(j)}_u}
\big(\lW \cW(\phi,\psi;K)\rW_{\psi,C_j(u;K)}\big)\cr
&=\log\sfrac{1}{Z(0)}\int e^{\phi J\ze}\ 
e^{\lW \cW(\phi,\psi+\ze;K)\rW_{\psi,C_j(u;K)}}\, d\mu_{C^{(j)}_u}(\ze)
}$$
Then
$$\eqalign{
\cW'(\phi,\psi;K)& = \cW''(\phi,\psi;K)-\cW''(\phi,0;K)\cr
\cG'(\phi) &= \cG(\phi) +\cW''(\phi,0;K)\cr
}\EQN\eqnNPwpwppgp$$
We now apply Theorems \thOSrengroupestimate\ and
\thOSrengroupdiffestimate\ of [FKTo3] to bound $\cW''$. 
By (I1) and parts (i) and (ii) of Lemma \lemNPpptyu, the hypotheses of 
Theorem \thOSrengroupestimate\ of [FKTo3], with $\mu=\abcst$, 
$\La=\sfrac{\la_0^{1-\upsilon}}{M^{j-1}}$ and $X=\big\| K\big\|_{1,\Si_{j}}$,
are fulfilled. Therefore $\cW''$ has a sectorized representative $w''$
obeying
$$\eqalign{
N_j(w''-\sfrac{1}{2}\phi JC^{(j)}_uJ\phi-w,\al,\,\| K\|_{1,\Si_{j}}) 
&\le \sfrac{\const}{\al}\,
\sfrac{N_j(w,64\al,\,\| K\|_{1,\Si_{j}})}{1-{\const \over\al}N_j(w,64\al,\,\| K\|_{1,\Si_{j}})}\cr
\v w_{0,2}'' \v_{1,\Si_j}  
&\le  \sfrac{\const}{\al^8\,\rho_{0;2}}\sfrac{ \fl_j}{M^j}\,
\sfrac{N_j(w,64\al,\,\| K\|_{1,\Si_{j}})^2}
          {1-{\const \over\al}N_j(w,64\al,\,\| K\|_{1,\Si_{j}})}\cr
\V w_{0,4}''\!-\!w_{0,4} -
\sfrac{1}{4}\! \smsum_{\ell=1}^\infty \!\!{\sst (-1)^\ell(12)^{\ell+1} }
L_\ell(w_{0,4};{\sst C^{(j)}_{u}, D_j(u;K)})
\V_{3,\Si_j} 
 &\le\sfrac{\const}{\al^{10}\rho_{0;4}}\,\fl_j\,
\sfrac{N_j(w,64\al,\,\| K\|_{1,\Si_{j}})^2}
           {1-{\const \over\al}N_j(w,64\al,\,\| K\|_{1,\Si_{j}})}\cr
}\EQN\eqnNPodI$$
The hypotheses of Theorem \thOSrengroupdiffestimate\
of [FKTo3], with, in addition, $Y=\abcst\big\| K'\big\|_{1,\Si_{j}}$
and $\veps=\const M^j\,\| K'\|_{1,\Si_{j}}\big|_{t=0}$,
are fulfilled. Hence 
$$\eqalign{
&N_j\big(\sfrac{d\hfill}{ds}\big[w''(K+sK')
          -\sfrac{1}{2}\phi JC^{(j)}_{u(K+sK')}J\phi\big]_{s=0}
,\al,\,\| K\|_{1,\Si_{j}}\big)\cr
&\hskip0.3in\le 
N_j\big(\sfrac{d\hfill}{ds}w(K+sK')\big|_{s=0}
,\al,\,\| K\|_{1,\Si_{j}}\big)
+\sfrac{\const}{\al}N_j\big(\,\sfrac{d\hfill}{ds}w(K+sK')\big|_{s=0}
,16\al,\,\| K\|_{1,\Si_{j}}\big)\cr
&\hskip0.6in+\sfrac{\const}{\al}\,
     \sfrac{N_j(w,64\al,\,\| K\|_{1,\Si_{j}})}
     {1-{\const\over\al^2}N_j(w,64\al,\,\| K\|_{1,\Si_{j}})}
     \Big\{ N_j\big(\,\sfrac{d\hfill}{ds}w(K+sK')\big|_{s=0}
,16\al,\,\| K\|_{1,\Si_{j}}\big)+M^jY\Big\}\cr
&\hskip0.6in+\sfrac{\const}{\al^2}\,
     \sfrac{N_j(w,64\al,\,\| K\|_{1,\Si_{j}})^2}
      {1-{\const\over\al^2}N_j(w,63\al,\,\| K\|_{1,\Si_{j}})}
\Big\{M^{j}Y+\veps\Big\}
}\EQN\eqnNPodIb$$

By (I1) and Corollary \corOSappMonoidIV.ii of [FKTo1],
$$
\sfrac{N_j(w,64\al,\,\| K\|_{1,\Si_{j}})}{1-{\const \over\al}N_j(w,64\al,\,\| K\|_{1,\Si_{j}})}
\le \sfrac{\fe_j(\| K\|_{1,\Si_{j}})}
{1-{\const \over\al}\fe_j(\| K\|_{1,\Si_{j}})}
\le \const \fe_j(\| K\|_{1,\Si_{j}})
$$
and
$$
\sfrac{N_j(w,64\al,\,\| K\|_{1,\Si_{j}})^2}
          {1-{\const \over\al}N_j(w,64\al,\,\| K\|_{1,\Si_{j}})}
\le \const \fe_j(\| K\|_{1,\Si_{j}})^2
\le \const \fe_j(\| K\|_{1,\Si_{j}})
$$
Therefore, by (\eqnNPodI) and (I1), recalling that $w$ vanishes when $\psi=0$,
$$\eqalignno{
N_j(w''-\sfrac{1}{2}\phi JC^{(j)}_uJ\phi,\al,\,\| K\|_{1,\Si_{j}}) 
&\le N_j(w,\al,\,\| K\|_{1,\Si_{j}})
+\sfrac{\const}{\al}\,\fe_j(\| K\|_{1,\Si_{j}})\cr
&\le\sfrac{1}{64} N_j(w,64\al,\,\| K\|_{1,\Si_{j}})
+\sfrac{\const}{\al}\,\fe_j(\| K\|_{1,\Si_{j}})\cr
&\le \fe_j(\| K\|_{1,\Si_{j}})
&\EQNO\eqnNPodII}$$
and
$$
\v w_{0,2}'' \v_{1,\Si_j}  
\le  \const\sfrac{\la_0^{1-\upsilon}}{\al^8}\sfrac{ \fl_j}{M^j}\,
\fe_j(\| K\|_{1,\Si_{j}})
\le  \sfrac{\la_0^{1-\upsilon}}{\al^7}\sfrac{ \fl_j}{M^j}\,
\fe_j(\| K\|_{1,\Si_{j}})
\EQN\eqnNPodIII$$
and, by (\eqnNPodIb) and (I1),
$$\eqalignno{
&N_j\big(\sfrac{d\hfill}{ds}\big[w''(K+sK')
            -\sfrac{1}{2}\phi JC^{(j)}_{u(K+sK')}J\phi\big]_{s=0}
,\al,\,\| K\|_{1,\Si_{j}}\big) \cr
&\hskip0.5in\le N_j\big(\sfrac{d\hfill}{ds}w(K+sK')\big|_{s=0}
,\al,\,\| K\|_{1,\Si_{j}}\big)
+\sfrac{\const}{\al}\,\fe_j(\| K\|_{1,\Si_{j}})\,M^j\| K'\|_{1,\Si_{j}}\cr
&\hskip0.5in\le\sfrac{1}{64} N_j\big(\sfrac{d\hfill}{ds}w(K+sK')\big|_{s=0}
,64\al,\,\| K\|_{1,\Si_{j}}\big)
+\sfrac{\const}{\al}\,\fe_j(\| K\|_{1,\Si_{j}})\,M^j\| K'\|_{1,\Si_{j}}\cr
&\hskip0.5in\le M^j\fe_j(\| K\|_{1,\Si_{j}})\,\| K'\|_{1,\Si_{j}}
&\EQNO\eqnNPodIIb}$$
By (\eqnNPwpwppgp), $w'(\phi,\psi;K)=w''(\phi,\psi;K)-w''(\phi,0;K)$ is
a $\Si_j$--sectorized representative for $\cW'$. (O1) now follows from
(\eqnNPodIII), (\eqnNPodII) and (\eqnNPodIIb).

\vskip.25in
In preparation for the verification of (O2), set
$$
\de F'_1(K)=w_{0,4}''(K)-w_{0,4}(K) -
\sfrac{1}{4} \smsum_{\ell=1}^\infty (-1)^\ell(12)^{\ell+1} 
{\rm Ant\,}L_\ell\big(w_{0,4}(K);C^{(j)}_{u(K)}, D_j(u(K);K)\big)
$$
By (\eqnNPodI), 
$$
\V \de F'_1(K)\V_{3,\Si_j} 
 \le\sfrac{\la_0^{1-\upsilon}}{\al^9}\,\fl_j\,\fe_j(\| K\|_{1,\Si_{j}}) 
$$
In particular
$$
\V \de F'_1(0)\V_{3,\Si_j} 
 \le\sfrac{\la_0^{1-\upsilon}}{\al^9}\,\fl_j\,\cb_j
\EQN\eqnNPodFI$$
Define
$$\eqalign{
\de F'_2(K)&= w_{0,4}''(K) - w_{0,4}''(0) \cr
\de F'_3&=\sfrac{1}{4} \smsum_{\ell=1}^\infty {\sst (-1)^\ell(12)^{\ell+1}}
         {\rm Ant\,} V_{\rm pp} \bigg(L_\ell\Big(w_{0,4}(0);C^{(j)}_{u(0)}, 
                  C^{(\ge j+1)}_{u(0)}\Big)^{\rm pp} \bigg)\cr
}$$
and
$$
{\de F'}^{(j+1)}(K)=\de F'_1(0)+\de F'_2(K)+\de F'_3
$$
Observe that $D_j(u(0);0) = C^{(\ge j+1)}_{u(0)}$ and, by Lemma \lemunordord.ii,
$$\eqalign{
L_\ell\big(w_{0,4}(0);C^{(j)}_{u(0)},  C^{(\ge j+1)}_{u(0)}\big)
&= V_{\rm pp} \bigg(L_\ell\Big(w_{0,4}(0);C^{(j)}_{u(0)}, 
                  C^{(\ge j+1)}_{u(0)}\Big)^{\rm pp} \bigg)\cr
&\hskip3cm+V_{\rm ph} \bigg(L_\ell\Big(w_{0,4}(0);C^{(j)}_{u(0)}, 
                  C^{(\ge j+1)}_{u(0)}\Big)^{\rm ph} \bigg) \cr
}$$ 
Therefore
$$
w''_{0,4}(K)
=w_{0,4}(0) + {\de F'}^{(j+1)}(K)
+ \sfrac{1}{4}  \smsum_{\ell=1}^\infty {\sst (-1)^\ell(12)^{\ell+1}}
{\rm Ant\,} V_{\rm ph}\bigg(L_\ell\Big(w_{0,4}(0);C^{(j)}_{u(0)}, 
                  C^{(\ge j+1)}_{u(0)}\Big)^{\rm ph} \bigg)
\EQN\eqnNPodFIa$$

We now estimate $\V {\de F'}^{(j+1)}(K)\V_{3,\Si_{j}}$.
By (\eqnNPodIIb), 
$$\eqalignno{
\V \de F'_2\V_{3,\Si_{j}}
&\le \VV \int_0^1\sfrac{d\hfill}{ds'} w''_{0,4}(sK+s'K)\Big|_{s'=0}\
ds\VV_{3,\Si_{j}}\cr 
&\le \sfrac{\la_0^{1-\upsilon}}{\al^4\IB^2}\, 
\fe_j({\sst \| K\|_{1,\Si_{j}}}) \, 
M^j \|K\|_{1,\Si_j}
&\EQNO\eqnNPodFII\cr
}$$
By Proposition \propparticleparticleladder, for $\ell \ge 1$
$$
\VV  V_{\rm pp} \Big(L_\ell\Big(w_{0,4}(0);C^{(j)}_{u(0)}, 
                  C^{(\ge j+1)}_{u(0)}\Big)^{\rm pp} \Big)
\VV_{3,\Si_{j}} 
\le \big(\const\vi_j\cb_j\big)^\ell\, \v w_{0,4}(0)\v_{3,\Si_{j}}^{\ell+1}
$$
The hypotheses of this Proposition are fulfilled by parts (i) and (ii) of 
Lemma \lemNPpptyu. Observe that, by (I1),
$$
\sfrac{M^{2j}}{\fl_j}\fe_j({\sst \| K\|_{1,\Si_{j}}})
(64\al)^4\big(\sfrac{\fl_j\IB}{M^j}\big)^2\ 
\sfrac{1}{\la_0^{1-\upsilon}}\sfrac{1}{\fl_j}\ 
\v w_{0,4}(K)\v_{3,\Si_{j}}
\le \fe_j({\sst \| K\|_{1,\Si_{j}}})
$$
so that
$$
\v w_{0,4}(K)\v_{3,\Si_{j}}\le\sfrac{\la_0^{1-\upsilon}}{(64\al)^4\IB^2} 
\fe_j({\sst \| K\|_{1,\Si_{j}}})
$$
In particular,
$$
\v w_{0,4}(0)\v_{3,\Si_{j}}\le\sfrac{\la_0^{1-\upsilon}}{(64\al)^4\IB^2} 
\cb_j
$$
Therefore,  by Corollary \:\corOSappMonoidIV.ii of [FKTo1],
$$\eqalignno{
\V \de F'_3\V_{3,\Si_{j}}
&\le \smsum_{\ell=1}^\infty \big(\const\vi_j\cb_j\big)^\ell\, 
       \v w_{0,4}(0)\v_{3,\Si_{j}}^{\ell+1}\cr
&\le \smsum_{\ell=1}^\infty \big(\const\vi_j\cb_j\big)^\ell\, 
       \big(\sfrac{\la_0^{1-\upsilon}}{\al^4} 
\cb_j\big)^{\ell+1}\cr
&\le  \const \big(\sfrac{\la_0^{1-\upsilon}}{\al^4}\big)^2\vi_j\ \cb_j^3
\ \frac{1}{1-\const \sfrac{\la_0^{1-\upsilon}}{\al^4}
            \vi_j\,\cb_j^2 }\cr
&\le  \const \sfrac{\la_0^{2-2\upsilon}}{\al^8} \vi_j\ \cb_j
&\EQNO\eqnNPodFIII\cr
}$$
Hence, by  (\eqnNPodFI), (\eqnNPodFII) and (\eqnNPodFIII), 
$$\eqalignno{
\V {\de F'}^{(j+1)}(K)\V_{3,\Si_{j}}
&\le \sfrac{\la_0^{1-\upsilon}}{\al^4}\Big\{
      \sfrac{\fl_j}{\al^5}
     + \sfrac{1}{\IB^2}  M^j\|K\|_{1,\Si_j}
     + \sfrac{\const\la_0^{1-\upsilon}}{\al^4}\vi_j
     \Big\}\,\fe_j({\sst \| K\|_{1,\Si_{j}}})\cr
&\le \sfrac{\la_0^{1-\upsilon}}{\al^4}\Big\{
      \sfrac{\vi_{j+1}}{\al^4}
     +\sfrac{1}{\IB^2}\,  M^j\|K\|_{1,\Si_j}
\Big\}\fe_j({\sst \| K\|_{1,\Si_{j}}})
&\EQNO\eqnNPFjplusI\cr
}$$

\vskip0.3cm
\goodbreak

To verify (O2), set ${F'}^{(i)}=F^{(i)}$ for all $0\le i\le j-1$, 
${F'}^{(j)}=\de F^{(j)}(0)$.
By (I2)
$$
w_{0,4}(0) = \sum_{i=2}^{j}{F'}_{\Si_j}^{(i)}
+\sfrac{1}{8} {\rm Ant\,}\Big( V_{\rm ph} 
\big( \cL^{(j)}(\vec p,\vec F')\big)\Big)_{\Si_j}
$$
Therefore, by (\eqnNPodFIa) and the Definition
\defcompLadder\ of iterated particle--hole ladders 
$$\eqalignno{
w'_{0,4}(K)
&=\sum_{i=2}^{j}{F'}_{\Si_j}^{(i)} +{\de F'}^{(j+1)}(K)
+\sfrac{1}{8} {\rm Ant\,}
\Big( V_{\rm ph} \big( \cL^{(j)}(\vec p,\vec F')\big)\Big)_{\Si_j}\cr
&\hskip.5in
+\sfrac{1}{4}  \smsum_{\ell=1}^\infty {\sst (-1)^\ell(12)^{\ell+1}}
{\rm Ant\,} V_{\rm ph} \bigg(L_\ell\Big(w_{0,4}(0);C^{(j)}_{u(0)}, 
                  C^{(\ge j+1)}_{u(0)}\Big)^{\rm ph} \bigg)\cr
&=\sum_{i=2}^{j}{F'}_{\Si_j}^{(i)}+{\de F'}^{(j+1)}(K)
+\sfrac{1}{8} {\rm Ant\,}\Big( V_{\rm ph} 
\big( \cL^{(j+1)}(\vec p,\vec F')\big)\Big)\cr
}$$
The estimate on $\V {\de F'}^{(j+1)}(K)\V_{3,\Si_{j}}$ required for (O2) was
verified in (\eqnNPFjplusI). By (I2)
$$
\V {F'}^{(j)}\V_{3,\Si_j}
=\V \de F^{(j)}(0)\V_{3,\Si_j}\le 
\sfrac{\la_0^{1-\upsilon}}{\al^7}\vi_j\cb_j
$$
and
$\ \V {F'}^{(i)}\V_{3,\Si_i}
=\V F^{(i)}\V_{3,\Si_i}\le 
\sfrac{\la_0^{1-\upsilon}}{\al^7}\vi_i\cb_i
\ $ for $2\le i\le j-1$.

\vskip.6cm
This leaves only the verification of (O3).
By (\eqnNPwpwppgp)
$$\eqalign{
\cG'(\phi;K) - \half\phi JC^{(\le j)}J\phi
&= \Big(\cG(\phi;K)  - \half\phi JC^{(<j)}J\phi \Big)\cr
&\hskip .5in  +  \Big(w'' (\phi,0;K ) - \half\phi JC^{(j)}_{u(K)}J\phi\Big)\cr
&\hskip.5in- \half \phi J\big(  C^{(j)}-\phi C^{(j)}_{u(K)}\big)J\phi
}$$
Now
$$
C^{(j)}(k)- C^{(j)}_{u(K)}(k)
=-\nu^{(j)}(k)\sfrac{\check u(k;K)}
{[\imath k_0-e(\k)][\imath k_0-e(\k)-\check u(k;K)]}
$$
By Lemma \lemNPpptyu.i,
$$
\big|\check u(k;K)+\check K(k)\nu^{(\ge j+2)}(k)\big|
\le\la_0^{1-\upsilon}\,|\imath k_0-e(\k)|
$$
On the support of $\nu^{(j)}$,  $\nu^{(\ge j+2)}(k)$ vanishes and
$\big|\check u(k;K)\big|\le\la_0^{1-\upsilon}\,|\imath k_0-e(\k)|$ so that
$$
\big|C^{(j)}(k)- C^{(j)}_{u(K)}(k)\big|
\le \sfrac{2\la_0^{1-\upsilon}}{|\imath k_0-e(\k)|}\nu^{(j)}(k)
$$   
On the support of $\nu^{(j)}$, $|\imath k_0-e(\k)|\le\sfrac{\sqrt{2M}}{M^j}$,
so that
$$\eqalign{
\TN C^{(j)}- C^{(j)}_{u(K)} \TN_\infty
&\le \int\sfrac{d^3 k}{(2\pi)^3}\sfrac{2\la_0^{1-\upsilon}}
                                     {|\imath k_0-e(\k)|}\nu^{(j)}(k)
\le \abcst\,\la_0^{1-\upsilon}
                \int_{|\imath x-y|\le {\sqrt{2M}\over M^j}} dx\,dy\ 
                         \sfrac{1}{|\imath x-y|}\cr
&\le\abcst\,\sfrac{\sqrt{M}}{M^j}\la_0^{1-\upsilon}
}$$
and
$$\eqalign{
 N\big(  \phi J(C^{(j)}-C^{(j)}_{u(K)})J\phi\big)
&=\sfrac{1}{\la_0^{1-\upsilon}}\TN J\big(C^{(j)}- C^{(j)}_{u(K)}\big)J \TN_\infty
\le\sfrac{1}{\la_0^{1-\upsilon}}\TN C^{(j)}- C^{(j)}_{u(K)} \TN_\infty\cr
&\le\abcst\, \sfrac{\sqrt{M}}{M^j}
=\abcst\, \sfrac{\sqrt{M}}{M^{j(3+\aleph)/4}}\sfrac{1}{\root{4}\of{\fl_jM^j}}\cr
&\le \sfrac{1/2}{\root{4}\of{\fl_jM^j}}\cr
}$$
if $j\ge 1$ and $M$ is big enough.
Therefore, by (\eqnNPodII) and Remark \remNPnvsnj.ii,
$$\eqalign{
&N\big( \cG'(\phi;K) -\half\phi JC^{(j)}J\phi-\cG(\phi;K)\big) \cr
&\hskip.3in\le N\Big( w'' (\phi,0;K ) - \half\phi JC^{(j)}_{u(K)}J\phi \Big) 
  + \half N\big( \phi J(C^{(j)}- C^{(j)}_{u(K)})J\phi\big)\cr
&\hskip.3in\le
   N_j\big(w''(K)-\half\phi JC^{(j)}_{u(K)}J\phi\,,\,\al\,,\,
        \| K\|_{1,\Si_j}\big)_\0
   +\sfrac{1/2}{\root{4}\of{\fl_jM^j}}\cr
&\hskip.3in\le \sfrac{2}{\root{4}\of{\fl_jM^j}}
}\EQN\eqnNPgdiffinout$$  
Consequently, by (I3),
$$\eqalign{
&N\big( \cG'(\phi;K) - \half\phi JC^{(\le j)}J\phi\big) \cr
&\hskip.3in\le N\Big( \cG(\phi;K)  - \half\phi JC^{(<j)}J\phi  \Big)
  + N\big( \cG'(\phi;K) - \half\phi JC^{(j)}J\phi-\cG(\phi;K)\big) \cr
&\hskip.3in\le 4\smsum_{i=2}^{j-1}  \sfrac{1}{\root{4}\of{\fl_iM^i}}
   + \sfrac{2}{\root{4}\of{\fl_jM^j}}\cr
}$$  

\vskip.3cm
\noindent By (\eqnNPwpwppgp), (\eqnNPodIIb) and Remark \remNPnvsnj.ii,
$$\eqalign{
&N\Big(\sfrac{d\hfill}{ds}\big[\cG'(\phi;{\sst K+sK'})-\cG(\phi;{\sst K+sK'})
-\sfrac{1}{2}\phi JC^{(j)}_{u(K+sK')}J\phi\big]_{s=0}\Big)\cr
&\hskip1in
=N\Big(\sfrac{d\hfill}{ds}\big[w''(\phi,0;{\sst K+sK'})
-\sfrac{1}{2}\phi JC^{(j)}_{u(K+sK')}J\phi\big]_{s=0}\Big)\cr
&\hskip1in
\le \sfrac{1}{\root{4}\of{\fl_jM^j}}\,
N_j\Big(\sfrac{d\hfill}{ds}\big[w''({\sst K+sK'})
-\sfrac{1}{2}\phi JC^{(j)}_{u(K+sK')}J\phi\big]_{s=0}\,,\,\al\,,\,
        \| K\|_{1,\Si_j}\Big)_\0\cr
&\hskip1in
\le  \sfrac{1}{\root{4}\of{\fl_jM^j}}\,M^j\,
\fe_j({\sst \|K\|_{1,\Si_j}})\|K'\|_{1,\Si_j}\big|_{t=0}\cr
&\hskip1in\le  \sfrac{1}{4}\,M^{j}\|K'\|_{1,\Si_j} \cr
}\EQN\eqnNPderivgprimeI$$
It follows directly from this bound and (I3) that
$$\eqalign{
N\big(\sfrac{d\hfill}{ds}\big[\cG(K+sK')-\cG_2(K+sK')\big]_{s=0}\big)
&\le M^j \|K'\|_{1,\Si_j} \cr
}$$
\vskip.3cm\noindent
To prove the last inequality of (O3), observe that, by parts (i) and (ii) of
 Lemma \lemNPpptyu\  and Lemma \lemOSNormMom\ of [FKTo3],
$$\eqalign{
\Big|\sfrac{d\hfill}{ds}C^{(j)}_{u(K+sK')}\big|_{s=0}\Big|
&=\Big|\sfrac{{d\hfill\over ds}\check u(k;K+sK')|_{s=0}}
{[\imath k_0-e(\k)-\check u(k;K)]^2}\Big|\nu^{(j)}(k)\cr
&\le \abcst\sfrac{\|K' \|_{1,\Si_{j}}}
{|\imath k_0-e(\k)|^2}\nu^{(j)}(k)
}$$
Hence
$$\eqalign{
\TN \sfrac{d\hfill}{ds}C^{(j)}_{u(K+sK')}\big|_{s=0} \TN_\infty
&\le \abcst\,\|K' \|_{1,\Si_j}
\int d^3k \ \sfrac{\nu^{(j)}(k)}{|\imath k_0-e(\k)|^2}\cr
&\le \abcst\,\|K' \|_{1,\Si_j}
\int_{{1\over\sqrt{M}}{1\over M^j}
              \le r\le \sqrt{2M}{1\over M^j}} dr\ \sfrac{1}{r} \cr
&= \abcst\,\|K' \|_{1,\Si_j}\ln\big(\sqrt{2}\,M\big)\cr
&\le  \sfrac{1}{4}M^j\|K' \|_{1,\Si_{j}}\cr
}\EQN\eqnNPderivgprimeIV$$
Combining (I3), (\eqnNPderivgprimeI) and (\eqnNPderivgprimeIV),
$$\eqalign{
&\TN\sfrac{d\hfill}{ds}G'_2(K+sK')\big|_{s=0}\TN_\infty\cr
&\hskip.5in\le 
\TN\sfrac{d\hfill}{ds}G_2(K+sK')\big|_{s=0}\TN_\infty\cr
&\hskip.9in+\la_0^{1-\upsilon}N\Big(\sfrac{d\hfill}{ds}\big[\cG'(\phi;{\sst K+sK'})-\cG(\phi;{\sst K+sK'})
-\sfrac{1}{2}\phi JC^{(j)}_{u(K+sK')}J\phi\big]_{s=0}\Big)
       \cr
&\hskip.9in
+\half\TN \sfrac{d\hfill}{ds}C^{(j)}_{u(K+sK')}\big|_{s=0} \TN_\infty\cr
&\hskip.5in\le  M^j\|K' \|_{1,\Si_{j}}
}$$
\endproof
\remark{\STM\remNPFnochangeI}{
Let $(\cW,\cG,u,\vec p)\in\cD^{(j)}_{\rm in}$ and
 $(\cW',\cG',u,\vec p)=\Om_j(\cW,\cG,u,\vec p)$.
\Item (i)
The data ${F'}^{(2)},\ \cdots,\ {F'}^{(j-1)}$ of
(O2) coincides with the data  ${F}^{(2)},\ \cdots,\ {F}^{(j-1)}$ of
(I2). 
\Item (ii)
By (\eqnNPgdiffinout),
$\ 
N\big( \cG'(\phi;K) - \half\phi JC^{(j)}J\phi-\cG(\phi;K)\big)
\le \sfrac{2}{\root{4}\of{\fl_jM^j}}
\ $.
\Item (iii)
By Remark \remOSchoiceofrep\ of [FKTo3],
the sectorized representative $w'$ of $\cW'$ and the function $\cG'(\phi)$
may be obtained from the sectorized representative $w$ of $\cW$
 and the function $\cG(\phi)$ by
$$\eqalign{
\lW w''(\phi,\psi;K)\rW_{\psi,D_{j}(u;K)_{\Si_j}}
&=\half\phi JC^{(j)}_{u}J\phi
+\Om_{C^{(j)}_{u,\Si_j}}(\lw w \rw_{\psi,C_j(u;K)_{\Si_j}})(\phi,\psi+ C^{(j)}_u J\phi)\cr
w'(\phi,\psi;K)&=w''(\phi,\psi;K)-w''(\phi,0;K)\cr
\cG'(\phi) &= \cG(\phi) +w''(\phi,0;K)\cr
}$$  
The covariances $C^{(j)}_{u,\Si_j}$, $C_j(u;K)_{\Si_j}$ and $D_j(u;K)_{\Si_j}$
are defined as follows. Let $C(k)$ be one of $C^{(j)}_{u}(k)$, $C_j(u;K)(k)$,
$D_j(u;K)(k)$, as specified just before Definition \defNPomegaj\ and 
let $c\big((\cdot,s),(\cdot,s') \big)$ be the Fourier transform of 
$\chi_s(k)\,C(k)\,\chi_{s'}(k)$ as in (\eqnNPcovFT) and (\eqnNPantisymmCov).
Then
$$\eqalign{
C_\Si({\sst(\xi,s),(\xi',s')}) 
&= \sum_{t\cap s \ne \emptyset \atop t'\cap s' \ne \emptyset}
c({\sst (\xi,t),(\xi',t')}) \cr
}$$

\Item (iv) If the input data  are analytic functions of $K$, then, by 
Remark \remOSchoiceofrep\ of [FKTo3], the output data are analytic functions of the input data and $K$.
\Item (v) If the input data  is real, then, by 
Remark \remOSrengrppreserves\ of [FKTo2], the output data is real.

}

\goodbreak
\titleb{Sector Refinement, ReWick ordering and Renormalization}\PG\pgNPIXc
We now implement an analog with estimates of the formal power series map
$$
\cO_j:\cD^{(j)}_{\rm out}\rightarrow\cD^{(j+1)}_{\rm in}
$$
that reWick orders. For each $(\cW,\cG,u,\vec p)\in\cD^{(j)}_{\rm out}$ we choose a sectorized representative 
$$\eqalign{
w(\phi,\psi;K) = \smsum_{m,n}\ \smsum_{s_1,\cdots,s_n\in\Si_j}\
\int {\sst d\eta_1\cdots d\eta_m\,d\xi_1\cdots d\xi_n}\ & 
w_{m,n}({\sst \eta_1,\cdots, \eta_m\,(\xi_1,s_1),\cdots ,(\xi_n,s_n)};K)\cr
& \hskip 1cm \phi({\sst \eta_1})\cdots \phi({\sst \eta_m})\
\psi({\sst (\xi_1,s_1)})\cdots \psi({\sst (\xi_n,s_n)\,})\cr
}$$
for $\cW$ satisfying (O1) and (O2).

We first solve the re-Wick ordering equation, that -- in the context of 
formal power series -- was discussed in Lemma \lemNPformalselfconsistent.
As in (\eqnNPKuprime), we set, for any function $q: \fK_{j+1}\rightarrow \cF_0(2,\Si_j)$,
$$\eqalign{
 \de K\big((\x,s),(\x',s');K';q\big) &= \int dx_0\ 
 \big(q(K')*\hat \nu^{(\ge j+1)}\big)\big((x_0,\x,s),(0,\x',s')\big)\cr
K(K';q) &= K'_{\Si_j}+\de K(K';q) \cr
u'(K';q) &= u(K(K';q))_{\Si_{j+1}} + q(K')_{\Si_{j+1}}*\hat\nu^{(\ge j+1)}\cr
E(K';q)&=C_{j+1}\big( u'(\,\cdot\,;q);K' \big) - D_j( u;K(K';q))\cr
}\EQN\eqnNPreWickDefs$$
Let $e\big((\cdot,s),(\cdot,s') \big)$ be
the Fourier transform of 
$\chi_s(k)\,\check E(k;K';q)\,\chi_{s'}(k)$ 
in the sense of Definition \defOSftcov\ of [FKTo2]. As in
Proposition \propOSfunctorialitySect\ of [FKTo3], $e$  defines a covariance 
on the vector space $V_{\Si_j}$, generated by the fields $\psi(\xi,s)$, by
$$\eqalign{
E_{\Si_j}\big(\psi({\sst\xi,s}),\psi({\sst\xi',s'});K';q \big) 
&= \sum_{t\cap s \ne \emptyset \atop t'\cap s' \ne \emptyset}
e\big((\xi,t),(\xi',t')\big) \cr
}$$
Set
$$
\tilde w(\phi,\psi;K';q)
=\int w(\phi,\psi+\psi';K(K';q))\ d\mu_{E_{\Si_j}(K';q)}(\psi')
\EQN\eqnNPtwrewick$$
and expand
$$\eqalign{
\tilde w(0,\psi;K';q) =\sum_{n\ge 0} \ \smsum_{s_1,\cdots,s_n\in\Si_j}\
\int {\sst d\xi_1\cdots d\xi_n}\ & 
\tilde w_{0,n}({\sst (\xi_1,s_1),\cdots ,(\xi_n,s_n)};K';q)
\psi({\sst (\xi_1,s_1)})\cdots \psi({\sst (\xi_n,s_n)\,})\cr
}$$
\lemma{\STM\lemreWick}{
For each $K'\in \fK_{j+1}$
there exists an antisymmetric, spin independent, particle number conserving 
function $q_0(K')\in\cF_0(2,\Si_j)$ that solves the equation 
$$
\half q(K') = \tilde w_{0,2}(K';q(K'))
$$
and fulfills
$$\eqalign{
\v q_0(K')\v_{1,\Si_j} 
&\le \sfrac{\la_0^{1-\upsilon}}{\al^6}\sfrac{\fl_j}{M^j}\,
\fe_j({\sst \|K'\|_{1,\Si_{j+1}}})\, \cr
\V \sfrac{d\hfill}{ds} q_0(K'+sK'')\big|_{s=0}\V_{1,\Si_j}
&\le\sfrac{\la_0^{1-\upsilon}}{\al}\,\fe_j({\sst \|K'\|_{1,\Si_{j+1}}})\,
 \, \| K''\|_{1,\Si_{j+1}} \cr
}$$
If $w$ and $u$ are analytic in $K$, then $q_0$ is jointly analytic in $K'$,
$w$ and $u$.
}
\prf The proof is an application the implicit function Theorem and is given 
following Lemma \: \lemNPcontrmap\ in Appendix \APappRewick.
\endproof

Define, for $K'\in\fK_{j+1}$,
$$\eqalign{
\de K(K') &=\de K\big(K';q_0(K')\big)\cr
\ren_{j,j+1}(K',\cW,u)=K(K')&= K(K';q_0)\cr
}\EQN\eqndefkkprime$$
If $w$ and $u$ are analytic in $K$, then $\de K$ and $\ren_{j,j+1}$
are analytic in $K'$, $w$ and $u$. If the output data is real and $K'$ is real, then $\ren_{j,j+1}(K',\cW,u)$ is real.
\lemma{\STM\lemNPdeKbnd}{
There is a constant $\abcst$, independent of $M$
and $j$, such that if $K'\in\fK_{j+1}$, then
$$\leqalignno{
\| \de K(K')\|_{1,\Si_j}
&\le \sfrac{\la_0^{1-\upsilon}}{\al^6}\sfrac{\fl_j}{M^j}\,
\fe_j({\sst \|K'\|_{1,\Si_{j+1}}})&(i) \cr
\big\| \sfrac{d\hfill}{ds} \de K(K'+sK'')\big|_{s=0}\big\|_{1,\Si_j}
&\le\sfrac{\la_0^{1-\upsilon}}{\al}\,\fe_j({\sst \|K'\|_{1,\Si_{j+1}}})\,
 \, \| K''\|_{1,\Si_{j+1}} \cr
}$$
$$\leqalignno{ 
\|K(K')\|_{1,\Si_j} &\le \abcst\sfrac{\fl_j}{\fl_{j+1}}\cb_{j-1}\|K'\|_{1,\Si_{j+1}}
        +\sfrac{\la_0^{1-\upsilon}}{\al^6}\,\sfrac{\fl_j}{M^j}
        \fe_j({\sst \|K'\|_{1,\Si_{j+1}}})&(ii)\cr
\big\|\sfrac{d\hfill}{ds}K(K'+sK'')\big|_{s=0}\big\|_{1,\Si_j} 
&\le \abcst\,M^\aleph\fe_j({\sst \|K'\|_{1,\Si_{j+1}}})\|K''\|_{1,\Si_{j+1}}
}$$
$$\leqalignno{ 
\fe_j({\sst \|K(K')\|_{1,\Si_j}})
&\le   \abcst\,\fe_{j+1}({\sst \|K'\|_{1,\Si_{j+1}}})&(iii)\cr
}$$

}
\prf The proof is given following Lemma \:\lemNPcontrmap\ in Appendix \APappRewick.
\endproof

\proposition{\STM\propNPctmmap}{
Let $(\cW,\cG,u,\vec p)\in \cD^{(j)}_{\rm out}$ and  $K'\in \fK_{j+1}$. Then 
$$
\ren_{j,j+1}(K',\cW,u) \in \fK_{j}
$$

}
\prf As $\ren_{j,j+1}(K',\cW,u)=K(K';q_0)$, the Proposition follows from
Lemma \lemRWintbnd.i and Lemma  \lemreWick.
\endproof

\vskip.25in\noindent
We define, for each $K'\in \fK_{j+1}$,
$$\eqalign{
\tilde w(\phi,\psi;K')
&=\tilde w(\phi,\psi;K';q_0(K'))\cr
w''(\phi,\psi;K') &=\tilde w\big(\phi,\psi;K'\big)
         -\tilde w\big(\phi,0;K'\big)\cr
         &\hskip.5in-\half\sum_{s_1,s_1\in\Si_j}\int d\xi_1 d\xi_2\ 
q_0({\sst(\xi_1,s_1),(\xi_2,s_2)};K')\,
           \psi({\sst(\xi_1,s_1)})\psi({\sst(\xi_2,s_2)}) \cr
}\EQN\eqnNPtwwpp$$
and let $\cW'$ be the Grassmann function having sectorized representative $w''$.
Also set
$$\eqalign{
\cG'(\phi;K') &= \cG(\phi;K(K'))  +\tilde w(\phi,0;K') -\tilde  w(0,0;K')\cr
u'(K') &=  u'(K';q_0)\cr
{p'}^{(i)}&=p^{(i)}\hskip1.5in\hbox{for all }2\le i\le j-1\cr
{p'}^{(j)}&= \de u(\de K(0))_{\Si_j}*\hat \nu^{(\ge j)}
    + q_0(0)*\hat\nu^{(\ge j+1)} 
    -\big[\de K(0)_{\rm ext}\big]_{\Si_j}*\hat \nu^{(\ge j)} \cr
\cr
}\EQN\eqnNPgup$$
Observe that
$$\eqalign{
\check u'(k;K') &=  \check u(k;K(K'))\,\tilde\nu^{(\ge j+1)}(k)
                      + \check q_0(k;K')\,\nu^{(\ge j+1)}(k)  \cr
\check{p'}^{(j)}(k)&=  \de\check u(k;\de K(0))\, \nu^{(\ge j)}(k)
    + \check q_0(k;0)\,\nu^{(\ge j+1)}(k) -\de\check K(\k;0)\,\nu^{(\ge j)}(k) \cr
}\EQN\eqnNPcheckup$$
We now define 
$$
\cO_j(\cW,\cG,u,\vec p)=(\cW',\cG',u',\vec p')
$$

\theorem{\STM\thmOuttoIn}{
Let $(\cW,\cG,u,\vec p)\in \cD^{(j)}_{\rm out}$. Then  
$\cO_j(\cW,\cG,u,\vec p)\in \cD^{(j+1)}_{\rm in}$.
}
\prf   Set $(\cW',\cG',u',\vec p')=\cO_j(\cW,\cG,u,\vec p)$. Let $w$ be the sectorized representative of $\cW$ and
$$
u(K)=  
\smsum_{i=2}^{j-1} p^{(i)}_{\Si_j}+\big[\de u(K) -K_{\rm ext}\big]_{\Si_j}
$$
be the decomposition of $u$ specified in (\eqnNPudecomp).  
Let $\tilde w,\ w''$ be as in (\eqnNPtwwpp),
$q_0$ be the function of Lemma \lemreWick\ and let $\de K(K'),\ K(K')$ be as in (\eqndefkkprime). 
\Item {\it Verification that $(\cW',\cG',u',\vec p')$ is an interaction quadruple of scale $j+1$:} 

We first check the properties of $\vec{p'}$. As ${p'}^{(i)}=p^{(i)}$, for all $2\le i\le j-1$, we need only discuss ${p'}^{(j)}$. By (\eqnNPcheckup),
$\check{p'}^{(j)}(k)$ is supported on the $j^{\rm th}$ neighbourhood
and, since $\de \check K(\k;0) = \check q_0\big((0,\k);0\big)\,
\nu^{(\ge j+1)}((0,\k))$, vanishes at $k_0=0$.
By Lemma \lemOSumu\ of [FKTo3], Remark \remOStoresectorI\  of [FKTo4], Lemma \lemOSsectorext\ of [FKTo4], (\eqnNPdeubound), 
Lemma \lemreWick, Lemma \lemNPdeKbnd.i  and Corollary \corOSappMonoidIV.i of [FKTo1],
$$\eqalign{
\V p'^{(j)}\V_{1,\Si_{j}} 
&\le \V\de u(\de K(0))_{\Si_j}*\hat \nu^{(\ge j)}(k)\V_{1,\Si_{j}}  
    +\V q_0(0)*\hat\nu^{(\ge j+1)}\V_{1,\Si_{j}} 
+\V\big[\de K(0)_{\rm ext}\big]_{\Si_j}*\hat \nu^{(\ge j)}\V_{1,\Si_j}  \cr
&\le \const \cb_j\Big[ \V\de u(\de K(0)) \V_{1,\Si_{j-1}}  
    +\V q_0(0)\V_{1,\Si_j} 
    +\|\de K(0) \|_{1,\Si_j} \Big]  \cr
&\le \const \Big[ \la_0^{1-\upsilon}\,\fe_j({\sst\| \de K(0) \|_{1,\Si_j} })\,
     \| \de K(0) \|_{1,\Si_j}
    +\sfrac{\la_0^{1-\upsilon}}{\al^6}\,\sfrac{\fl_j}{M^j}\cb_j  \Big]  \cr
&\le \const \sfrac{\la_0^{1-\upsilon}}{\al^6}\,\sfrac{\fl_j}{M^j}\cb_j
\Big[ \la_0^{1-\upsilon}\,
\fe_j\Big(\sfrac{\la_0^{1-\upsilon}}{\al^6}\,\sfrac{\fl_j}{M^j}\cb_j\Big)
    +1  \Big]  \cr
&\le \sfrac{\la_0^{1-\upsilon}}{\al^5} \sfrac{\fl_j}{M^j}\cb_j \cr
}\EQN\eqnNPpboundal$$
Thus, (\eqnNPpbound) holds for $\vec p'$.
\vskip.5cm
\noindent 
Next, we construct the decomposition  (\eqnNPudecomp) for $u'$. Set
$$\eqalign{
\de u'(K')&= \de u( K(L))_{\Si_j}
    + q_0(L)*\hat\nu^{(\ge j+1)} -\de K(L)_{\rm ext}*\hat \nu^{(\ge j)} 
     \Big|^{L=K'}_{L=0} \cr
}$$
By construction, $\de u'\in\cF_0(2,\Si_j)$ and $\de u'(0)=0$. Since 
$\de\check u\big((0,\k);K'\big)$ vanishes,
$$
\de\check u'\big((0,\k);K'\big)
 = \check q_0\big((0,\k);L\big)\,\nu^{(\ge j+1)}\big((0,\k)\big) 
   -\de\check K(\k;L)\,\nu^{(\ge j)}\big((0,\k)\big)\Big|^{L=K'}_{L=0}
=0
$$
by (\eqnNPreWickDefs) and (\eqndefkkprime).
Observe that, by Remark \remOSresector.iii of [FKTo4],
$$\eqalign{
\de u'(K')_{\Si_{j+1}}&= \de u( K(L))_{\Si_{j+1}}
    + q_0(L)_{\Si_{j+1}}*\hat\nu^{(\ge j+1)} 
    -\big[\de K(L)_{\rm ext}\big]_{\Si_{j+1}} 
     \Big|^{L=K'}_{L=0} \cr
p'^{(j)}_{\Si_{j+1}}&= \de u( K(L))_{\Si_{j+1}}
    + q_0(L)_{\Si_{j+1}}*\hat\nu^{(\ge j+1)} 
    -\big[\de K(L)_{\rm ext}\big]_{\Si_{j+1}}
     \Big|_{L=0}\cr
}$$
Since $K(K')=K'_{\Si_j}+\de K(K')$
$$\eqalign{
u'(K') & =  u(K(K'))_{\Si_{j+1}}
    + q_0(K')_{\Si_{j+1}}*\hat\nu^{(\ge j+1)} \cr
& = \smsum_{i=2}^{j-1} p^{(i)}_{\Si_{j+1}} \ 
     +\big[\de u(K(K')) -K(K')_{\rm ext}\big]_{\Si_{j+1}}  
     + q_0(K')_{\Si_{j+1}}*\hat\nu^{(\ge j+1)} \cr
& = \smsum_{i=2}^{j-1} p^{(i)}_{\Si_{j+1}}
  +  \Big[\de u(K(L))_{\Si_{j+1}}
  + q_0(L)_{\Si_{j+1}}*\hat\nu^{(\ge j+1)} 
  - [\de K(L)_{\rm ext}]_{\Si_{j+1}}
     \Big]_{L=K'} \cr
& \hskip 1cm  - [K'_{\rm ext}]_{\Si_{j+1}} \cr
& = \smsum_{i=2}^{j-1} {p'}^{(i)}_{\Si_{j+1}} +{p'}^{(j)}_{\Si_{j+1}}
+\big[\de u'(K')-K'_{\rm ext}\big]_{\Si_{j+1}}
}$$
and $u'(K')$ has the desired form.

\vskip.5cm
\noindent 
We next bound $\V  \cD^{(1,0,0)}_{1,2}\sfrac{d\hfill}{ds} \de u'(K'+sK'')\big|_{s=0}\,  \V_{1,\Si_{j}}$, for $K',K''\in\fK_{j+1}$. By definition
$$\eqalign{
&\V  \cD^{(1,0,0)}_{1,2}\sfrac{d\hfill}{ds} \de u'(K'+sK'')\big|_{s=0}\,  \V_{1,\Si_{j}}\cr
&\hskip1.5cm\le \V  \cD^{(1,0,0)}_{1,2}\sfrac{d\hfill}{ds} 
\de u(K'_{\Si_j}+sK''_{\Si_j}+\de K(K'+sK''))_{\Si_j}\big|_{s=0}\,  \V_{1,\Si_{j}} \cr
&\hskip 2cm  
+\V   \cD^{(1,0,0)}_{1,2}\sfrac{d\hfill}{ds}
q_0(K'+sK'')\big|_{s=0}*\hat\nu^{(\ge j+1)}\V_{1,\Si_j} \cr
&\hskip 2.5cm  
  +\V  \cD^{(1,0,0)}_{1,2}\sfrac{d\hfill}{ds}
\de K(K'+sK'')_{\rm ext}*\hat \nu^{(\ge j)}\big|_{s=0}\,
 \V_{1,\Si_j}  \cr
}\EQN\eqnNPdelicateA$$
By Remark \remOSdiffnorm\ of [FKTo3] and Lemma \lemOSsectorext\ of [FKTo4], 
Lemma \lemNPdeKbnd.i and Remark \remOSscalednorms.iv of [FKTo3],
the third term is bounded by
$$\eqalign{
\V  \cD^{(1,0,0)}_{1,2}\sfrac{d\hfill}{ds}
\de& K(K'+sK'')_{\rm ext}*\hat \nu^{(\ge j)}\big|_{s=0}\,
 \V_{1,\Si_j}
\le \sfrac{\partial\hfill}{\partial t_0}
\V  \sfrac{d\hfill}{ds}
\de K(K'+sK'')_{\rm ext}*\hat \nu^{(\ge j)}\big|_{s=0}\,
 \V_{1,\Si_j}\cr
&\le \abcst\sfrac{\partial\hfill}{\partial t_0}
\Big(\cb_j
\big\|\sfrac{d\hfill}{ds}\de K(K'+sK'')\big|_{s=0}\big\|_{1,\Si_j}\Big)\cr
&\le \abcst\sfrac{\partial\hfill}{\partial t_0}
\Big(\cb_j
\sfrac{\la_0^{1-\upsilon}}{\al}\,\fe_j({\sst \|K'\|_{1,\Si_{j+1}}})\,
 \, \| K''\|_{1,\Si_{j+1}}\Big)\cr
&\le \abcst\sfrac{\la_0^{1-\upsilon}}{\al}\| K''\|_{1,\Si_{j+1}}
\sfrac{\partial\hfill}{\partial t_0}\Big(
\,\fe_j({\sst \|K'\|_{1,\Si_{j+1}}})\,\Big)\cr
&\le \abcst\sfrac{\la_0^{1-\upsilon}}{\al}M^j
\,\fe_j({\sst \|K'\|_{1,\Si_{j+1}}})\,\| K''\|_{1,\Si_{j+1}}
+\sum_{\de_0=r_0}\infty\,t^\de\cr
&\le \Big(\abcst\sfrac{M^2}{\al}\Big)\,\la_0^{1-\upsilon}M^{(j+1)-3}
\,\fe_{j+1}({\sst \|K'\|_{1,\Si_{j+1}}})\,\| K''\|_{1,\Si_{j+1}}
+\sum_{\de_0=r_0}\infty\,t^\de\cr
}\EQN\eqnNPdelicateI$$
Similarly, by Remark \remOSdiffnorm\ and Lemma \lemOSumu\ of [FKTo3], 
and Lemma \lemreWick, the second term is bounded by
$$\eqalign{
\V   \cD^{(1,0,0)}_{1,2}\sfrac{d\hfill}{ds}&
q_0(K'+sK'')\big|_{s=0}*\hat\nu^{(\ge j+1)}\V_{1,\Si_j}
\le \sfrac{\partial\hfill}{\partial t_0}
\V  \sfrac{d\hfill}{ds}q_0(K'+sK'')\big|_{s=0}*\hat \nu^{(\ge j+1)}\, \V_{1,\Si_j}\cr
&\le \abcst\sfrac{\partial\hfill}{\partial t_0}
\Big(\cb_j
\V  \sfrac{d\hfill}{ds}q_0(K'+sK'')\big|_{s=0} \V_{1,\Si_j}\Big)\cr
&\le \abcst\sfrac{\partial\hfill}{\partial t_0}
\Big(\cb_j
\sfrac{\la_0^{1-\upsilon}}{\al}\,\fe_j({\sst \|K'\|_{1,\Si_{j+1}}})\,
 \, \| K''\|_{1,\Si_{j+1}}\Big)\cr
&\le \Big(\abcst\sfrac{M^2}{\al}\Big)\,\la_0^{1-\upsilon}M^{(j+1)-3}
\,\fe_{j+1}({\sst \|K'\|_{1,\Si_{j+1}}})\,\| K''\|_{1,\Si_{j+1}}
+\sum_{\de_0=r_0}\infty\,t^\de\cr
}\EQN\eqnNPdelicateII$$
By the chain rule,
$$\eqalign{
&\sfrac{d\hfill}{ds} 
\de u(K'_{\Si_j}+sK''_{\Si_j}+\de K(K'+sK''))\big|_{s=0}\cr
&\hskip.5cm=\sfrac{d\hfill}{ds} 
\de u(K'_{\Si_j}+\de K(K')+sK''_{\Si_j})\big|_{s=0}
+\sfrac{d\hfill}{ds}\de u(K'_{\Si_j}+\de K(K')+s\sfrac{d\de K(K'+xK'')}{dt}\big|_{x=0}) 
\big|_{s=0}\cr
&\hskip.5cm=\sfrac{d\hfill}{ds} 
\de u(K(K')+sK''_{\Si_j})\big|_{s=0}
+\sfrac{d\hfill}{ds}\de u(K(K')+s\sfrac{d\hfill}{dx}\de K(K'+xK'')\big|_{x=0}) 
\big|_{s=0}
}\EQN\eqnNPdelicateB$$
We bound these two contributions to the first term of the right hand side of (\eqnNPdelicateA) separately. By Remark \remOStoresectorI\   of [FKTo4], (\eqnNPdeubound), with $K$ replaced by $K(K')$,
Proposition \propOSindresectorI.ii  of [FKTo4]
and Lemma \lemNPdeKbnd.iii,
$$\eqalignno{
&\V  \cD^{(1,0,0)}_{1,2}\sfrac{d\hfill}{ds} 
\de u(K(K')+sK''_{\Si_j})_{\Si_j}\big|_{s=0}\,  \V_{1,\Si_{j}}  \cr
&\hskip1cm\le \abcst\,\cb_{j-1}\V\cD^{(1,0,0)}_{1,2}\sfrac{d\hfill}{ds} 
\de u(K(K')+sK''_{\Si_j})\big|_{s=0}\,  \V_{1,\Si_{j-1}}  \cr
&\hskip1cm\le \abcst\,\la_0^{1-\upsilon}M^{j-3}
\,\fe_j\big(\|K(K')\|_{1,\Si_j}\big) \,  \|K''_{\Si_j}\|_{1,\Si_j}
+\smsum_{\de_0=r_0}\infty\, t^\de  \cr
&\hskip1cm\le \abcst\,\la_0^{1-\upsilon}M^{j-3}
\,\fe_{j+1}\big(\|K'\|_{1,\Si_{j+1}}\big) \, \sfrac{\fl_j}{\fl_{j+1}}\cb_{j-1} \|K''\|_{1,\Si_{j+1}}
+\smsum_{\de_0=r_0}\infty\, t^\de  \cr
 &\hskip1cm\le \Big(\abcst\sfrac{1}{M}\sfrac{\fl_j}{\fl_{j+1}}\Big)\,\la_0^{1-\upsilon}
M^{(j+1)-3}\,\fe_{j+1}\big(\|K'\|_{1,\Si_{j+1}}\big) \|K''\|_{1,\Si_{j+1}}
+\smsum_{\de_0=r_0}\infty\, t^\de \cr
&&\EQNO\eqnNPdelicateIII\cr
}$$
Similarly, by Remark \remOStoresectorI\  of [FKTo4], 
(\eqnNPdeubound), parts (i) and (iii) of Lemma \lemNPdeKbnd\ and Corollary \corOSappMonoidIV.ii
of [FKTo1],
$$\eqalignno{
&\V \cD^{(1,0,0)}_{1,2}\sfrac{d\hfill}{ds}\de u{(K(K')
                  +s\sfrac{d\hfill}{dx}\de K(K'+xK'')\big|_{x=0})}_{\Si_j}
\big|_{s=0}\,  \V_{1,\Si_{j}}  \cr
&\hskip1cm\le \abcst\,\cb_{j-1}\V \cD^{(1,0,0)}_{1,2}\sfrac{d\hfill}{ds}\de u(K(K')
                  +s\sfrac{d\hfill}{dx}\de K(K'+xK'')\big|_{x=0}) 
\big|_{s=0}\,  \V_{1,\Si_{j-1}}  \cr
&\hskip1cm\le \abcst\,\la_0^{1-\upsilon}M^{j-3}
\,\fe_j\big(\|K(K')\|_{1,\Si_j}\big) \ 
\big\|\sfrac{d\hfill}{dx}\de K(K'+xK'')\big|_{x=0}\big\|_{1,\Si_j}
+\smsum_{\de_0=r_0}\infty\, t^\de  \cr
&\hskip1cm\le \abcst\,\la_0^{1-\upsilon}M^{j-3}
\,\fe_{j+1}\big(\|K'\|_{1,\Si_{j+1}}\big) \ 
\sfrac{\la_0^{1-\upsilon}}{\al}\,\fe_j({\sst \|K'\|_{1,\Si_{j+1}}})\,
 \| K''\|_{1,\Si_{j+1}}
+\smsum_{\de_0=r_0}\infty\, t^\de  \cr
 &\hskip1cm\le\Big(\abcst\sfrac{\la_0^{1-\upsilon}}{\al}\Big)
\,\la_0^{1-\upsilon}M^{(j+1)-3}
\,\fe_{j+1}\big(\|K'\|_{1,\Si_{j+1}}\big) \ 
\| K''\|_{1,\Si_{j+1}}
+\smsum_{\de_0=r_0}\infty\, t^\de \cr
& &\EQNO\eqnNPdelicateIV\cr
}$$
Substituting (\eqnNPdelicateB) into (\eqnNPdelicateA) and applying the bounds
(\eqnNPdelicateI), (\eqnNPdelicateII), (\eqnNPdelicateIII) and
(\eqnNPdelicateIV), we have
$$\eqalign{
&\V  \cD^{(1,0,0)}_{1,2}\sfrac{d\hfill}{ds} \de u'(K'+sK'')\big|_{s=0}\,  \V_{1,\Si_{j}}\cr
&\hskip0.5cm\le \abcst_0\Big(\sfrac{M^2}{\al}
                        +\sfrac{1}{M}\sfrac{\fl_j}{\fl_{j+1}}   
                        +\sfrac{\la_0^{1-\upsilon}}{\al}\Big)
\,\la_0^{1-\upsilon}M^{(j+1)-3}
\,\fe_{j+1}\big(\|K'\|_{1,\Si_{j+1}}\big) \ 
\| K''\|_{1,\Si_{j+1}}
+\smsum_{\de_0=r_0}\infty\, t^\de
}$$
with an $M$--independent constant $\abcst_0$. If $M$ is sufficiently large
$$
\abcst_0\sfrac{1}{M}\sfrac{\fl_j}{\fl_{j+1}}
=\abcst_0\sfrac{1}{M^{1-\aleph}}
\le \half
$$
If $\al$ is sufficiently large and $\la_0$ is sufficiently small, depending on $M$,
$$
\abcst_0\Big(\sfrac{M^2}{\al}
                        +\sfrac{\la_0^{1-\upsilon}}{\al}\Big)\le \half
$$
and
$$\eqalign{
&\V  \cD^{(1,0,0)}_{1,2}\sfrac{d\hfill}{ds} \de u'(K'+sK'')\big|_{s=0}\,  \V_{1,\Si_{j}}
\le 
\la_0^{1-\upsilon}M^{(j+1)-3}
\,\fe_{j+1}({\sst\|K'\|_{1,\Si_{j+1}}}) \ 
\| K''\|_{1,\Si_{j+1}}
+\!\!\smsum_{\de_0=r_0}\infty\, t^\de
}$$
The remaining bound required in (\eqnNPdeubound), namely,
$$
\V \sfrac{d\hfill}{ds}\de u'(K'+sK'')\big|_{s=0}\,  \V_{1,\Si_j}
\le\la_0^{1-\upsilon}\,\fe_{j+1}\big(\|K'\|_{1,\Si_{j+1}}\big) 
\,\|K''\|_{1,\Si_{j+1}}
$$
is now a consequence of Corollary \corOSIntUp\ of [FKTo4]. 
\vskip.3cm
\noindent 
The remaining requirements of Definition \defNPintquad\ are easily verified.

\Item {\it Preparation for the verification of (I1), (I2) and (I3):} 

Clearly
$$
w'(K')=w''_{\Si_{j+1}}(K')
\EQN\eqnNPwpwpp$$
is a $\Si_{j+1}$--sectorized representative of $\cW'$. Write 
$$\eqalign{
\tilde w(\phi,\psi;K') = \smsum_{m,n}\ \smsum_{s_1,\cdots,s_n\in\Si_j}\
\int {\sst d\eta_1\cdots d\eta_m\,d\xi_1\cdots d\xi_n}\ & 
\tilde w_{m,n}({\sst \eta_1,\cdots, \eta_m\,(\xi_1,s_1),\cdots ,(\xi_n,s_n)};K')\cr
& \hskip .5cm \phi({\sst \eta_1})\cdots \phi({\sst \eta_m})\
\psi({\sst (\xi_1,s_1)})\cdots \psi({\sst (\xi_n,s_n)\,})\cr
w''(\phi,\psi;K') = \smsum_{m,n}\ \smsum_{s_1,\cdots,s_n\in\Si_j}\
\int {\sst d\eta_1\cdots d\eta_m\,d\xi_1\cdots d\xi_n}\ & 
w''_{m,n}({\sst \eta_1,\cdots, \eta_m\,(\xi_1,s_1),\cdots ,(\xi_n,s_n)};K')\cr
& \hskip .5cm \phi({\sst \eta_1})\cdots \phi({\sst \eta_m})\
\psi({\sst (\xi_1,s_1)})\cdots \psi({\sst (\xi_n,s_n)\,})\cr
w'(\phi,\psi;K') = \smsum_{m,n}\ \smsum_{s_1,\cdots,s_n\in\Si_j}\
\int {\sst d\eta_1\cdots d\eta_m\,d\xi_1\cdots d\xi_n}\ & 
w'_{m,n}({\sst \eta_1,\cdots, \eta_m\,(\xi_1,s_1),\cdots ,(\xi_n,s_n)};K')\cr
& \hskip .5cm \phi({\sst \eta_1})\cdots \phi({\sst \eta_m})\
\psi({\sst (\xi_1,s_1)})\cdots \psi({\sst (\xi_n,s_n)\,})\cr
}$$
By Lemma \lemreWick, $\tilde w_{0,2}(K')=\half q_0(K')$
and hence, by (\eqnNPtwwpp),
$w''_{0,2}=0$. Consequently, by (\eqnNPwpwpp), $w'_{0,2}=0$.

By (\eqnNPtwrewick) and Proposition \propBII\ of [FKTr1],
$$\eqalign{
\tilde w(\phi,\psi;K')=\lw w(\phi,\psi;K(K'))\rw_{\psi,-E_{\Si_j}(K';q_0)}\cr
}\EQN\eqnNPwickwtw$$
By Lemma \lemRWintbnd.iii, 
$\sqrt{\la_0^{1-\upsilon}\fl_j}\sqrt{\sfrac{\fl_j\IB}{M^j}}$ is an 
integral bound for $E_{\Si_j}$. Hence
by Corollary \corwicknorm.ii\ of [FKTr1], with 
$\ 
f(\psi)=w(\phi,\psi;K(K'))
\ $ and 
$\ 
f'(\psi) =\lw  w(\phi,\psi;K(K'))\rw_{\psi,-E_{\Si_j}}
\ $,
$$\eqalign{
N_j\Big(\tilde w(K')-w\big(K(K')\big),\sfrac{\al}{2},\,X\Big) 
&\le \sfrac{8\la_0^{1-\upsilon}}{\al^2}\fl_j\
   N_j(w(K(K')),\al,\,X) \cr
}\EQN\eqnNPreWickBnd
$$
for all $X\in\fN_{d+1}$. In particular
$$
N_j\big(\tilde w(K'),\sfrac{\al}{2},\,X\big)
\le \ \sfrac{3}{2}\, N_j\big(w(K(K')),\al,\,X\big)
\EQN\eqnNPreWickBndbis$$
To get a similar bound on $\sfrac{d\hfill}{ds} \tilde w (K'+sK'')\big|_{s=0}$,
observe that, by (\eqnNPwickwtw)
$$\eqalign{
\sfrac{d\hfill}{ds} \tilde w (K'+sK'')\big|_{s=0}\ =\ 
   &\lW \sfrac{d\hfill}{ds} w({\sst K(K'+sK'')})\rW_{\psi,-E_{\Si_j}(K';q_0)}\big|_{s=0}\cr
&+\sfrac{d\hfill}{ds}
       \lw w({\sst K(K')})\rw_{\psi,-E_{\Si_j}(K'+sK'';q_0)}\big|_{s=0}
}\EQN\eqnNPderivwickwtw$$
By Corollary \corwicknorm\ of [FKTr1], (O1), Lemma \lemNPdeKbnd, parts
(ii) and (iii), and Corollary \corOSappMonoidIV.ii of [FKTo1], the first term
is bounded by
$$\eqalign{
&N_j\Big( 
  \lW\sfrac{d\hfill}{ds}
        w({\sst K(K'+sK'')})\rW_{\psi,-E_{\Si_j}(K';q_0)}\big|_{s=0}
               \, ,\,\sfrac{\al}{2}\,,\,{\sst \|K'\|_{1,\Si_{j+1}}} \Big) \cr
&\hskip0.4in\le  N_j\Big( 
  \sfrac{d\hfill}{ds}w({\sst K(K'+sK'')})\big|_{s=0}\, ,
   \,\al\,,\,{\sst \|K'\|_{1,\Si_{j+1}}} \Big) \cr
&\hskip0.4in\le \fe_{j}({\sst \|K'\|_{1,\Si_{j+1}}}) N_j\Big( 
  \sfrac{d\hfill}{ds}w({\sst K(K'+sK'')})\big|_{s=0}\, ,\,\al\,,
  \,{\sst \|K(K')\|_{1,\Si_{j}}} \Big) \cr
&\hskip0.4in\le M^{j}
   \fe_{j}({\sst \|K'\|_{1,\Si_{j+1}}})\fe_j({\sst \|K(K')\|_{1,\Si_{j}}})
   \big\|\sfrac{d\hfill}{dx}K(K'+xK'')\big\|_{1,\Si_j}\big|_{x=0} \cr
&\hskip0.4in\le\abcst\,\,M^{j+\aleph}
 \fe_{j+1}({\sst \|K'\|_{1,\Si_{j+1}}})^3\|K''\|_{1,\Si_{j+1}}
   \cr
&\hskip0.4in\le \abcst\,M^{j+\aleph}\fe_{j+1}({\sst \|K'\|_{1,\Si_{j+1}}}) \|K''\|_{1,\Si_{j+1}} \cr
}\EQN\eqnNPderivwprimeI$$
In preparation for bounding the second term of (\eqnNPderivwickwtw), 
observe that, by Remark \scaleIntConsts\ of [FKTr1] and Lemma \:\lemRWintbnd
$$
\sfrac{d\hfill}{ds}E_{\Si_j}({\sst K'+sK''};q_0)\big|_{s=0}
=\sfrac{d\hfill}{ds}E_{\Si_j}({\sst K'+sK'';q_0(K')})\big|_{s=0}
+\sfrac{d\hfill}{ds}E_{\Si_j}({\sst K';q_0(K')+s{d\hfil\over dx}
q_0(K'+xK'')|_{x=0}})\big|_{s=0}
$$
has integral bound
$$
\const\sqrt{\fl_j\|K''\|_{1,\Si_{j+1}}\big|_{t=0}}
\ +\ \const
\sqrt{\fl_j\sfrac{\la_0^{1-\upsilon}}{\al}\|K''\|_{1,\Si_{j+1}}\big|_{t=0}}
\le \const\sqrt{\fl_j\|K''\|_{1,\Si_{j+1}}\big|_{t=0}}
$$
since, by Lemma \lemreWick,
$$
\V \sfrac{d\hfill}{dx} q_0(K'+xK'')\big|_{x=0}\V_{1,\Si_j}
\le\sfrac{\la_0^{1-\upsilon}}{\al}\,\fe_j({\sst \|K'\|_{1,\Si_{j+1}}})\,
 \, \| K''\|_{1,\Si_{j+1}}
$$
Consequently, by Corollary \corwicknorm.iii of [FKTr1], (O1), Lemma \lemNPdeKbnd,
parts (ii) and (iii), and Corollary \corOSappMonoidIV.ii of [FKTo1],
$$\eqalignno{
& N_j\Big( 
  \sfrac{d\hfill}{ds}\lW
        w({\sst K(K')})
       \rW_{\psi,-E_{\Si_j}(K'+sK'';q_0)}\big|_{s=0}
               \, ,\,\sfrac{\al}{2}\,,\,{\sst \|K'\|_{1,\Si_{j+1}}} \Big) \cr
&\hskip0.2in\le\sfrac{\const}{(\al-1)^2}\,N_j\big(w({\sst K(K')})
       \, ,\,\al\,,\,{\sst \|K'\|_{1,\Si_{j+1}}} \Big)
        M^j\|K''\|_{1,\Si_{j+1}} \cr
&\hskip0.2in\le\sfrac{\const}{(\al-1)^2}
      \fe_{j}({\sst \|K'\|_{1,\Si_{j+1}}})\,N_j\big( w({\sst K(K')})
       \, ,\,\al\,,\,{\sst \|K(K')\|_{1,\Si_{j}}} \Big)
        M^j\|K''\|_{1,\Si_{j+1}} \cr
&\hskip0.2in\le\sfrac{\const}{(\al-1)^2}
     \fe_{j}({\sst \|K'\|_{1,\Si_{j+1}}})
     \fe_j({\sst \|K(K')\|_{1,\Si_{j}}})
      \  M^j\|K''\|_{1,\Si_{j+1}} \cr
&\hskip0.2in\le\sfrac{\const}{(\al-1)^2}
     \fe_{j+1}({\sst \|K'\|_{1,\Si_{j+1}}})
       \ M^j\|K''\|_{1,\Si_{j+1}} 
&\EQNO\eqnNPderivwprimeII}$$
Combining (\eqnNPderivwickwtw), (\eqnNPderivwprimeI) and (\eqnNPderivwprimeII),
$$
N_j\Big(\sfrac{d\hfill}{ds} \tilde w (K'+sK'')\big|_{s=0}
               \, ,\,\sfrac{\al}{2}\,,\,{\sst \|K'\|_{1,\Si_{j+1}}} \Big)
\le \abcst\,M^{j+\aleph}\fe_{j+1}({\sst \|K'\|_{1,\Si_{j+1}}}) \|K''\|_{1,\Si_{j+1}}
\EQN\eqnNPderivwtildeest$$

\goodbreak
\Item  {\it Verification of (I3):} 

By the definition, (\eqnNPgup), of $\cG'$
$$\eqalign{
N\big(\cG'(\phi;K') - \cG(\phi;K(K')) \big) 
&=N\big(\tilde w (\phi,0;K' )-\tilde w (0,0;K' )\big)\cr
&\le \sfrac{1}{\root{4}\of{\fl_jM^j}}\,N_j\big( \tilde w (\phi,0;K' ) 
   \, ,\,\sfrac{\al}{2}\,,\,0 \big)_\0\cr
&\le \sfrac{3/2}{\root{4}\of{\fl_jM^j}}\,
   N_j\big(w(K(K'))\,,\,\al\,,\,0\big)_\0 
  \cr
&\le \sfrac{3/2}{\root{4}\of{\fl_jM^j}}\,
   N_j\big(w(K(K'))\,,\,\al\,,\,
        \| K(K')\|_{1,\Si_j}\big)_\0
   \cr
&\le \sfrac{2}{\root{4}\of{\fl_jM^j}}
    \cr
}\EQN\eqnNPgdiffouttoin$$
by Remark \remNPnvsnj.ii, (\eqnNPreWickBndbis) and (O1).
Therefore, by (O3), 
$$\eqalign{
&N\big( \cG'(\phi;K') - \half\phi JC^{(<j+1)}J\phi\big) \cr
&\hskip.3in\le N\Big( \cG(\phi;K(K'))  - \half\phi JC^{(\le j)}J\phi \Big)
  + N\big(\cG'(\phi;K') - \cG(\phi;K(K')) \big)  \cr
&\hskip.3in\le 4\smsum_{i=0}^{j} \sfrac{1}{\root{4}\of{\fl_iM^i}}
   \cr
}$$  

\vskip.25in
\noindent From (\eqnNPgup),
$$\eqalign{
N\Big(\sfrac{d\hfill}{ds}&\big[\cG'(\phi;{\sst K'+sK''})-
\cG'_2(\phi;{\sst K'+sK''})\big]_{s=0}\Big)\cr
&\le N\Big(\sfrac{d\hfill}{ds}\big[\cG(\phi;{\sst K(K'+sK'')})-
\cG_2(\phi;{\sst K(K'+sK'')})\big]_{s=0}\Big)
+N\Big(\sfrac{d\hfill}{ds} \tilde w (K'+sK'')\big|_{s=0\atop\psi=0}\Big)\cr
&\le N\Big(\sfrac{d\hfill}{ds}\big[ \cG\big(\phi;{\sst K(K')+sL}\big)
-\cG_2\big(\phi;{\sst K(K')+sL}\big)\big]_{s=0}\Big)\cr
&\hskip1in+\sfrac{1}{\root{4}\of{\fl_jM^j}}
N_j\Big(\sfrac{d\hfill}{ds} \tilde w (K'+sK'')\big|_{s=0}
               \, ,\,\sfrac{\al}{2}\,,\,{\sst \|K'\|_{1,\Si_{j+1}}} \Big)_\0\cr
}$$
where $L={d\hfill\over dx}K(K'+xK'')|_{x=0}$ obeys 
$ 
\| L\|_{1,\Si_j}
\le\abcst\,M^\aleph\,\fe_j({\sst \|K'\|_{1,\Si_{j+1}}})\,
 \, \| K''\|_{1,\Si_{j+1}}
$, by Lemma \lemNPdeKbnd.ii. Hence, by (O3) and (\eqnNPderivwtildeest)
$$\eqalign{
N\Big(\sfrac{d\hfill}{ds}&\big[\cG'(\phi;{\sst K'+sK''})-
\cG'_2(\phi;{\sst K'+sK''})\big]_{s=0}\Big)\cr
&\le M^j\| L \|_{1,\Si_j}+\abcst\,M^{j+\aleph}\fe_{j+1}({\sst \|K'\|_{1,\Si_{j+1}}}) \|K''\|_{1,\Si_{j+1}}\cr
&\le  \abcst\,M^{j+\aleph}\fe_j({\sst \|K'\|_{1,\Si_{j+1}}})\|K''\|_{1,\Si_{j+1}}\cr
&\le  \sfrac{1}{2}\,M^{j+1}\|K''\|_{1,\Si_{j+1}}+\sum_{\de\ne 0}\infty\, t^\de \cr
}\EQN\eqnNPderivgprimeI$$
Similarly,
$$\eqalign{
&\TN\sfrac{d\hfill}{ds}G'_2(\phi;K'+sK'')\big|_{s=0}\TN_\infty\cr
&\hskip.5in\le 
\TN\sfrac{d\hfill}{ds}G_2(\phi;{\sst K(K'+sK'')})\big|_{s=0}\TN_\infty
+\la_0^{1-\upsilon}
N\Big(\sfrac{d\hfill}{ds} \tilde w (K'+sK'')\big|_{s=0\atop\psi=0}\Big)\cr
&\hskip.5in\le M^j\|L \|_{1,\Si_j}
+\abcst\,\la_0^{1-\upsilon}\,\sfrac{1}{\root{4}\of{\fl_jM^j}}
\,M^{j+\aleph}\fe_{j+1}({\sst \|K'\|_{1,\Si_{j+1}}}) \|K''\|_{1,\Si_{j+1}}\cr
&\hskip.5in\le \sfrac{1}{2}\,M^{j+1}\|K''\|_{1,\Si_{j+1}}+\sum_{\de\ne 0}\infty\, t^\de
}$$

\Item  {\it Verification of (I2):}

Observe that by (\eqnNPreWickBnd), (O1) and Lemma \lemNPdeKbnd.iii
$$\eqalign{
\sfrac{M^{2j}}{\fl_j}\,\sfrac{\al^4}{16}\,\Big(\sfrac{\fl_j \IB}{M^j} \Big)^2
\sfrac{1}{\la_0^{1-\upsilon}\fl_j}\,&
\V \tilde w_{0,4}(K')-w_{0,4}\big(K(K')\big)\V_{3,\Si_j} \cr 
&\le \sfrac{8\la_0^{1-\upsilon}}{\al^2}\fl_j\
   N_j\big(w(K(K')),\al,\,\|K(K')\|_{1,\Si_j}\big) \cr
&\le \sfrac{8\la_0^{1-\upsilon}}{\al^2}\fl_j\
   \fe_j({\sst \|K(K')\|_{1,\Si_j}})\cr
&\le \abcst \sfrac{\la_0^{1-\upsilon}}{\al^2}\fl_j\,\fe_{j+1}({\sst
\|K'\|_{1,\Si_{j+1}}}) \cr
 }$$
so that, by Remark \remOStoresectorI\ of [FKTo4] and Corollary 
\corOSappMonoidIV.ii of [FKTo1]
$$
\VV \Big( \tilde w_{0,4}(K')-w_{0,4}\big(K(K')\big)\Big)_{\Si_{j+1}}\VV_{3,\Si_{j+1}} 
\le \abcst \sfrac{\la_0^{2-2\upsilon}}{\al^6}\fl_j\,
\fe_{j+1}({\sst\|K'\|_{1,\Si_{j+1}}}) 
\EQN\eqnNPdifffourlegged$$
Set
$\ 
{F'}^{(i)}=F^{(i)}
\ $ for all $2\le i\le j$ and
$$
{\de F'}^{(j+1)}(K')={\de F}^{(j+1)}(K(K'))_{\Si_{j+1}}
   +\big( \tilde w_{0,4}(K')-w_{0,4}(K(K'))\big)_{\Si_{j+1}}
$$
By Definition \defcompLadder, $\cL^{(j+1)}({\vec p}, \vec F)$ depends only on 
$p^{(2)},\cdots, p^{(j-1)}$ and $F^{(2)},\cdots, F^{(j)}$. In particular
$ \cL^{(j+1)}({\vec p}^{\,\prime}, \vec F') =\cL^{(j+1)}({\vec p}, \vec F)$.
Therefore
$$
w'_{0,4}(K')
=\tilde w_{0,4}(K')_{\Si_{j+1}}
=\de F^{\prime(j+1)}(K')+\smsum_{i=2}^{j}F^{\prime(i)}_{\Si_{j+1}}
 +\sfrac{1}{8} {\rm Ant\,}\Big( V_{\rm ph} \big( \cL^{(j+1)}
({\vec p}^{\,\prime}, \vec F')\big) \Big)_{\Si_{j+1}}
$$
and
$$
\V {F'}^{(i)}\V_{3,\Si_i}\le \sfrac{\la_0^{1-\upsilon}}{\al^7}\vi_i\cb_i
\qquad\qquad\hbox{for all }2\le i\le j
$$
Furthermore, by Remark \remOStoresectorI\ of [FKTo4], 
(\eqnNPdifffourlegged), (O2), Lemma \lemNPdeKbnd\ and Corollary 
\:\corOSappMonoidIV.ii of [FKTo1]
$$\eqalignno{
\V {\de F'}&^{(j+1)}(K')\V_{3,\Si_{j+1}}
\le \abcst\, \cb_j \,\V \de F^{(j+1)}(K(K'))\V_{3,\Si_j} 
+ \abcst \sfrac{\la_0^{2-2\upsilon}}{\al^6}\fl_j\,
\fe_{j+1}({\sst\|K'\|_{1,\Si_{j+1}}})  \cr
&\le \abcst  \sfrac{\la_0^{1-\upsilon}}{\al^4}\Big\{
      \sfrac{\vi_{j+1}}{\al^4}
     +\sfrac{1}{\IB^2}  M^j\|K(K')\|_{1,\Si_j} 
\Big\}\fe_j({\sst \| K(K')\|_{1,\Si_{j}}}) \cr
&\hskip 6cm + \abcst \sfrac{\la_0^{2-2\upsilon}}{\al^6}\fl_j\,
\fe_{j+1}({\sst\|K'\|_{1,\Si_{j+1}}}) \cr
&\le \sfrac{\la_0^{1-\upsilon}}{\al^4}\Big\{
      \abcst  \sfrac{\vi_{j+1}}{\al^4}
     +\abcst  \sfrac{1}{\IB^2} \sfrac{\fl_j}{\fl_{j+1}} M^j\|K'\|_{1,\Si_{j+1}}
     +\const \sfrac{\la_0^{1-\upsilon}}{\al^2}\,\fl_j
     \Big\}\fe_{j+1}({\sst\|K'\|_{1,\Si_{j+1}}})  \cr
&\le \sfrac{\la_0^{1-\upsilon}}{\al^4}\Big\{
      \sfrac{\vi_{j+1}}{\al^3}
        +\abcst \sfrac{1}{M^{1-\aleph}} \sfrac{1}{\IB^2}  
             M^{j+1}\|K'\|_{1,\Si_{j+1}}
      \Big\}\fe_{j+1}({\sst\|K'\|_{1,\Si_{j+1}}})  
&\EQNO\eqnNPgoodFprimeest
}$$
since $\ \const\,\la_0^{1-\upsilon} \fl_j\le \sfrac{\vi_{j+1}}{2\al}$,
by the hypothesis $\al<\sfrac{1}{\la_0^{\upsilon/10}}$ of 
Theorem \theoremNPinduction\ and the requirement of Definition \defNPrhomn\ 
that $0<\upsilon<\sfrac{1}{4}$. In particular
$$
\V {\de F'}^{(j+1)}(K')\V_{3,\Si_{j+1}}
\le \sfrac{\la_0^{1-\upsilon}}{\al^4}\Big\{
      \sfrac{\vi_{j+1}}{\al^3}
       +\sfrac{1}{\IB^2}  M^{j+1}\|K'\|_{1,\Si_{j+1}}
      \Big\}\fe_{j+1}({\sst\|K'\|_{1,\Si_{j+1}}})  
$$

\Item  {\it Verification of (I1):}

Set
$$\eqalign{
\tilde \om_{m,n} 
&=   \smsum_{s_1,\cdots,s_n\in\Si_j}\
\int {\sst d\eta_1\cdots d\eta_m\,d\xi_1\cdots d\xi_n}\  
\tilde w_{m,n}({\sst \eta_1,\cdots, \eta_m\,(\xi_1,s_1),\cdots ,(\xi_n,s_n)}) \cr
& \hskip 7cm \phi({\sst \eta_1})\cdots \phi({\sst \eta_m})\
\psi({\sst (\xi_1,s_1)})\cdots \psi({\sst (\xi_n,s_n)\,})   \cr
\om'_{m,n}    
&=   \smsum_{s_1,\cdots,s_n\in\Si_{j+1}}\
\int {\sst d\eta_1\cdots d\eta_m\,d\xi_1\cdots d\xi_n}\  
 w'_{m,n}({\sst \eta_1,\cdots, \eta_m\,(\xi_1,s_1),\cdots ,(\xi_n,s_n)}) \cr
& \hskip 7cm \phi({\sst \eta_1})\cdots \phi({\sst \eta_m})\
\psi({\sst (\xi_1,s_1)})\cdots \psi({\sst (\xi_n,s_n)\,})   \cr   \cr
}$$
Then
$$\eqalign{
w''&=\smsum_{m,n\ge 1} \tilde \om_{m,n} 
   + \smsum_{n\ge 4}  \tilde\om_{0,n}\cr
w'&=\smsum_{m,n\ge 1}  \om'_{m,n} 
   + \smsum_{n\ge 4}  \om'_{0,n}\cr 
}$$
By (O2), (\eqnNPgoodFprimeest), Remark \remOStoresectorI\ of 
[FKTo4] and Theorem \theoremcompLadder\ 
(with $\rho =\la_0^{1-\upsilon},\ \veps =\sfrac{\aleph}{\scriptscriptstyle n_0}$) 
$$\eqalign{
\V w'_{0,4}&(K')\V_{3,\Si_{j+1}}
 \le \V \de F^{\prime(j+1)}(K')\V_{3,\Si_{j+1}}
 +\smsum_{i=2}^{j} \V F^{\prime(i)}_{\Si_{j+1}}\V_{3,\Si_{j+1}} \cr
& \hskip 2.5cm +\sfrac{1}{8} \V {\rm Ant\,}\Big( V_{\rm ph} \big( \cL^{(j+1)}
({\vec p}^{\,\prime}, \vec F')\big) \Big)_{\Si_{j+1}} \V_{3,\Si_{j+1}} \cr
& \le \abcst \sfrac{\la_0^{1-\upsilon}}{\al^4}\Big\{
      \smsum_{i=2}^{j+1} \sfrac{\vi_{i}}{\al^3}
        +\sfrac{1}{M^{1-\aleph}} \sfrac{1}{\IB^2}  
             M^{j+1}\|K'\|_{1,\Si_{j+1}}
      \Big\}\fe_{j+1}({\sst\|K'\|_{1,\Si_{j+1}}})  
            +\const \la_0^{2-2\upsilon}\cb_{j+1}\cr
& \le \abcst \sfrac{\la_0^{1-\upsilon}}{\al^4}\Big\{
      \sfrac{1}{\al^3}
        +\sfrac{1}{M^{1-\aleph}} \sfrac{1}{\IB^2}  
             M^{j+1}\|K'\|_{1,\Si_{j+1}}
      \Big\}\fe_{j+1}({\sst\|K'\|_{1,\Si_{j+1}}}) \cr
}$$
Consequently, by Proposition \propOSthreetoonenorm \ of [FKTo4]
$$
\V w'_{0,4}(K')\V_{1,\Si_{j+1}} 
\le \abcst \sfrac{\la_0^{1-\upsilon}}{\al^4\,\fl_{j+1}}\Big\{
      \sfrac{1}{\al^3}
        +\sfrac{1}{M^{1-\aleph}} \sfrac{1}{\IB^2}  
             M^{j+1}\|K'\|_{1,\Si_{j+1}}
      \Big\}\fe_{j+1}({\sst\|K'\|_{1,\Si_{j+1}}}) 
$$
Therefore, by Corollary \:\corOSappMonoidIV\ of [FKTo1]  
$$\eqalignno{
N_{j+1}&(\om'_{0,4}(K'),64\al,\,\|K'\|_{1,\Si_{j+1}})  \cr
&=2^{24}\,\al^{4}\,\IB^2\,\,\sfrac{\fl_{j+1}}{\la_0^{1-\upsilon}}\,
\fe_{j+1}({\sst \|K'\|_{1,\Si_{j+1}}})
\Big( \V w'_{0,4}(K')\V_{1,\Si_{j+1}} + 
\sfrac{1}{\fl_{j+1}}\V w'_{0,4}(K')\V_{3,\Si_{j+1}} \Big) \cr
& \le \abcst \Big\{ \sfrac{\IB^2}{\al^3}
        +\sfrac{1}{M^{1-\aleph}}M^{j+1}\|K'\|_{1,\Si_{j+1}}
      \Big\}\fe_{j+1}({\sst\|K'\|_{1,\Si_{j+1}}})   \cr
& \le \half \fe_{j+1}({\sst\|K'\|_{1,\Si_{j+1}}}) 
&\EQNO\eqnNPfourlegest
}$$
By (\eqnNPreWickBndbis), (O1) and Lemma \lemNPdeKbnd.iii 
$$\eqalign{
N_j( w''(K'),\sfrac{\al}{2},0)
&\le \sfrac{3}{2}
\, N_j(w(K(K')),\al,0) \cr
&\le \sfrac{3}{2}
\, N_j(w(K(K')),\al,\,\|K(K')\|_{1,\Si_j}) \cr
&\le \sfrac{3}{2}\,\fe_j(\sst{\|K(K')\|_{1,\Si_j}})  \cr
&\le  \abcst\,\fe_{j+1}({\sst\|K'\|_{1,\Si_{j+1}}})\cr
}$$
so that, by Corollary \corOSirrelevantresect\ of [FKTo4] and 
Corollary \:\corOSappMonoidIV\ of [FKTo1],
$$\eqalign{
&N_{j+1}(w'(K')-\om'_{0,4}(K'),64\al,\|K'\|_{1,\Si_{j+1}})\cr
&\hskip1in
\le \sfrac{1}{M^{(1-\aleph)/8}}\,\fe_{j+1}({\sst\|K'\|_{1,\Si_{j+1}}})\,
N_j\big( w''(K')-\tilde\om_{0,4}(K'),\sfrac{\al}{2},\|K'\|_{1,\Si_{j+1}}\big)\cr
&\hskip1in
\le \sfrac{1}{M^{(1-\aleph)/8}}\,\fe_{j+1}({\sst\|K'\|_{1,\Si_{j+1}}})^2\,
N_j\big( w''(K'),\sfrac{\al}{2},0\big)\cr
&\hskip1in
\le \abcst\,\sfrac{1}{M^{(1-\aleph)/8}}
\,\fe_{j+1}({\sst\|K'\|_{1,\Si_{j+1}}})^3\cr
&\hskip1in \le \half \fe_{j+1}({\sst\|K'\|_{1,\Si_{j+1}}}) 
}\EQN\eqnNPnonfourlegest$$
Combining (\eqnNPfourlegest) and (\eqnNPnonfourlegest), we get
$$
N_{j+1}(w'(K'),64\al,\|K'\|_{1,\Si_{j+1}})
\le \fe_{j+1}({\sst\|K'\|_{1,\Si_{j+1}}}) 
$$

\vskip.25in\noindent
By (\eqnNPtwwpp) and (\eqnNPderivwtildeest),
$$
N_j\Big(\sfrac{d\hfill}{ds} w'' (K'+sK'')\big|_{s=0}
               \, ,\,\sfrac{\al}{2}\,,\,{\sst \|K'\|_{1,\Si_{j+1}}} \Big)
\le \abcst\,M^{j+\aleph}\fe_{j+1}({\sst \|K'\|_{1,\Si_{j+1}}}) \|K''\|_{1,\Si_{j+1}}
$$
Therefore by Corollary \corOSirrelevantresect\ of [FKTo4] and 
Corollary \:\corOSappMonoidIV\ of [FKTo1],
$$\eqalign{
&N_{j+1}\Big(\sfrac{d\hfill}{ds} w' (K'+sK'')\big|_{s=0},64\al,\|K'\|_{1,\Si_{j+1}}\Big)\cr
&\hskip1in
\le \abcst\,\fe_{j+1}({\sst\|K'\|_{1,\Si_{j+1}}})\,
N_j\Big(\sfrac{d\hfill}{ds} w'' (K'+sK'')\big|_{s=0},\sfrac{\al}{2},\|K'\|_{1,\Si_{j+1}}\Big)\cr
&\hskip1in
\le \abcst\,M^{j+\aleph}
\,\fe_{j+1}({\sst\|K'\|_{1,\Si_{j+1}}})^2\|K''\|_{1,\Si_{j+1}}\cr
&\hskip1in \le M^{j+1}
\,\fe_{j+1}({\sst\|K'\|_{1,\Si_{j+1}}})\|K''\|_{1,\Si_{j+1}}
}$$

\endproof

\remark{\STM\remNPFnochangeII}{
Let $(\cW,\cG,u,\vec p)\in \cD^{(j)}_{\rm out}$ and
 $(\cW',\cG',u',\vec p')=\cO_j(\cW,\cG,u,\vec p)\in \cD^{(j+1)}_{\rm in}$.
\Item (i)
The data ${F'}^{(2)},\ \cdots,\ {F'}^{(j)}$ of
(I2) coincides with the data  ${F}^{(2)},\ \cdots,\ {F}^{(j)}$ of
(O2). Also, by (\eqnNPgup), ${p'}^{(i)}=p^{(i)}$ for all $2\le i\le j-1$.
\Item (ii)
By (\eqnNPgdiffouttoin),
$$
N\Big(\cG'(\phi;K') - \cG\big(\phi;\ren_{j,j+1}(K',\cW,u)\big) \Big) 
\le \sfrac{2}{\root{4}\of{\fl_jM^j}}
$$
\Item (iii)
If the output data  are analytic functions of $K$, then, by 
Lemma \lemreWick, the input data are analytic functions of the output 
data and $K$.

\Item (iv) If the output data  is real, then the input data is real.

}

\vfill\eject

\chap{ The Recursive Construction of the}
\null\vskip-.6in
\centerline{\tafontt Green's Functions}\PG\pgNPX

In this section, we construct the data of Theorem \theoremNPinduction, 
recursively in $j$.

\Item {\it Initialization at $j=j_0$.}\PG\pgNPXa 

\noindent We set
\item{$\circ$}
$\de e_{j_0}(K)(\k)=\check K(\k)$ for $K\in\fK_{j_0}$.
\item{$\circ$} $p^{(2)}=p^{(3)}=\cdots=p^{(j_0-1)}=0$
\item{$\circ$} $F^{(2)}=\cdots=F^{(j_0)}=0$ 

\noindent and define $\cW_{j_0}$, $\cG^\rg_{j_0}$ and $u_{j_0}$
as follows.
$$\eqalign{
u_{j_0}(K)&=- \big[K_{\rm ext}\big]_{\Si_{j_0}}\cr
\cW_{j_0}(K)&=\tilde\Om_{C^{(\le j_0)}_{u_{j_0}(K)}}(\cV{\sst (\psi)})(\phi,\psi)
-\tilde\Om_{C^{(\le j_0)}_{u_{j_0}(K)}}(\cV{\sst (\psi)})(\phi,0)\cr
\cG^\rg_{j_0}(K)&=\tilde\Om_{C^{(\le j_0)}_{u_{j_0}(K)}}(\cV{\sst (\psi)})(\phi,0)\cr
}$$
Clearly, $\big(\cW_{j_0},\cG^\rg_{j_0},u_{j_0},
({\sst p^{(2)},\cdots,p^{({j_0}-1)}})\big)$
is an interaction quadruple at scale ${j_0}$. Next, we verify that it is in
$\cD^{(j_0)}_\out$. Let 
$\tilde w(\phi,\psi;K)$ be the $\Si_{j_0}$--sectorized representative for 
$$
\tilde\Om_{C_{u(K)}^{(\le j_0)}}\big(\cV{\sst (\psi)}\big) (\phi,\psi) 
 -\half\phi JC^{(\le j_0)}_{u_{j_0}(K)}J\phi
=\cW_{j_0}(K)(\phi,\psi) +\cG^\rg_{j_0}(K)(\phi) 
-\half\phi JC^{(\le j_0)}_{u_{j_0}(K)}J\phi
$$ 
chosen in Theorem \thmNPsetupinduction\ and set
$$\eqalign{
w(\phi,\psi;K) &=\tilde w(\phi,\psi;K) -\tilde w(\phi,0;K)\cr
&= \smsum_{m,n}\ \smsum_{s_1,\cdots,s_n\in\Si_{j_0}}\
\int {\sst d\eta_1\cdots d\eta_m\,d\xi_1\cdots d\xi_n}\ 
w_{m,n}({\sst \eta_1,\cdots, \eta_m\,(\xi_1,s_1),\cdots ,(\xi_n,s_n)};K)\cr
& \hskip 5cm \phi({\sst \eta_1})\cdots \phi({\sst \eta_m})\
\psi({\sst (\xi_1,s_1)})\cdots \psi({\sst (\xi_n,s_n)\,})\cr
}$$
 By Theorem \thmNPsetupinduction,
$$
N_{j_0}\big(w(K),\al,\|K\|_{1,\Si_{j_0}}\big) 
\le N_{j_0}\big(\tilde w(K),\al,\|K\|_{1,\Si_{j_0}}\big)
\le \const\al^4\la_0^\upsilon\fe_{j_0}\big(\|K\|_{1,\Si_{j_0}}\big)
$$
and
$$\eqalign{
N_{j_0}\big(\sfrac{d\hfill}{ds}w(K+sK')\big|_{s=0}
,\al,\,\| K\|_{1,\Si_{j_0}}\big) 
&\le N_{j_0}\big(\sfrac{d\hfill}{ds}\tilde w(K+sK')\big|_{s=0}
,\al,\,\| K\|_{1,\Si_{j_0}}\big) \cr
&\le M^{j_0} \,\fe_{j_0}\big(\|K\|_{1,\Si_{j_0}}\big) \,  \|K'\|_{1,\Si_{j_0}}
}$$
In particular,
$$\eqalign{
N_{j_0}\big(w(K),\al,\|K\|_{1,\Si_{j_0}}\big) 
&\le \fe_{j_0}\big(\|K\|_{1,\Si_{j_0}}\big) \cr
\V w_{0,2}(K)\V_{1,\Si_{j_0}}
&\le \sfrac{1}{M^{j_0}\IB\al^2\,\rho_{0;2}}
N_{j_0}\big(w(K),\al,\|K\|_{1,\Si_{j_0}}\big)\cr
&\le\const\,\al^2\la_0\,\fe_{j_0}\big(\|K\|_{1,\Si_{j_0}}\big)\cr
&\le\sfrac{\la_0^{1-\upsilon}}{\al^{7}} \,\sfrac{\fl_{j_0}}{M^{j_0}}\,
 \fe_{j_0}\big(\|K\|_{1,\Si_{j_0}}\big)\cr
}$$
so that (O1) is satisfied. Similarly,
$$\eqalign{
\V w_{0,4}(K)\V_{3,\Si_{j_0}}
&\le \sfrac{1}{\IB^2\al^4\,\rho_{0;4}(\la_0)}
N_{j_0}\big(w(K),\al,\|K\|_{1,\Si_{j_0}}\big)\cr
&\le\sfrac{\la_0^{1-\upsilon}}{\al^{9}} 
 \fe_{j_0}\big(\|K\|_{1,\Si_{j_0}}\big)\cr
}$$
 Setting
$\ 
\de F^{(j_0+1)}(K)=w_{0,4}(K)
\ $, (O2) is satisfied. 
Observe that
$$
\cG^\rg_{j_0}(K)-\half\phi JC^{(\le j_0)}J\phi
= \tilde w(\phi,0;K)
-\half\phi J\big[C^{(\le j_0)}-C^{(\le j_0)}_{u_{j_0}(K)}\big]J\phi
$$
Now
$$
C^{(\le j_0)}(k)- C^{(\le j_0)}_{u_{j_0}(K)}(k)
=\sfrac{\check K(\k)[U(\k)-\nu^{(>j_0)}(k)]}
{[\imath k_0-e(\k)][\imath k_0-e(\k)+\check K(\k)]}
$$
On the support of $U(\k)-\nu^{(>j_0)}(k)$, 
$|\imath k_0-e(\k)|\ge\sfrac{1}{\sqrt{M}\, M^{j_0}}$
and $|\check K(\k)|\le \abcst \la_0^{1-\upsilon}\sfrac{\fl_{j_0+1}}{M^{j_0+1}}$
so that
$$\eqalign{
 N\big(  \phi J\big[C^{(\le j_0)}-C^{(\le j_0)}_{u_{j_0}(K)}\big]J\phi\big)
&=\sfrac{1}{\la_0^{1-\upsilon}}
\TN J\big(C^{(\le j_0)}- C^{(\le j_0)}_{u_{j_0}(K)}\big)J \TN_\infty\cr
&\le \abcst\, \sfrac{\fl_{j_0+1}}{M^{j_0+1}}
\int\sfrac{d^3 k}{(2\pi)^3}\sfrac{U(\k)-\nu^{(>j_0)}(k)}
                             {|\imath k_0-e(\k)|^2}\cr
&\le \abcst\,\sfrac{\fl_{j_0+1}}{M^{j_0+1}}
                \int_{ |\imath x-y|\ge {1\over \sqrt{M}\,M^{j_0}}\atop |y|\le\abcst} dx\,dy\ 
                         \sfrac{1}{|\imath x-y|^2}\cr
&\le \abcst\,\sfrac{\fl_{j_0+1}}{M^{j_0+1}}
\ln\big(\sqrt{M}\,M^{j_0}\big) \cr
&\le \sfrac{1}{\root{4}\of{\fl_{j_0}M^{j_0}}}\cr
}$$
if $M$ is big enough. Therefore, by  Remark  \remNPnvsnj.ii and 
Theorem \thmNPsetupinduction
$$\eqalign{
&N\big(\cG^\rg_{j_0}(K)-\half\phi JC^{(\le j_0)}J\phi\big)\cr
&\hskip1in\le \sfrac{1}{\root{4}\of{\fl_{j_0}M^{j_0}}}
    N_{j_0}\big(\tilde w(K),\al,\|K\|_{1,\Si_{j_0}}\big)_\0
+  N\big(  \phi J\big[C^{(\le j_0)}-C^{(\le j_0)}_{u_{j_0}(K)}\big]J\phi\big)\cr
&\hskip1in\le \sfrac{2}{\root{4}\of{\fl_{j_0}M^{j_0}}}\cr
}$$
Similarly,
 $$\eqalign{
N\big(
\sfrac{d\hfill}{ds}\big[\cG^\rg_{j_0}(K+sK')-&\cG_{j_0,2}(K+sK')\big]_{s=0}\big)
=N\big(\sfrac{d\hfill}{ds}\tilde w(\phi,0;K+sK')\big|_{s=0}\big)\cr
&\le \sfrac{1}{\root{4}\of{\fl_{j_0}M^{j_0}}}
    N_{j_0}\big(\sfrac{d\hfill}{ds}\tilde w(K+sK')\big|_{s=0},\al,\|K\|_{1,\Si_{j_0}}\big)_\0\cr
&\le M^{j_0} \|K'\|_{1,\Si_{j_0}} \cr
}$$
and, since $\sfrac{d\hfill}{ds}C^{(\le j_0)}_{u_{j_0}(K+sK')}(k)\big|_{s=0}
=-\sfrac{\check K'(\k)[U(\k)-\nu^{(>j_0)}(k)]}
{[\imath k_0-e(\k)+\check K(\k)]^2}$,
$$\eqalign{
&\TN \sfrac{d\hfill}{ds}G_{j_0,2}(K+sK')\big|_{s=0}\TN_\infty
\le \TN \sfrac{d\hfill}{ds}\tilde w_{2,0}(K+sK'))\big|_{s=0}\TN_\infty
+\half \TN J\sfrac{d\hfill}{ds}
C^{(\le j_0)}_{u_{j_0}(K+sK')}\big|_{s=0}J\TN_\infty
\cr 
&\hskip0.5in
\le \la_0^{1-\upsilon}
N\big( \sfrac{d\hfill}{ds}\tilde w(\phi,0;K+sK')\big|_{s=0}\big)
+\abcst\, \sup_{\k}\big|\check K'(\k)\big|
\int\sfrac{d^3 k}{(2\pi)^3}\sfrac{U(\k)-\nu^{(>j_0)}(k)}
                             {|\imath k_0-e(\k)|^2}
\cr 
&\hskip0.5in
\le \la_0^{1-\upsilon}M^{j_0} \|K'\|_{1,\Si_{j_0}}
+\abcst\, \sup_{\k}\big|\check K'(\k)\big|
 \int_{ |\imath x-y|\ge {1\over \sqrt{M}\,M^{j_0}}\atop |y|\le\abcst} dx\,dy\ 
                         \sfrac{1}{|\imath x-y|^2}
\cr 
&\hskip0.5in
\le \la_0^{1-\upsilon}M^{j_0} \|K'\|_{1,\Si_{j_0}}
+\abcst\, \|K'\|_{1,\Si_{j_0}}
 \ln\big(\sqrt{M}\,M^{j_0}\big)
\cr 
&\hskip0.5in
\le M^{j_0} \|K'\|_{1,\Si_{j_0}} \cr
}$$
if $M$ is big enough, for all $K\in\fK_{j_0}$ and all $K'$.
This completes the verification that (O3) is satisfied.

\noindent As pointed out in Remark \remNPOandRconditions, conditions
(O1--3) imply conditions (R1--3) of Theorem \theoremNPinduction.
For $j=j_0$, (R4) is vacuous. Condition (R5) was verified formally as
(\eqnNPgenfnrengrp).  The analyticity and reality conditions of 
Theorem \theoremNPinduction\ follow from Theorem \thmNPsetupinduction. 

\vskip.3cm

\Item {\it Recursive step $j\rightarrow j+1$.}\PG\pgNPXb 

\noindent
Fix $j\ge j_0$ and assume that 
\item{$\circ$} maps $\de e_{j'},\ \ren_{i,j'}$, $j_0\le i\le j'\le j$
\item{$\circ$} $p^{(2)},\ \cdots,\ p^{(j-1)}$
\item{$\circ$} $F^{(2)},\ \cdots,\ F^{(j)}$

\noindent and output data 
$\big(\cW_j,\cG^\rg_{j},u_{j},({\sst p^{(2)},\cdots,p^{(j-1)}})\big)$ at scale $j$ have been constructed and fulfill the conclusions of Theorem 
 \theoremNPinduction.

\noindent Define $\cW_{j+1},\cG^\rg_{j+1},u_{j+1}$ and $p^{(j)}$ by
$$
\big(\cW_{j+1},\cG^\rg_{j+1},u_{j+1},({\sst p^{(2)},\cdots,p^{({j})}})\big)
=\Om_{j+1}\circ\cO_{j}
\big(\cW_j,\cG^\rg_{j},u_{j},({\sst p^{(2)},\cdots,p^{({j}-1)}})\big)
$$
By Theorems \thmOuttoIn\ and \thmIntoOut, the left hand side is an output 
datum of scale $j+1$ and, by Remark  \remNPOandRconditions,
satisfy conditions (R1--3). By Remarks \remNPFnochangeII.i and
\remNPFnochangeI.i, the $F^{(j+1)}$ of (O2) may be appended to 
$F^{(2)},\ \cdots,\ F^{(j)}$ so that (R2) is satisfied.
The analyticity and reality conditions of Theorem \theoremNPinduction,
follow from Remarks \remNPFnochangeI.iv,v and \remNPFnochangeII.iii,iv.

\noindent
Define $\ren_{j,j+1}$ to be the map $\ren_{j,j+1}(\ \cdot\ ,\cW_j,u_j)$ 
of (\eqndefkkprime). By Remarks \remNPFnochangeII.ii and
\remNPFnochangeI.ii, (R4) is satisfied. Define, for $j_0\le i\le j$,
$$
\ren_{i,j+1} =\ren_{i,j}\circ\ren_{j,j+1}
$$
and
$$
\de e_{j+1}(K)(\k)=\big[\ren_{j_0,j+1}(K)\big]^{\check{}}(\k)
$$
Then the algebraic conditions of Definition \defNPprojectivesyst\ are fulfilled.
The analyticity and reality of $\ren_{j,j+1}$ was observed following (\eqndefkkprime).
That the estimates are also fulfilled is proven in

\lemma{\STM\lemNPprojSysBnds}{
\Item i) For all $K\in\fK_{j+1}$,
$$\eqalign{
\big\|\ren_{i,j+1}(K)\big\|_{1,\Si_i}
&\le \la_0^{1-\upsilon}\sfrac{\fl_i}{M^i}+\smsum_{\de\ne 0}\infty\, t^\de\cr
\tn\de \hat e_{j+1}(K)\tn_{1,\infty}
&\le \la_0^{1-\upsilon}\cr
}$$
\Item ii) There is a universal constant $\abcst$ such that, for all 
$j_0\le i\le j+1$ and $K,K'\in\fK_{j+1}$, 
$$
\big\|\sfrac{d\hfill}{ds} \ren_{i,j+1}(K+sK')\big|_{s=0}\big\|_{1,\Si_i}
\le\abcst^{j+1-i}\ \sfrac{\fl_i}{\fl_{j+1}}\| K'\|_{1,\Si_{j+1}}
+\smsum_{\de\ne 0}\infty\, t^\de
$$

\Item iii) For all $j_0\le j'\le j+1$, 
$$\eqalign{
\big\|\ren_{i,j+1}(0)-\ren_{i,j'}(0)\big\|_{1,\Si_i}
&\le  \la_0^{1-\upsilon}\sfrac{1}{2^{j'}}+\smsum_{\de\ne 0}\infty\, t^\de\cr
\tn\de e_{j+1}(0)-\de e_{j'}(0)\tn_{1,\infty}
&\le  \la_0^{1-\upsilon}\sfrac{1}{2^{j'}}\cr
}$$

}
\prf i) We prove, by induction on $j-i$ that
$$
\ren_{i,j+1}(K)=\sum_{\ell=i}^{j+1} P_{\ell,j+1}(K)_{\Si_i}
\EQN\eqnNPrenijInda$$
with
$$
P_{\ell,j+1}(K)\in \fK_{\ell}\qquad
P_{j+1,j+1}(K)=K
$$
and
$$
\|P_{\ell,j+1}(K)\|_{1,\Si_\ell} \le
\sfrac{\la_0^{1-\upsilon}}{\al^{5}}\,\sfrac{\fl_{\ell}}{M^{\ell}}
        \fe_{j}({\sst \|K\|_{1,\Si_{j+1}}})
\EQN\eqnNPrenijIndb$$
for $\ell<j+1$.
This will then imply
$$\deqalign{
\big\| \ren_{i,j+1}(K)\big\|_{1,\Si_i}
&\le \sum_{\ell=i}^{j+1} \big\|P_{\ell,j+1}(K)_{\Si_i}\big\|_{1,\Si_i}\cr
& \le \abcst\,\sum_{\ell=i}^{j+1}\sfrac{\fl_i}{\fl_\ell}\,\cb_{i-1}
\,\big\| P_{\ell,j+1}(K) \big\|_{1,\Si_\ell}\quad&
\hbox{by Proposition \propOSindresectorI.ii of [FKTo4]}\cr
&\le\abcst\,\sfrac{\fl_i}{\fl_{j+1}}\,\cb_{i-1}\,\big\| K \big\|_{1,\Si_{j+1}}
+ \abcst\,\sum_{\ell=i}^{j}
\,\sfrac{\la_0^{1-\upsilon}}{\al^{5}}\,\sfrac{\fl_i}{M^{\ell}}
        \cb_{i-1}\fe_{j}({\sst \|K\|_{1,\Si_{j+1}}})\hidewidth\cr
&\le\abcst\,\sfrac{\fl_i}{\fl_{j+1}}\,\cb_{i-1}\,\big\| K \big\|_{1,\Si_{j+1}}
+ \abcst\,\sum_{\ell=i}^{j}
\,\sfrac{\la_0^{1-\upsilon}}{\al^{5}}\,\sfrac{\fl_i}{M^{\ell}}
        \fe_{j}({\sst \|K\|_{1,\Si_{j+1}}})\hidewidth\cr
&&\hbox{by Example \exOSappMonoidI\ of [FKTo1]}\cr
&\le\abcst\,\sfrac{\fl_i}{\fl_{j+1}}\,\cb_{i-1}\,\big\| K \big\|_{1,\Si_{j+1}}
+ \abcst\,\sfrac{\la_0^{1-\upsilon}}{\al^{5}}\,\sfrac{\fl_i}{M^{i}}
        \fe_{j}({\sst \|K\|_{1,\Si_{j+1}}})\hidewidth\cr
&\le\abcst\,\sfrac{\fl_i}{\fl_{j+1}}\,\cb_{i-1}\,\big\| K \big\|_{1,\Si_{j+1}}
+ \sfrac{\la_0^{1-\upsilon}}{\al^{4}}\,\sfrac{\fl_i}{M^{i}}
        \fe_{j}({\sst \|K\|_{1,\Si_{j+1}}})\hidewidth\cr
}\EQN\eqnNPrenijbnd$$
In particular, setting $t=0$,
$$\eqalign{
\big\| \ren_{i,j+1}(K)\big\|_{1,\Si_i}\Big|_{t=0}
&\le \abcst\,\sfrac{\fl_i}{\fl_{j+1}}\,\la_0^{1-\upsilon}\sfrac{\fl_{j+2}}{M^{j+2}}
+ \sfrac{\la_0^{1-\upsilon}}{\al^{4}}\,\sfrac{\fl_i}{M^{i}}
        \frac{1}{1-M^{j}\la_0^{1-\upsilon}\sfrac{\fl_{j+2}}{M^{j+2}}}\cr
&\le \la_0^{1-\upsilon} \sfrac{\fl_i}{M^{j+2}}
+ 2\sfrac{\la_0^{1-\upsilon}}{\al^{4}}\,\sfrac{\fl_i}{M^{i}}\cr
&\le \la_0^{1-\upsilon} \sfrac{\fl_i}{M^{i}}\cr
}\EQN\eqnNPrenijsupbnd$$
by Definitions \defNPCTmSpace\ and \defNPFancynormdomain.iii, if $M$ is 
big enough. Substituting $i=j_0$ and using
$$
\TN\smsum_{s,s'\in\Si_i}\varphi\big((\,\cdot\,,s),(\,\cdot\,,s')\big)
          \TN_{1,\infty}\le\sfrac{\abcst}{\fl_i}\|\varphi\|_{1,\Si_i}
$$
which applies to any translation invariant sectorized function 
on $\big(\bbbr^2\times\Si_i\big)^2$,
also gives the desired bound on $\de e_{j+1}$.

\vskip.25in
Now suppose that (\eqnNPrenijInda) and (\eqnNPrenijIndb)
hold for $\ren_{i+1,j+1}(K)$. Then, defining
$\de K^{(i+1)}(K')=\ren_{i,i+1}(K')-K'_{\Si_i}$,
$$\eqalignno{
\ren_{i,j+1}(K)
&=\ren_{i,i+1}\Big(\ren_{i+1,j+1}(K)\Big)\cr
&=\Big(\ren_{i+1,j+1}(K)\big)_{\Si_i}
+\de K^{(i+1)}\Big(\ren_{i+1,j+1}(K)\Big)\cr
&=\sum_{\ell=i+1}^{j+1} P_{\ell,j+1}(K)_{\Si_i}
+\de K^{(i+1)}\Big(\ren_{i+1,j+1}(K)\Big)\cr
&=\sum_{\ell=i}^{j+1} P_{\ell,j+1}(K)_{\Si_i}\cr
}$$
if we choose
$$
P_{i,j+1}(K)=\de K^{(i+1)}\Big(\ren_{i+1,j+1}(K)\Big)
$$
By Lemma \lemNPdeKbnd.i,
$$\eqalign{ 
\|\de K^{(i+1)}(K')\|_{1,\Si_i} 
&\le 
        \sfrac{\la_0^{1-\upsilon}}{\al^6}\,\sfrac{\fl_i}{M^i}
        \fe_i({\sst \|K'\|_{1,\Si_{i+1}}})
}$$
By the inductive hypothesis, (\eqnNPrenijbnd) applies when $i$ is replaced
by $i+1$, so
$$\eqalign{ 
\|P_{i,j+1}(K)\|_{1,\Si_i} 
&\le 
        \sfrac{\la_0^{1-\upsilon}}{\al^6}\,\sfrac{\fl_i}{M^i}
        \fe_i({\sst \|\ren_{i+1,j+1}(K)\|_{1,\Si_{i+1}}})\cr
&\le 
        \sfrac{\la_0^{1-\upsilon}}{\al^6}\,\sfrac{\fl_i}{M^i}
        \fe_i\big({\sst \abcst\,\sfrac{\fl_{i+1}}{\fl_{j+1}}\,\cb_{i}\,\| K\|_{1,\Si_{j+1}}
+ \sfrac{\la_0^{1-\upsilon}}{\al^{4}}\,\sfrac{\fl_{i+1}}{M^{i+1}}
        \fe_{j}({\sst \|K\|_{1,\Si_{j+1}}})}\big)\cr
&\le 
        \sfrac{\la_0^{1-\upsilon}}{\al^6}\,\sfrac{\fl_i}{M^i}
       \frac{\cb_i}{1- \abcst\,{\fl_{i+1}\over\fl_{j+1}}M^i\,\cb_{i}\,\| K\|_{1,\Si_{j+1}}
- {\la_0^{1-\upsilon}\over\al^{4}}\,{\fl_{i+1}\over M}
        \fe_{j}({\sst \|K\|_{1,\Si_{j+1}}})}\cr
&\le 
        \sfrac{\la_0^{1-\upsilon}}{\al^6}\,\sfrac{\fl_i}{M^i}
       \sfrac{\cb_i}{1- \abcst\,M^{j}\,\cb_{j}\,\| K\|_{1,\Si_{j+1}}
- \la_0^{1-\upsilon}\fe_{j}(\|K\|_{1,\Si_{j+1}})}\cr
}$$
By Lemma \lemOSappMonoidIV.ii and Corollary \corOSappMonoidIV.ii
of [FKTo1],
$$\eqalign{
\|P_{i,j+1}(K)\|_{1,\Si_i}
&\le\abcst\  \sfrac{\la_0^{1-\upsilon}}{\al^6}\,\sfrac{\fl_i}{M^i}\ \sfrac{\cb_i}{1- \abcst\,M^{j}\,\cb_{j}\,\| K\|_{1,\Si_{j+1}}}
\ \sfrac{1}{1-\la_0^{1-\upsilon}\fe_{j}(\|K\|_{1,\Si_{j+1}})}\cr
&\le\abcst\  \sfrac{\la_0^{1-\upsilon}}{\al^6}\,\sfrac{\fl_i}{M^i}\ \sfrac{\cb_{j}}{1-\abcst M^{j}\cb_{j}\|K\|_{1,\Si_{j+1}}}
\ \fe_{j}({\sst \|K\|_{1,\Si_{j+1}}})\cr
}$$
 Corollary \corOSappMonoidIV.i of [FKTo1], with $\mu=\abcst$ and
$X=M^{j}\|K\|_{1,\Si_{j+1}}$, yields
$$
\sfrac{\cb_{j}}{1-\abcst M^{j}\cb_{j}\|K\|_{1,\Si_{j+1}}}
\le\abcst\, \fe_{j}({\sst \|K\|_{1,\Si_{j+1}}})
$$
which, by  Corollary \corOSappMonoidIV.ii of [FKTo1], implies
$$
\|P_{i,j+1}(K)\|_{1,\Si_i} \le
\sfrac{\la_0^{1-\upsilon}}{\al^{5}}\,\sfrac{\fl_i}{M^i}
        \fe_{j}({\sst \|K\|_{1,\Si_{j+1}}})
$$
as desired.

\Item ii) We use induction on $j-i$. Introduce the local notation 
$$
\tn K\tn_{j+1}=\| K\|_{1,\Si_{j+1}}\big|_{t=0}
$$
As $\ren_{i,j+1}$ is the identity map for $i=j+1$, the case $i=j+1$ is trivial.
As
$$\eqalign{
\sfrac{d\hfill}{ds} \ren_{i,j+1}(K+sK')\big|_{s=0}
&=\sfrac{d\hfill}{ds} \ren_{i,i+1}\Big(\ren_{i+1,j+1}(K)
+s\sfrac{d\hfill}{ds'}\ren_{i+1,j+1}(K+s'K')\big|_{s'=0}
\Big)\Big|_{s=0}
}$$
we have, by Lemma \lemNPdeKbnd.ii,
$$\eqalign{
&\big\|\sfrac{d\hfill}{ds}\ren_{i,j+1}(K+sK')\big|_{s=0}\big\|_{1,\Si_i}\cr
&\hskip.5in\le\abcst\,M^{\aleph}
\fe_{i}\big(\|\ren_{i+1,j+1}(K)\|_{1,\Si_{i+1}}\big)\ 
\big\|\sfrac{d\hfill}{ds'}\ren_{i+1,j+1}(K+s'K')\big|_{s'=0}
\big\|_{1,\Si_{i+1}}
}$$
Setting $t=0$,
$$\eqalign{
&\TN\sfrac{d\hfill}{ds}\ren_{i,j+1}(K+sK')\big|_{s=0}\TN_i\cr
&\hskip.5in\le\abcst\,M^{\aleph}
\sfrac{1}{1-M^i\tn\ren_{i+1,j+1}(K)\tn_{i+1}}
\TN\sfrac{d\hfill}{ds}\ren_{i+1,j+1}(K+sK')\big|_{s=0}\TN_{i+1}\cr
&\hskip.5in\le\abcst\,M^{\aleph}
\sfrac{1}{1-M^i\la_0^{1-\upsilon}{\fl_{i+1}\over M^{i+1}}}
\TN\sfrac{d\hfill}{ds}\ren_{i+1,j+1}(K+sK')\big|_{s=0}\TN_{i+1}
\qquad\hbox{by (\eqnNPrenijsupbnd)}\cr
&\hskip.5in\le\abcst\,M^{\aleph}
\TN\sfrac{d\hfill}{ds}\ren_{i+1,j+1}(K+sK')\big|_{s=0}\TN_{i+1}\cr
}$$
By induction
$$
\TN\sfrac{d\hfill}{ds} \ren_{i,j+1}(K+sK')\big|_{s=0}\TN_i
\le \big(\abcst\,M^\aleph\big)^{j+1-i}\tn K'\tn_{j+1}
$$
as desired.
\Item iii)
$$
\ren_{i,j+1}(0)-\ren_{i,j'}(0)
=\ren_{i,j'}\big(\ren_{j',j+1}(0)\big)-\ren_{i,j'}(0)
$$
Hence, by part (ii),
$$
\big\|\ren_{i,j+1}(0)-\ren_{i,j'}(0)\big\|_{1,\Si_i}
\le \abcst^{j'-j_0}\ \sfrac{\fl_{j_0}}{\fl_{j'}}
\| \ren_{j',j+1}(0)\|_{1,\Si_{j'}}+\smsum_{\de\ne 0}\infty\, t^\de
$$
By (\eqnNPrenijsupbnd)
$$\eqalign{
\big\|\ren_{i,j+1}(0)-\ren_{i,j'}(0)\big\|_{1,\Si_i}
&\le \abcst^{j'-j_0}\ \sfrac{\fl_{j_0}}{\fl_{j'}}
\la_0^{1-\upsilon}\sfrac{\fl_{j'}}{M^{j'}}+\smsum_{\de\ne 0}\infty\, t^\de\cr
&\le \la_0^{1-\upsilon}\big(\sfrac{\abcst}{M}\big)^{j'}
+\smsum_{\de\ne 0}\infty\, t^\de\cr
&\le \la_0^{1-\upsilon}\sfrac{1}{2^{j'}}+\smsum_{\de\ne 0}\infty\, t^\de\cr
}$$
and, setting $i=j_0$,
$$\eqalign{
\tn\de e_{j+1}(0)-\de e_{j'}(0)\tn_{1,\infty}
&\le\sfrac{\abcst}{\fl_{j_0}}
\big\|\ren_{j_0,j+1}(0)-\ren_{j_0,j'}(0)\big\|_{1,\Si_{j_0}}\Big|_{t=0}\cr
&\le \abcst\,\abcst^{j'-j_0}\ 
\la_0^{1-\upsilon}\sfrac{1}{M^{j'}}\cr
&\le \la_0^{1-\upsilon}\sfrac{1}{2^{j'}}\cr
}$$

\endproof

\noindent This completes the proof of Theorem \theoremNPinduction.

\vfill\eject

\appendix{\APappRewick}{Self--consistent ReWick Ordering}\PG\pgNPB

In this appendix, we prove Lemma \lemreWick\ and parts (i) and (ii) of Lemma \lemNPdeKbnd. We view any fixed $\sQ\in\cF_0(2,\Si_j)$ as the constant function
$K'\mapsto \sQ$ on $\fK_{j+1}$. In this sense, the definitions of (\eqnNPreWickDefs) apply. For example,
$$
 \de K\big((\x,s),(\x',s');\sQ\big) = \int dx_0\ 
 \big(\sQ*\hat \nu^{(\ge j+1)}\big)\big((x_0,\x,s),(0,\x',s')\big)
$$

\lemma{\STM\lemRWintbnd}{ Assume that $K'\in\fK_{j+1}$ and  $\sQ\in\cF_0(2,\Si_j)$
obeys
$$
\v \sQ\v_{1,\Si_j}\le \sfrac{\la_0^{1-\upsilon}}{\al}\sfrac{\fl_{j}}{M^{j}}
\fe_j({\sst \|K'\|_{1,\Si_{j+1}}})
$$
Then
\Item i) $ K(K';\sQ)\in\fK_{j} $

\Item ii) There are constants $\const$, independent of $j$ but possibly
depending on $M$, and $\abcst$, independent of $M$ and $j$, such that 
$$\eqalign{
\fe_j({\sst \| K(K';\sQ)\|_{1,\Si_j}})
&\le   \const\,\fe_{j}({\sst \|K'\|_{1,\Si_{j+1}}})\cr
\fe_j({\sst \| K(K';\sQ)\|_{1,\Si_j}})
&\le   \abcst\,\fe_{j+1}({\sst \|K'\|_{1,\Si_{j+1}}})\cr
}$$

\Item iii)
{\parindent=.25in
\item{$\bullet$}
$\sqrt{\la_0^{1-\upsilon}\fl_j}\sqrt{\sfrac{\fl_j\IB}{M^j}}$ is an 
integral bound for $E_{\Si_j}(K';\sQ)$ 
\item{$\bullet$}
$
\const\sqrt{\sfrac{\fl_j}{M^j}\Big(
      M^{j}\v \sQ'\v_{1,\Si_j}+
      \sfrac{\partial\ \cb_j
                \sv \sQ'\sv_{1,\Si_j}}{\partial\ t_0}
      \Big)_{t=0}}
$
is an integral bound 
for $\sfrac{d\hfill}{ds}E_{\Si_j}(K';\sQ+s\sQ')\big|_{s=0}$. 
In particular, if $\fd\in\fN_{d+1}$ is independent of $t_0$ and
$\ \v \sQ'\v_{1,\Si_j}\le \fd\cb_j$, 
then $\ \const\sqrt{\fl_j\fd_{\0}}\ $ is an integral bound 
for $\sfrac{d\hfill}{ds}E_{\Si_j}(K';\sQ+s\sQ')\big|_{s=0}$.
\item{$\bullet$}
$\const\sqrt{\fl_j\|K''\|_{1,\Si_{j+1}}\eval{t=0}}$ is an integral bound 
for $\sfrac{d\hfill}{ds}E_{\Si_j}(K'+sK'';\sQ)\big|_{s=0}$.

}}
\prf i) 
Observe that
$$\eqalign{
 \de \check K'(\k;\sQ) &= \check \sQ\big((0,\k)\big)\,
\nu^{(\ge j+1)}((0,\k))
}$$
By Proposition \propOSindresectorI.ii of [FKTo4] and Lemma \lemOSumu\ of
[FKTo3],
$$\eqalign{
\| K'_{\Si_j}+\de K(\sQ)&\|_{1,\Si_j} 
\le \|K'_{\Si_j}\|_{1,\Si_j}+\V \sQ*\hat \nu^{(\ge j+1)}\V_{1,\Si_j} \cr
&\le \abcst\sfrac{\fl_j}{\fl_{j+1}}\cb_{j-1}\|K'\|_{1,\Si_{j+1}}
        +\abcst\,\cb_{j+1}\V \sQ\V_{1,\Si_j}\cr
&\le \abcst\, M^\aleph\cb_{j-1}\|K'\|_{1,\Si_{j+1}}
        +\abcst\,\cb_{j+1}\sfrac{\la_0^{1-\upsilon}}{\al}\sfrac{\fl_{j}}{M^{j}}
\fe_j({\sst \|K'\|_{1,\Si_{j+1}}})\cr
&\le \abcst\,M^\aleph\la_0^{1-\upsilon}\sfrac{\fl_{j+2}}{M^{j+2}}
        +\abcst\,\sfrac{\la_0^{1-\upsilon}}{\al}\sfrac{\fl_{j}}{M^{j}}
        +\smsum_{\de\ne 0}\infty\, t^\de\cr
&\le \la_0^{1-\upsilon}\sfrac{\fl_{j+1}}{M^{j+1}}
        +\smsum_{\de\ne 0}\infty\, t^\de\cr
}\EQN\eqnNPwkKofKp$$
if $M$ is large enough and $\al$ is large enough, depending on $M$.
By definition, 
${\rm supp}\,\check K' \subset {\rm supp}\, \nu^{(\ge j+2)}(0,\k)
 \subset {\rm supp}\, \nu^{(\ge j+1)}(0,\k)$, and by construction
${\rm supp}\,\check \de K  \subset {\rm supp}\, \nu^{(\ge j+1)}(0,\k)$, so
$\check K(K';\sQ)$ fulfills the required support property.

\Item ii)
By (\eqnNPwkKofKp)
$$\eqalign{
\fe_j({\sst \| K(K';\sQ)\|_{1,\Si_j}})
&=\frac{\cb_j}{1-M^j\| K'_{\Si_j}+\de K(\sQ)\|_{1,\Si_j}}\cr
&\le\frac{\cb_{j}}{1-M^j\big[
      \abcst\, M^\aleph\cb_{j-1}\|K'\|_{1,\Si_{j+1}}
        +\abcst\,\cb_{j+1}\sfrac{\la_0^{1-\upsilon}}{\al}\sfrac{\fl_{j}}{M^{j}}
\fe_j({\sst \|K'\|_{1,\Si_{j+1}}})\big]}\cr
&\le\frac{\cb_j}{1- M^{j+1}\cb_j\|K'\|_{1,\Si_{j+1}}
        -\abcst \sfrac{\la_0^{1-\upsilon}}{\al}\,\cb_{j+1}\,
        \fe_j({\sst \|K'\|_{1,\Si_{j+1}}})}\cr
&\le\frac{\cb_j}{1- M^{j+1}\cb_j\|K'\|_{1,\Si_{j+1}}
        -\la_0^{1-\upsilon}\,\fe_j({\sst \|K'\|_{1,\Si_{j+1}}})}\cr
}$$
if $\al$ is large enough, since 
$$ 
\cb_{j+1}\fe_j({\sst \|K'\|_{1,\Si_{j+1}}})
\le M^{r+r_0}\cb_j\fe_j({\sst \|K'\|_{1,\Si_{j+1}}})
\le\const\fe_j({\sst \|K'\|_{1,\Si_{j+1}}})
\EQN\eqnNPwkce
$$
By Lemma \lemOSappMonoidIV.ii and Corollary \corOSappMonoidIV.ii
of [FKTo1],
$$\eqalign{
\fe_j({\sst \| K(K';\sQ)\|_{1,\Si_j}})
&\le\abcst\ \sfrac{\cb_j}{1- M^{j+1}\cb_j\|K'\|_{1,\Si_{j+1}}}
\ \sfrac{1}{1-\la_0^{1-\upsilon}\,\fe_j({\sst \|K'\|_{1,\Si_{j+1}}})}\cr
&\le\abcst\ \sfrac{\cb_j}{1- M^{j+1}\cb_j\|K'\|_{1,\Si_{j+1}}}
\ \fe_j({\sst \|K'\|_{1,\Si_{j+1}}})\cr
}$$
 Corollary \corOSappMonoidIV.i of [FKTo1], with $\mu=M$ and
$X=M^{j}\|K'\|_{1,\Si_{j+1}}$, yields
$$
 \sfrac{\cb_j}{1- M^{j+1}\cb_j\|K'\|_{1,\Si_{j+1}}}
\le\const \fe_j({\sst \|K'\|_{1,\Si_{j+1}}})
$$
which, by  Corollary \corOSappMonoidIV.ii of [FKTo1], implies
the first bound.
On the other hand, Corollary \corOSappMonoidIV.i of [FKTo1], 
with $\mu=1$ and $X=M^{j+1}\|K'\|_{1,\Si_{j+1}}$, yields
$$
 \sfrac{\cb_j}{1- M^{j+1}\cb_j\|K'\|_{1,\Si_{j+1}}}
\le \sfrac{\cb_{j+1}}{1- M^{j+1}\cb_{j+1}\|K'\|_{1,\Si_{j+1}}}
\le\abcst\, \fe_{j+1}({\sst \|K'\|_{1,\Si_{j+1}}})
$$
which, by  Corollary \corOSappMonoidIV.ii of [FKTo1], implies
the second bound.

\Item iii)
Set
$$\eqalign{
V(K';\sQ)&=u'(K';\sQ)+K'_{\rm ext}*\hat \nu^{(\ge j+3)}\cr
v(K';\sQ)&=u(K(K';\sQ))+K(K';\sQ)_{\rm ext}*\hat \nu^{(\ge j+2)}\cr
}$$
Then
$$\eqalign{
&E(K';\sQ)=C_{j+1}\big( u'(\ \cdot\ ;\sQ);K' \big)- D_{j}\big( u;K(K';\sQ)\big)\cr
&=\sfrac{\nu^{(\ge j+1)}(k)}
{\imath k_0 -e(\k) -\check u'(k;K';\sQ) - \check K'(\k)\nu^{(\ge j+3)}(k)}
-\sfrac{\nu^{(\ge j+1)}(k)}
{\imath k_0 -e(\k) -\check u(k;K(K';\sQ)) 
- \check K(\k;K';\sQ)\nu^{(\ge j+2)}(k)}\cr
&=\sfrac{\nu^{(\ge j+1)}(k)}{\imath k_0 -e(\k) -\check V(k;K';\sQ)}
-\sfrac{\nu^{(\ge j+1)}(k)}{\imath k_0 -e(\k) -\check v(k;K';\sQ)}
}\EQN\eqnNPmomspE$$
Also
$$\eqalign{
\check V(k;K';\sQ)&-\check v(k;K';\sQ)\cr
&=\left[\check u'(K';\sQ) + \check K'\nu^{(\ge j+3)}\right]
-\left[\check u(K(K';\sQ)) 
+\big(\check K'_{\Si_j}+\de\check K'(\sQ)\big)\nu^{(\ge j+2)}\right]\cr
&=\left[\check u(K(K';\sQ))+\check \sQ\,\nu^{(\ge j+1)}
 + \check K'\nu^{(\ge j+3)}\right]\cr
&\hskip1in
-\left[\check u(K(K';\sQ)) 
+\big(\check K'_{\Si_j}+\de\check K'(\sQ)\big)\nu^{(\ge j+2)}\right]\cr
&=- \check K'(\k)\nu^{(j+2)}(k)+\check \sQ(k)\,\nu^{(\ge j+1)}(k)
-\check \sQ\big((0,\k)\big)\,
\nu^{(\ge j+1)}((0,\k))\nu^{(\ge j+2)}(k)\cr
}\EQN\eqnNPwkdecomp$$
For the last equality, we used that 
$\check K'(\k)=\check K'_{\Si_j}(\k)$,
by Definitions \:\defOSindresector\ of [FKTo4] and \:\defOSsectrepr\ of [FKTo3],
since $\check K'(\k)$ vanishes outside the support of $\nu^{(\ge j+2)}((0,\k))$.
By Definition \defNPCTmSpace, Lemma \lemOSNormMom\ of [FKTo3]
and Definition \defOSscales\ of [FKTo2],
$$\eqalign{
\big| \check K'(\k)\nu^{(j+2)}(k)\big|
&\le 2\la_0^{1-\upsilon}\sfrac{\fl_{j+2}}{M^{j+2}}\nu^{(j+2)}(k)
\le 2\la_0^{1-\upsilon}\sfrac{\fl_{j+2}}{M^{j+2}}\sfrac{|\imath k_0-e(\k)|}
{{1\over\sqrt{M}}{1\over M^{j+2}}}\cr
&=2\sqrt{M}\la_0^{1-\upsilon}\fl_{j+2}|\imath k_0-e(\k)|
\le\sfrac{1}{10}\la_0^{1-\upsilon}\fl_j|\imath k_0-e(\k)|
}\EQN\eqnNPwka$$
Similarly, using Lemma \lemOSumu\ of [FKTo3],
$$\eqalign{
\big|\check \sQ(k)\,\nu^{(\ge j+1)}(k)-\check \sQ\big((0,\k)\big)\,
                  \nu^{(\ge j+1)}((0,\k))\big|\nu^{(j+2)}(k)
&\le 2|k_0|\,\sfrac{\partial\hfill}{\partial t_0}
                \V \sQ*\hat\nu^{(\ge j+1)}\V_{1,\Si_j}\Big|_{t=0}\cr
&\le \abcst\,|k_0|\,\sfrac{\partial\hfill}{\partial t_0}\big(\cb_{j+1}
                \V \sQ\V_{1,\Si_j}\big)\Big|_{t=0}
}\EQN\eqnNPwkb$$
and 
$$\eqalign{
\big| \check \sQ(k)\big(1-\nu^{(\ge j+2)}(k)\big)\,\nu^{(\ge j+1)}(k)\big|
&\le 2\V \sQ\V_{1,\Si_j}\nu^{(j+1)}(k)
\le 2\V \sQ\V_{1,\Si_j}\sfrac{|\imath k_0-e(\k)|}
{{1\over\sqrt{M}}{1\over M^{j+1}}}\cr
&=2M^{j+{3\over 2}}\V \sQ\V_{1,\Si_j}|\imath k_0-e(\k)|
}\EQN\eqnNPwkc$$
Combining (\eqnNPwkdecomp)--(\eqnNPwkc)
$$\eqalignno{
|\check V(k;K';\sQ)-\check v(k;K';\sQ)|&\le 
\Big[\sfrac{1}{10}\la_0^{1-\upsilon}\fl_{j+2}
      +\abcst\sfrac{\partial\ \cb_{j+1}
                \sv \sQ\sv_{1,\Si_j}}{\partial\ t_0}
      +2M^{j+{3\over 2}}\V \sQ\V_{1,\Si_j}\Big]_{t=0}
|\imath k_0-e(\k)|\cr
&\le \sfrac{1}{4}\la_0^{1-\upsilon}\fl_j\ |\imath k_0-e(\k)| &\EQNO\eqnNPwkd\cr
}$$
if $\al$ is large enough. Lemma \lemNPpptyu.i implies
$$
|\check v(k;K';\sQ)|
=\big|\check u(k;K(K';\sQ)) 
+ \check K(\k;K';\sQ)\nu^{(\ge j+2)}(k)\big|
\le \la_0^{1-\upsilon}|\imath k_0-e(\k)|
\EQN\eqnNPwke$$
as well as
$$\eqalignno{
\big|\sfrac{d\hfill}{ds}\check v(k;K';\sQ+s\sQ')|_{s=0}\big|
&=\big|\sfrac{d\hfill}{ds}\check u(k;K'_{\Si_j}+\de K(\sQ)+s\de K(\sQ'))|_{s=0}
+ \de\check K'(\k;\sQ')\nu^{(\ge j+2)}(k)\big|\cr
&\le 4 M^{j+{3\over 2}}\|\de K(\sQ')\|_{1,\Si_j}|\imath k_0-e(\k)|\cr
&\le \abcst\, M^{j+{3\over 2}}\v \sQ'\v_{1,\Si_j}\,|\imath k_0-e(\k)|
&\EQNO\eqnNPwkf\cr
}$$
and
$$\eqalignno{
\big|\sfrac{d\hfill}{ds}\check v(k;K'+sK'';\sQ)|_{s=0}\big|
&=\big|\sfrac{d\hfill}{ds}\check u(k;K'_{\Si_j}+\de K(\sQ)+sK''_{\Si_j})|_{s=0}
+ \check K''(\k)\nu^{(\ge j+2)}(k)\big|\cr
&\le 4 M^{j+{3\over 2}}\|K''_{\Si_j}\|_{1,\Si_j}|\imath k_0-e(\k)|\cr
&\le \abcst\sfrac{\fl_j}{\fl_{j+1}}
 M^{j+{3\over 2}}\cb_{j-1}\|K''\|_{1,\Si_{j+1}}|\imath k_0-e(\k)|\cr
&\le \abcst
 M^{j+{3\over 2}+\aleph}\cb_{j-1}\|K''\|_{1,\Si_{j+1}}|\imath k_0-e(\k)|
&\EQNO\eqnNPwki\cr
}$$
From (\eqnNPwkdecomp)
$$
\sfrac{d\hfill}{ds}\big[\check V(k;K';\sQ+s\sQ')-\check v(k;K';\sQ+s\sQ')\big]
\!=\!\check \sQ'(k)\,\nu^{(\ge j+1)}(k)
-\check \sQ'\big((0,\k)\big)
\nu^{(\ge j+1)}((0,\k))\nu^{(\ge j+2)}(k)
$$
so that
$$
\Big|\sfrac{d\hfill}{ds}\big[\check V(k;K';\sQ+s\sQ')-\check v(k;K';\sQ+s\sQ')\big]\Big|\le
\Big[\abcst\sfrac{\partial\ \cb_{j+1}
                \sv \sQ'\sv_{1,\Si_j}}{\partial\ t_0}
      +2M^{j+{3\over 2}}\V \sQ'\V_{1,\Si_j}\Big]_{t=0}
\ |\imath k_0-e(\k)|
\EQN\eqnNPwkg$$
by  (\eqnNPwkb) and (\eqnNPwkc). Similarly,
$$
\sfrac{d\hfill}{ds}\big[\check V(k;K'+sK'';\sQ)-\check v(k;K'+sK'';\sQ)\big]
=-\check K''(\k)\nu^{(\ge j+2)}(k)
$$
so that
$$\eqalign{
\Big|\sfrac{d\hfill}{ds}\big[\check V(k;K'+sK'';\sQ)-\check v(k;K'+sK'';\sQ)\big]\Big|
&\le 2\V K''\V_{1,\Si_{j+1}}\nu^{(j+2)}(k)\cr
&\le 2\V K''\V_{1,\Si_{j+1}}\sfrac{|\imath k_0-e(\k)|}
{{1\over\sqrt{M}}{1\over M^{j+2}}}\cr
&=2M^{j+{5\over 2}}\V K''\V_{1,\Si_{j+1}}|\imath k_0-e(\k)|
}\EQN\eqnNPwkh$$
as in (\eqnNPwkc).

Using (\eqnNPwkd) and (\eqnNPwke)
$$\eqalign{
\big|E(K';\sQ)\big|
&=\Big|\sfrac{\nu^{(\ge j+1)}(k)}{\imath k_0 -e(\k) -\check V(k;K';\sQ)}
-\sfrac{\nu^{(\ge j+1)}(k)}{\imath k_0 -e(\k) -\check v(k;K';\sQ)}\Big|\cr
&=\Big|\sfrac{\check V(k;K';\sQ)-\check v(k;K';\sQ)}
{[\imath k_0 -e(\k) -\check V(k;K';\sQ)]
\ [\imath k_0 -e(\k) -\check v(k;K';\sQ)]}\Big|
\nu^{(\ge j+1)}(k)\cr
&\le \sfrac{\la_0^{1-\upsilon}\fl_j}{|\imath k_0 -e(\k)|}
}$$
The integral bound for $E_{\Si_j}(K';\sQ)$ now follows from Proposition \propOScontrintboundsectors\ of [FKTo3].
Similarly,
$$\eqalign{
\big|\sfrac{d\hfill}{ds}E_{\Si_j}(K';\sQ+s\sQ')\big|_{s=0}\big|
&=\bigg|\sfrac{{d\hfill\over ds}{\check V(k;K';\sQ+s\sQ')|}_{s=0}}
         {[\imath k_0 -e(\k) -\check V(k;K';\sQ)]^2}
-\sfrac{{d\hfill\over ds}{\check v(k;K';\sQ+s\sQ')}|_{s=0}}
          {[\imath k_0 -e(\k) -\check v(k;K';\sQ)]^2}\bigg|
\nu^{(\ge j+1)}(k)\cr
&\le \const\sfrac{ M^{j}\v \sQ'\v_{1,\Si_j}+
      {\partial\ \cb_j\sv \sQ'\sv_{1,\Si_j}\over
         \partial\ t_0}\big|_{t=0}}
{|\imath k_0 -e(\k)|}
}$$
and the first integral bound for $\sfrac{d\hfill}{ds}E_{\Si_j}(K';\sQ+s\sQ')\big|_{s=0}$
 also follows from Proposition \propOScontrintboundsectors\ of [FKTo3].

If 
if $\fd\in\fN_{d+1}$ is independent of $t_0$ and
$\ \v \sQ'\v_{1,\Si_j}\le \fd\cb_j\ $, 
then
$$
 \Big(M^{j}\v \sQ'\v_{1,\Si_j}+
      \sfrac{\partial\ \cb_j\sv \sQ'\sv_{1,\Si_j}}{\partial\ t_0}
  \Big)_{t=0}\le \abcst M^j\fd_\0
$$
and the second integral bound for $\sfrac{d\hfill}{ds}E_{\Si_j}(K';\sQ+s\sQ')\big|_{s=0}$
 follows from the first.

Finally, using (\eqnNPwki) and (\eqnNPwkh),
$$\eqalign{
\big|\sfrac{d\hfill}{ds}E(K'+sK'';\sQ)\big|_{s=0}\big|
&=\bigg|\sfrac{{d\hfill\over ds}{\check V(k;K'+sK'';\sQ)|}_{s=0}}
         {[\imath k_0 -e(\k) -\check V(k;K';\sQ)]^2}
-\sfrac{{d\hfill\over ds}{\check v(k;K'+sK'';\sQ)}|_{s=0}}
          {[\imath k_0 -e(\k) -\check v(k;K';\sQ)]^2}\bigg|
\nu^{(\ge j+1)}(k)\cr
&\le \const\sfrac{ M^{j}\| K''\|_{1,\Si_{j+1}}}
{|\imath k_0 -e(\k)|}
}$$
and the integral bound for $\sfrac{d\hfill}{ds}E_{\Si_j}(K'+sK'';\sQ)\big|_{s=0}$
 also follows from Proposition \propOScontrintboundsectors\ of [FKTo3].
\endproof

Recall that $\ \tilde w_{0,2}(K';\sQ)\in\cF_0(2,\Si_j)\ $ was defined, following
(\eqnNPreWickDefs), to be the 
coefficient of $\psi({\sst (\xi_1,s_1)})\psi({\sst (\xi_2,s_2)\,})$
in
$\ 
\lW  w(K(K';\sQ))\rW_{-E_{\Si_j}(K';\sQ)}
\ $.

\lemma{\STM\lemRWrhsbounds}{ Assume that $K'\in\fK_{j+1}$,
$\fd\in\fN_{d+1}$ is independent of $t_0$ 
and
$$\eqalign{
\V \sQ\V_{1,\Si_j}&\le \sfrac{\la_0^{1-\upsilon}}{\al}\sfrac{\fl_{j}}{M^{j}}
\fe_j\big({\sst \|K'\|_{1,\Si_{j+1}}}\big)\cr
\V \sQ'\V_{1,\Si_j}&\le \fd\cb_j\cr
}$$
Then
$$\eqalign{
\v\tilde w_{0,2}(K';\sQ)\v_{1,\Si_j}
&\le \sfrac{\la_0^{1-\upsilon}}{\al^{6.5}}\sfrac{\fl_{j}}{M^{j}}
\fe_j\big({\sst \|K'\|_{1,\Si_{j+1}}}\big)\cr
\V\sfrac{d\hfill}{ds}\tilde w_{0,2}(K';\sQ+s\sQ')\eval{s=0}\V_{1,\Si_j}
&\le\sfrac{\la_0^{1-\upsilon}}{\al}\fd\,
\fe_j\big({\sst \|K'\|_{1,\Si_{j+1}}}\big) \cr
\V\sfrac{d\hfill}{ds}\tilde w_{0,2}(K'+sK'';\sQ)\eval{s=0}\V_{1,\Si_j}
&\le \sfrac{\la_0^{1-\upsilon}}{\al^{1.5}}\fe_j({\sst \|K'\|_{1,\Si_{j+1}}})
\|K''\|_{1,\Si_{j+1}} \cr
}$$
}
\prf We use the notation of \S\CHsectors\ of [FKTo3]. Set $\tilde \al=\sfrac{\al}{2\la_0^{(1-\upsilon)/2}}$,
$\ib =\sqrt{\sfrac{\IB\,\fl_j}{M^j}}$, $X=\|K(K';\sQ)\|_{1,\Si_j}$ and
$\cb = \cst{}{1}\,M^j\,\fe_j\big(X\big)$,
where $\cst{}{1}$ is the constant of Lemma \lemOSconcreteintconst\ of [FKTo3].  Let, for a sectorized Grassmann function $v=\smsum_{n} v_{n}$ with
$v_{n} \in \bbbc\otimes \bigwedge^nV_\Si$,
$$
N(v;\tilde\al)=\sfrac{1}{\ib^2}\cb\,
\smsum_{n}\tilde\al^n\,\ib^n\,\v v_{n} \v_{1,\Si} 
$$  
Observe that, if $V=\smsum_{m,n} V_{m,n}$ with
$V_{m,n} \in A_m\otimes \bigwedge^nV_\Si$ and $V_{0,2}=0$, and if 
$v=\smsum_{n} V_{0,n}$, then
$$
N(v;2\tilde \al) 
\le \sfrac{\cst{}{1}}{\IB\la_0^{1-\upsilon}}\, N_j\big(V,\al,X\big)
$$

Set, using the notation of Definition \defOStens\ of [FKTo3],
$$\eqalign{
W(K';\sQ)&=w(K(K';\sQ))\eval{\phi=0}\cr
W_2(K';\sQ)&= Gr\big(w(K(K';\sQ))_{0,2}\big)\cr
W_4(K';\sQ)&=W(K';\sQ)-W_2(K';\sQ)\cr
}$$
Then
$$\eqalign{
&\fe_j(X)\v\tilde w_{0,2}(K';\sQ)\v_{1,\Si_j}
= \sfrac{1}{\cst{}{1}\tilde\al^2M^j}
N\big(Gr(\tilde w_{0,2}(K';\sQ));\tilde \al\big)\cr
&\hskip.1in\le\fe_j(X)
         \v w(K(K';\sQ))_{0,2}\v_{1,\Si_j}
+\sfrac{1}{\cst{}{1}\tilde\al^2M^j}
N\big(Gr(\tilde w_{0,2}(K';\sQ))-W_2(K';\sQ);\tilde \al\big)\cr
&\hskip.1in\le\fe_j(X)
         \v w(K(K';\sQ))_{0,2}\v_{1,\Si_j}
\!+\sfrac{1}{\cst{}{1}\tilde\al^2M^j}
N\big(\lW W_4(K';\sQ)\rW_{-E_{\Si_j}(K';\sQ)}\!-W_4(K';\sQ);\tilde \al\big)\cr
&\hskip.1in\le\fe_j(X)
         \v w(K(K';\sQ))_{0,2}\v_{1,\Si_j}
+\const\sfrac{\la_0^{1-\upsilon}\fl_j}{\tilde\al^4M^j}
N\big( W_4(K';\sQ);2\tilde \al\big)\cr
}$$
by Corollary \corwicknorm\ of [FKTr1] and Lemma \lemRWintbnd.iii.
By the observation above
$$\eqalign{
\sfrac{\la_0^{1-\upsilon}\fl_j}{\tilde\al^4M^j}
N\big( W_4(K';\sQ);2\tilde \al\big)
&\le\const\sfrac{\fl_j}{\tilde\al^4M^j}
N_j\big( w(K(K';\sQ))+\half\phi C^{(j)}\phi, \al,X\big)\cr
&\le\const\sfrac{\la_0^{2(1-\upsilon)}\fl_j}{\al^4M^j}
N_j\big( w(K(K';\sQ))+\half\phi C^{(j)}\phi, \al,X\big)\cr
}$$
Hence, by (O1),
$$\eqalign{
\v\tilde w_{0,2}(K';\sQ)\v_{1,\Si_j}
&\le\sfrac{\la_0^{1-\upsilon}\fl_j}{\al^7M^j}\fe_j(X)^2
+\const \sfrac{\la_0^{2(1-\upsilon)}\fl_j}{\al^4M^j}\fe_j(X)\cr
&\le\const\sfrac{\la_0^{1-\upsilon}\fl_j}{\al^7M^j}
\fe_j({\sst \|K'\|_{1,\Si_{j+1}}})\cr
&\le\sfrac{\la_0^{1-\upsilon}\fl_j}{\al^{6.5}M^j}
\fe_j({\sst \|K'\|_{1,\Si_{j+1}}})\cr
}$$
by  Corollary \corOSappMonoidIV.ii of [FKTo1]\ and Lemma \lemRWintbnd.ii.
\vskip.25in

We now prove the bound on $\V\sfrac{d\hfill}{ds}\tilde w_{0,2}(K';\sQ+s\sQ')\eval{s=0}\V_{1,\Si_j}$.
This time we use $\al'=\sfrac{\al}{2}$ and, for any sectorized 
Grassmann function $v=\smsum_{n} v_{n}$ with 
$v_{n} \in \bbbc\otimes \bigwedge^nV_\Si$,
$$
N'(v;\al')=\sfrac{1}{\ib^2}\cb\,
\smsum_{n}{\al'}^n\,\ib^n\,\v v_{n} \v_{1,\Si} 
$$  
The other notation is as in the first part of this proof.
This time, if $V=\smsum_{m,n} V_{m,n}$ with
$V_{m,n} \in A_m\otimes \bigwedge^nV_\Si$ ($V_{0,2}$ need not vanish), and if 
$v=\smsum_{n} V_{0,n}$, then
$$
N'(v;2\al') 
\le \sfrac{\cst{}{1}}{\IB}\la_0^{1-\upsilon}\, N_j\big(V,\al,X\big)
$$
Hence
$$\eqalign{
&\fe_j(X)\V\sfrac{d\hfill}{ds}\tilde w_{0,2}(K';\sQ+s\sQ')\eval{s=0}\V_{1,\Si_j}
= \sfrac{1}{\cst{}{1}{\al'}^2M^j}
N'\big(\sfrac{d\hfill}{ds}Gr(\tilde w_{0,2}(K';\sQ+s\sQ'))\eval{s=0};\al'\big)\cr
&\hskip0.5in\le\sfrac{1}{\cst{}{1}{\al'}^2M^j}
N'\big(\sfrac{d\hfill}{ds}\lW W(K';\sQ+s\sQ')\rW_{-E_{\Si_j}(K';\sQ+s\sQ')}\eval{s=0};
 \al'\big)\cr
&\hskip0.5in\le\sfrac{1}{\cst{}{1}{\al'}^2M^j}
N'\big(\lW \sfrac{d\hfill}{ds}W(K';\sQ+s\sQ')\eval{s=0}\rW_{-E_{\Si_j}(K';\sQ)};
 \al'\big)\cr
&\hskip1in+\sfrac{1}{\cst{}{1}{\al'}^2M^j}
N'\big(\sfrac{d\hfill}{ds}\lW W(K';\sQ)\rW_{-E_{\Si_j}(K';\sQ+s\sQ')}\eval{s=0};
 \al'\big)\cr
&\hskip0.5in\le\sfrac{1}{\cst{}{1}{\al'}^2M^j}
N'\big( \sfrac{d\hfill}{ds}W(K';\sQ+s\sQ')\eval{s=0};
 2\al'\big)\cr
&\hskip1in+\sfrac{1}{\cst{}{1}{\al'}^2M^j}
\sfrac{1}{(\al'-1)^2}\ \const\sfrac{\fd_\0M^j}{\IB}N'\big(W(K';\sQ); 2\al'\big)\cr
&\hskip0.5in\le\const\sfrac{\la_0^{1-\upsilon}}{\al^2M^j}
N_j\big( \sfrac{d\hfill}{ds}W(K';\sQ+s\sQ')\eval{s=0},\al,X\big)\cr
&\hskip1in+\const\sfrac{\la_0^{1-\upsilon}\fd_\0}{\al^4}
N_j\big(W(K';\sQ), \al,X\big)\cr
}$$
In the second last inequality, we used Corollary \corwicknorm.i,iii
of [FKTr1] and Lemma \lemRWintbnd.iii. Since
$$\eqalign{
\sfrac{d\hfill}{ds}W(K';\sQ+s\sQ')\eval{s=0}
&=\sfrac{d\hfill}{ds}w\big(K'_{\Si_j}+\de K(\sQ+s\sQ')\big)\eval{s=0\atop\phi=0}\cr
&=\sfrac{d\hfill}{ds}w\big(K'_{\Si_j}+\de K(\sQ)+s\de K(\sQ')\big)
                   \eval{s=0\atop\phi=0}\cr
}$$
(O1) implies that
$$
N_j\big( \sfrac{d\hfill}{ds}W(K';\sQ+s\sQ')\eval{s=0},\al,X\big)
\le M^j \fe_j(X) \|\de K(\sQ') \|_{1,\Si_j}
\le \abcst  M^j \cb_{j+1}\fe_j(X) \|\sQ' \|_{1,\Si_j}
$$
(O1) also implies that 
$$
N_j\big(W(K';\sQ), \al,X\big)
=N_j\big( w\big(K(K';\sQ)\big)\eval{\phi=0},\al,X\big)
\le\fe_j(X) 
$$
Hence
$$\eqalign{
\fe_j(X)\V\sfrac{d\hfill}{ds}\tilde w_{0,2}(K';\sQ+s\sQ')\eval{s=0}\V_{1,\Si_j}
&\le \const\!\Big[\sfrac{\la_0^{1-\upsilon}}{\al^2M^j}
M^j \cb_{j+1}\fe_j(X)\ \fd\cb_j
+\sfrac{\la_0^{1-\upsilon}\fd_\0}{\al^4}\fe_j(X)\Big]\cr
&\le\const \sfrac{\la_0^{1-\upsilon}}{\al^2}\fd\fe_j({\sst \|K'\|_{1,\Si_{j+1}}})
\le\sfrac{\la_0^{1-\upsilon}}{\al}\fd\fe_j({\sst \|K'\|_{1,\Si_{j+1}}})
}$$
by Lemma \lemRWintbnd.ii, (\eqnNPwkce) and
Corollary \corOSappMonoidIV.ii of [FKTo1].

\vskip.25in

Finally, we prove the bound on 
$\V\sfrac{d\hfill}{ds}\tilde w_{0,2}(K'+sK'';\sQ)\eval{s=0}\V_{1,\Si_j}$.
We have
$$\eqalign{
&\fe_j(X)\V\sfrac{d\hfill}{ds}\tilde w_{0,2}(K'+sK'';\sQ)\eval{s=0}\V_{1,\Si_j}
= \sfrac{1}{\cst{}{1}{\al'}^2M^j}
N'\big(\sfrac{d\hfill}{ds}Gr(\tilde w_{0,2}(K'+sK'';\sQ))\eval{s=0};\al'\big)\cr
&\hskip0.5in\le\sfrac{1}{\cst{}{1}{\al'}^2M^j}
N'\big(\sfrac{d\hfill}{ds}\lW W(K'+sK'';\sQ)\rW_{-E_{\Si_j}(K'+sK'';\sQ)}\eval{s=0};
 \al'\big)\cr
&\hskip0.5in\le\sfrac{1}{\cst{}{1}{\al'}^2M^j}
N'\big(\lW \sfrac{d\hfill}{ds}W(K'+sK'';\sQ)\eval{s=0}\rW_{-E_{\Si_j}(K';\sQ)};
 \al'\big)\cr
&\hskip1in+\sfrac{1}{\cst{}{1}{\al'}^2M^j}
N'\big(\sfrac{d\hfill}{ds}\lW W(K';\sQ)\rW_{-E_{\Si_j}(K'+sK'';\sQ)}\eval{s=0};
 \al'\big)\cr
&\hskip0.5in\le\sfrac{1}{\cst{}{1}{\al'}^2M^j}
N'\big( \sfrac{d\hfill}{ds}W(K'+sK'';\sQ)\eval{s=0};
 2\al'\big)\cr
&\hskip1in+\sfrac{1}{\cst{}{1}{\al'}^2M^j}
\sfrac{1}{(\al'-1)^2}\ \const\sfrac{M^j\|K''\|_{1,\Si_{j+1}}}{\IB}N'\big(W(K';\sQ); 2\al'\big)\cr
&\hskip0.5in\le\const\sfrac{\la_0^{1-\upsilon}}{\al^2M^j}
N_j\big( \sfrac{d\hfill}{ds}W(K'+sK'';\sQ)\eval{s=0},\al,X\big)\cr
&\hskip1in+\const\sfrac{\la_0^{1-\upsilon}}{\al^4}\|K''\|_{1,\Si_{j+1}}
N_j\big(W(K';\sQ), \al,X\big)\cr
}$$
In the second last inequality, we used Corollary \corwicknorm.i,iii
of [FKTr1] and Lemma \lemRWintbnd.iii. Since
$$\eqalign{
\sfrac{d\hfill}{ds}W(K'+sK'';\sQ)\eval{s=0}
&=\sfrac{d\hfill}{ds}w\big(K'_{\Si_j}
           +\de K(\sQ)+sK''_{\Si_j}\big)\eval{s=0\atop\phi=0}\cr
}$$
(O1) implies that
$$
N_j\big( \sfrac{d\hfill}{ds}W(K'+sK'';\sQ)\eval{s=0},\al,X\big)
\le M^j \fe_j(X) \| K''_{\Si_j}\|_{1,\Si_j}
\le \abcst  M^{j+\aleph}\cb_{j-1}\fe_j(X) \|K''\|_{1,\Si_{j+1}}
$$
and, as we have already observed, 
$$
N_j\big( w\big(K(K';\sQ)\big)\eval{\phi=0},\al,X\big)
\le\fe_j(X) 
$$
Hence
$$\eqalign{
\fe_j(X)\V\sfrac{d\hfill}{ds}\tilde w_{0,2}(K'+sK'';\sQ)\eval{s=0}\V_{1,\Si_j}
&\le \const\!\Big[\sfrac{\la_0^{1-\upsilon}}{\al^2}\cb_{j-1}
+\sfrac{\la_0^{1-\upsilon}}{\al^4}\Big]\|K''\|_{1,\Si_{j+1}}\fe_j(X)\cr
&\le\const \sfrac{\la_0^{1-\upsilon}}{\al^2}\fe_j({\sst \|K'\|_{1,\Si_{j+1}}})
\|K''\|_{1,\Si_{j+1}}\cr
&\le\sfrac{\la_0^{1-\upsilon}}{\al^{1.5}}\fe_j({\sst \|K'\|_{1,\Si_{j+1}}})
\|K''\|_{1,\Si_{j+1}}
}$$
by Lemma \lemRWintbnd.ii.
\endproof

We now solve $ q(K') = 2\tilde w_{0,2}(q(K');K')$ by a standard contraction
mapping argument. Define
$$\eqalign{
q^{(0)}&= 0\cr
q^{(1)}&= 2\tilde w_{0,2}(0;K')\cr
q^{(n+1)}&= 2\tilde w_{0,2}(q^{(n)};K')\qquad\qquad n\ge 1\cr
}$$
We use the shorthand notation $\fe_j=\fe_j({\sst \|K'\|_{1,\Si_{j+1}}})$.
\lemma{\STM\lemNPcontrmap}{Let $K'\in\fK_{j+1}$. Then
$$
\V q^{(n)}-q^{(n-1)}\V_{1,\Si_j}\le 
\Big(\ka\sfrac{\la_0^{1-\upsilon}}{\al}\Big)^{n-1}
\Big(2\sfrac{\la_0^{1-\upsilon}}{\al^{6.5}}\sfrac{\fl_{j}}{M^{j}}\Big)
\fe_j
$$
}
\prf
The proof is by induction on $n$. By Lemma \lemRWrhsbounds
$$
\V q^{(1)}\V_{1,\Si_j}\le 2\sfrac{\la_0^{1-\upsilon}}{\al^{6.5}}\sfrac{\fl_{j}}{M^{j}}
\fe_j
$$
and the conclusion of the Lemma is true for $n=1$.
If the Lemma is satisfied for some $n$, then,
by Lemma \lemRWrhsbounds\ with 
$$
\fd=\Big(\ka\sfrac{\la_0^{1-\upsilon}}{\al}\Big)^{n-1}\Big(2\sfrac{\la_0^{1-\upsilon}}{\al^{6.5}}\sfrac{\fl_{j}}{M^{j}}\Big)
\sfrac{1}{1-M^j\|K'\|_{1,\Si_{j+1}}}$$
we have
$$\eqalign{
\V q^{(n+1)}-q^{(n)}\V_{1,\Si_j}
&=2\,\V\, \tilde w_{0,2}(q^{(n)};K')-\tilde w_{0,2}(q^{(n-1)};K')\,\V_{1,\Si_j}\cr
&\le 2\sfrac{\la_0^{1-\upsilon}}{\al}\fd\fe_j\cr
&\le 2\sfrac{\la_0^{1-\upsilon}}{\al}
\Big(\ka\sfrac{\la_0^{1-\upsilon}}{\al}\Big)^{n-1}\Big(2\sfrac{\la_0^{1-\upsilon}}{\al^{6.5}}\sfrac{\fl_{j}}{M^{j}}\Big)\fe_j^2\cr
&\le 
\Big(\ka\sfrac{\la_0^{1-\upsilon}}{\al}\Big)^{n}
\Big(2\sfrac{\la_0^{1-\upsilon}}{\al^{6.5}}\sfrac{\fl_{j}}{M^{j}}\Big)\fe_j
}$$

\endproof

\proof{ of Lemma \lemreWick} 
Fix any $K'\in\fK_{j+1}$.
By Corollary \corOSappMonoidIV.ii\ there is a constant $\ka$ such that
$\fe_j^2\le\sfrac{\ka}{2}\fe_j$.
If $\al$ is small enough, Lemma \lemNPcontrmap\ implies that every 
$$
\V q^{(n)}\V_{1,\Si_j}\le
\frac
{2\sfrac{\la_0^{1-\upsilon}}{\al^{6.5}}\sfrac{\fl_{j}}{M^{j}}}
{1-\ka\sfrac{\la_0^{1-\upsilon}}{\al}}
\fe_j
\le 4\sfrac{\la_0^{1-\upsilon}}{\al^{6.5}}\sfrac{\fl_{j}}{M^{j}}
\fe_j
$$
and that the sequence $\{q^{(n)}\}_{n\ge 1}$ converges to a $q_0(K')$ also obeying
$$
\V q_0(K')\V_{1,\Si_j}
\le 4\sfrac{\la_0^{1-\upsilon}}{\al^{6.5}}\sfrac{\fl_{j}}{M^{j}}\fe_j
\le \sfrac{\la_0^{1-\upsilon}}{\al^6}\sfrac{\fl_{j}}{M^{j}}\fe_j
\EQN\eqnNPQ$$

Fix any $K''$ and denote $\sQ_0=q_0(K')$ and 
$\sQ'=\sfrac{d\hfill}{ds}q_0(K'+sK'') \big|_{s=0}$.
  Applying $\sfrac{d\hfill}{ds}\hskip.1in\big|_{s=0}$ to 
$ q_0(K'+sK'') = 2\tilde w_{0,2}(q_0(K'+sK'');K'+sK'')$ yields
$$\eqalign{
\sQ'&=\sfrac{d\hfill}{ds}q_0(K'+sK'') \big|_{s=0}\cr
&= 2\sfrac{d\hfill}{ds}\tilde w_{0,2}(q_0(K'+sK'');K')\big|_{s=0}+
2\sfrac{d\hfill}{ds}\tilde w_{0,2}(q_0(K');K'+sK'')\big|_{s=0}\cr
&= 2\sfrac{d\hfill}{ds}\tilde w_{0,2}(\sQ_0+s\sQ';K')\big|_{s=0}+
2\sfrac{d\hfill}{ds}\tilde w_{0,2}(\sQ_0;K'+sK'')\big|_{s=0}\cr
}$$
As, for fixed $j$, $\tilde w_{0,2}(K';\sQ)$ is analytic in $\sQ$ and $K'$ and as $\fe_j\|K''\|_{1,\Si_{j+1}}$ has only finitely many
finite coefficients, there is some finite $\be$ such that
$$
\v \sQ'\v_{1,\Si_j}\le \be\fe_j\|K''\|_{1,\Si_{j+1}}
$$
Choose a $\be$ that is
within $\sfrac{\la_0^{1-\upsilon}}{2\al^{1.5}}$ of the infimum of all $\be$'s
that work.
By Lemma \lemRWrhsbounds, with 
$\fd=\be\sfrac{\|K''\|_{1,\Si_{j+1}}}{1-M^j\|K'\|_{1,\Si_{j+1}}}$,
$$\eqalign{
\v \sQ'\v_{1,\Si_j}
&\le  2\sfrac{\la_0^{1-\upsilon}}{\al}\ 
\be\sfrac{\|K''\|_{1,\Si_{j+1}}}{1-M^j\|K'\|_{1,\Si_{j+1}}}\fe_j 
+2\sfrac{\la_0^{1-\upsilon}}{\al^{1.5}}
\fe_j({\sst \|K'\|_{1,\Si_{j+1}}})\|K''\|_{1,\Si_{j+1}} \cr
&\le  \Big[\ka\sfrac{\la_0^{1-\upsilon}}{\al}\be
+2\sfrac{\la_0^{1-\upsilon}}{\al^{1.5}}\Big]
\fe_j\|K''\|_{1,\Si_{j+1}} \cr
&\le  \Big[\sfrac{1}{4}\be
+2\sfrac{\la_0^{1-\upsilon}}{\al^{1.5}}\Big]
\fe_j\|K''\|_{1,\Si_{j+1}} \cr
}$$
if $\al$ is large enough. Thus 
$\v \sQ'\v_{1,\Si_j}\le \be'\fe_j\|K''\|_{1,\Si_{j+1}}$
with $\be'=\sfrac{1}{4}\be+2\sfrac{\la_0^{1-\upsilon}}{\al^{1.5}}$.
If $\be\ge 4\sfrac{\la_0^{1-\upsilon}}{\al^{1.5}}$, then
$$
\be'-\be=
2\sfrac{\la_0^{1-\upsilon}}{\al^{1.5}}-\sfrac{3}{4}\be
\le -\sfrac{\la_0^{1-\upsilon}}{\al^{1.5}}
$$
which violates the requirement that $\be$ that is
within $\sfrac{\la_0^{1-\upsilon}}{2\al^{1.5}}$ of the infimum of all $\be$'s
that work. Hence
$$
\v \sQ'\v_{1,\Si_j}
\le  4\sfrac{\la_0^{1-\upsilon}}{\al^{1.5}}\fe_j\|K''\|_{1,\Si_{j+1}}
\le  \sfrac{\la_0^{1-\upsilon}}{\al}\fe_j\|K''\|_{1,\Si_{j+1}}
\EQN\eqnNPQprime$$
\endproof

\proof{ of Lemma \lemNPdeKbnd} (i)
By (\eqnNPQ), (\eqnNPQprime) and Lemma \lemOSumu\ of [FKTo3],
$$
\big\|\de K(K')\big\|_{1,\Si_j}\le\abcst\,\cb_{j+1}\v q_0(K')\v_{1,\Si_j}
\le\const \sfrac{\la_0^{1-\upsilon}}{\al^{6.5}}\sfrac{\fl_{j}}{M^{j}}\fe_j
\le \sfrac{\la_0^{1-\upsilon}}{\al^6}\sfrac{\fl_{j}}{M^{j}}\fe_j
$$
and
$$\eqalign{
\big\| \sfrac{d\hfill}{ds} \de K(K'+sK'')\big|_{s=0}\big\|_{1,\Si_j}
&\le\abcst\,\cb_{j+1}\V \sfrac{d\hfill}{ds}q_0(K'+sK'') \big|_{s=0}\V_{1,\Si_j}
\cr
&\le\const \sfrac{\la_0^{1-\upsilon}}{\al^{1.5}}\sfrac{\fl_{j}}{M^{j}}
\fe_j\|K''\|_{1,\Si_{j+1}}\cr
&\le \sfrac{\la_0^{1-\upsilon}}{\al}\sfrac{\fl_{j}}{M^{j}}
\fe_j\|K''\|_{1,\Si_{j+1}}
}$$
\Item (ii) By Proposition \propOSindresectorI.ii of [FKTo4] and part (i),
$$\eqalign{
\| K(K')\|_{1,\Si_j} 
&\le \abcst\sfrac{\fl_j}{\fl_{j+1}}\cb_{j-1}\|K'\|_{1,\Si_{j+1}}
        +\| \de K(K')\|_{1,\Si_j} \cr
&\le \abcst\sfrac{\fl_j}{\fl_{j+1}}\cb_{j-1}\|K'\|_{1,\Si_{j+1}}
        +\sfrac{\la_0^{1-\upsilon}}{\al^6}\sfrac{\fl_{j}}{M^{j}}\fe_j \cr
}$$
and
$$\eqalign{
\big\|\sfrac{d\hfill}{ds}K(K'+sK'')\big|_{s=0}\big\|_{1,\Si_j} 
&=\big\|\sfrac{d\hfill}{ds}
\big[K'_{\Si_j}+sK''_{\Si_j}+\de K(K'+sK'')\big]_{s=0}\big\|_{1,\Si_j} \cr
&\le \|K''_{\Si_j}\|_{1,\Si_j}
+\big\|\sfrac{d\hfill}{ds}\de K(K'+sK'')\big|_{s=0}\big\|_{1,\Si_j} \cr
&\le \abcst\,M^\aleph\cb_{j-1}\|K''\|_{1,\Si_{j+1}}
+\sfrac{\la_0^{1-\upsilon}}{\al}\,\fe_j({\sst \|K'\|_{1,\Si_{j+1}}}) 
\|K''\|_{1,\Si_{j+1}}\cr
&\le \abcst\,M^\aleph\fe_j({\sst \|K'\|_{1,\Si_{j+1}}})\|K''\|_{1,\Si_{j+1}}\cr
}$$

\Item (iii) is contained in Lemma \lemRWintbnd.ii.

\endproof

\vfill\eject

\titlea{References}\PG\pgNPIIref

\item{[FKTf1]} J. Feldman, H. Kn\"orrer, E. Trubowitz, 
{\bf A Two Dimensional Fermi Liquid, Part 1: Overview}, preprint.
\smallskip%
\item{[FKTf3]} J. Feldman, H. Kn\"orrer, E. Trubowitz, 
{\bf A Two Dimensional Fermi Liquid, Part 3: The Fermi Surface}, preprint.
\smallskip%
\item{[FKTl]} J. Feldman, H. Kn\"orrer, E. Trubowitz, 
 {\bf Particle--Hole Ladders}, preprint.
\smallskip%
\item{[FKTo1]} J. Feldman, H. Kn\"orrer, E. Trubowitz, 
{\bf Single Scale Analysis of Many Fermion Systems, Part 1: Insulators}, preprint.
\smallskip%
\item{[FKTo2]} J. Feldman, H. Kn\"orrer, E. Trubowitz, 
{\bf Single Scale Analysis of Many Fermion Systems, Part 2: The First Scale}, preprint.
\smallskip%
\item{[FKTo3]} J. Feldman, H. Kn\"orrer, E. Trubowitz, 
{\bf Single Scale Analysis of Many Fermion Systems, Part 3: Sectorized Norms}, preprint.
\smallskip%
\item{[FKTo4]} J. Feldman, H. Kn\"orrer, E. Trubowitz, 
{\bf Single Scale Analysis of Many Fermion Systems, Part 4: Sector Counting}, preprint.
\smallskip%
\item{[FKTr1]} J. Feldman, H. Kn\"orrer, E. Trubowitz, 
{\bf Convergence of Perturbation Expansions in Fermionic Models, Part 1: Nonperturbative Bounds}, preprint.
\smallskip%
\item{[FMRT]} J. Feldman, J. Magnen, V. Rivasseau and E. Trubowitz,
{\bf Fermionic Many-Body Models}, in {\it Mathematical Quantum Theory I: Field Theory and Many-Body Theory}, J. Feldman, R. Froese and L. Rosen eds, 
CRM Proceedings \& Lecture Notes {\bf 7}, 29-56 (1994).
\smallskip%
\item{[FW]} A.L. Fetter and J.D. Walecka, {\it Quantum Theory of Many-Particle
Systems}, McGraw-Hill, 1971.

\vfill\eject

\hoffset=-0.2in
\titlea{Notation}\PG\pgNPIInot
\vfil
\titleb{Norms}
\centerline{
\vbox{\offinterlineskip
\hrule
\halign{\vrule#&
         \strut\hskip0.05in\hfil#\hfil&
         \hskip0.05in\vrule#\hskip0.05in&
          #\hfil\hfil&
         \hskip0.05in\vrule#\hskip0.05in&
          #\hfil\hfil&
           \hskip0.05in\vrule#\cr
height2pt&\omit&&\omit&&\omit&\cr
&Norm&&Characteristics&&Reference&\cr
height2pt&\omit&&\omit&&\omit&\cr
\noalign{\hrule}
height2pt&\omit&&\omit&&\omit&\cr
&$\tn\ \cdot\ \tn_{1,\infty}$&&no derivatives, external positions, acts on 
functions&&Definition \defNPSymmNorm&\cr
height4pt&\omit&&\omit&&\omit&\cr
&$\|\ \cdot\ \|_{1,\infty}$&&derivatives, external positions, acts on functions&&Definition \defNPSymmNorm&\cr
height4pt&\omit&&\omit&&\omit&\cr
&$\tn\ \cdot\ \tn_{\infty}$&&no derivatives, external positions, acts on 
functions&&Definition \defNPsectGrnorm&\cr
height4pt&\omit&&\omit&&\omit&\cr
&$\v \ \cdot\ \v_{p,\Si}$&&derivatives, external positions, 
all but $p$ sectors summed
&&Definition \defNPsectnorm&\cr
height4pt&\omit&&\omit&&\omit&\cr
&$\tn \ \cdot\ \tn_{1,\Si}$&&no derivatives, all but $1$ sector summed
&&(\eqnOVtriponesi)&\cr
height4pt&\omit&&\omit&&\omit&\cr
&$\tn \ \cdot\ \tn_{3,\Si}$&&no derivatives, all but $3$ sectors summed
&& (\eqnOVtripthreesi) &\cr
height4pt&\omit&&\omit&&\omit&\cr
&$\|\ \cdot\ \|_{1,\Si}$&& like $\v \ \cdot\ \v_{1,\Si}$, but for functions
on $\big(\bbbr^2\times\Si\big)^2$
&&[Def'n \defOSzerosectorext, FKTo4]&\cr
height4pt&\omit&&\omit&&\omit&\cr
&$\v \varphi \v_j$&&$\rho_{m;n}\cases{
\v \varphi \v_{1,\Si_j} + \sfrac{1}{\fl_j}\,\v \varphi \v_{3,\Si_j}
+ \sfrac{1}{\fl_j^2}\,\v \varphi \v_{5,\Si_j} 
    & if $m=0$ \cr
\sfrac{\fl_j}{M^{2j}}\,\v \varphi \v_{1,\Si_j} & if $m\ne0$} $
&&Definition \defNPsectnorm&\cr
height4pt&\omit&&\omit&&\omit&\cr
&$N_j(w,\al,X)$&&$\sfrac{M^{2j}}{\fl_j}\,\fe_j(X) 
\smsum_{m,n\ge 0}\,
\al^{n}\,\big(\sfrac{\fl_j\IB}{M^j}\big)^{n/2} \,\v w_{m,n}\v_j$
&&Definition \defNPsectGrnorm&\cr
height4pt&\omit&&\omit&&\omit&\cr
&$N(\cG)$&&$
\smsum_{m> 0}\,\sfrac{1}{\la_0^{(1-\upsilon)\max\{m-2,2\}/2}}
\,\tn G_m\tn_\infty$
&&Definition \defNPsectGrnorm&\cr
height4pt&\omit&&\omit&&\omit&\cr
}\hrule}}
\vfill\goodbreak
\titleb{Spaces}
\centerline{
\vbox{\offinterlineskip
\hrule
\halign{\vrule#&
         \strut\hskip0.05in\hfil#\hfil&
         \hskip0.05in\vrule#\hskip0.05in&
          #\hfil\hfil&
         \hskip0.05in\vrule#\hskip0.05in&
          #\hfil\hfil&
           \hskip0.05in\vrule#\cr
height2pt&\omit&&\omit&&\omit&\cr
&Not'n&&Description&&Reference&\cr
height2pt&\omit&&\omit&&\omit&\cr
\noalign{\hrule}
height2pt&\omit&&\omit&&\omit&\cr
&$\cE$&&counterterm space&&Definition \defNPCTMSpace&\cr
height2pt&\omit&&\omit&&\omit&\cr
&$\fK_j$&&space of future counterterms for scale $j$
&&Definition \defNPCTmSpace&\cr
height2pt&\omit&&\omit&&\omit&\cr
&$\cB$&&$\bbbr \times \bbbr^d \times \{\uparrow, \downarrow\}\times\{0,1\}$
 viewed as position space&&before Def \defbubbleprop&\cr
height2pt&\omit&&\omit&&\omit&\cr
&$\check \cB$&&$\bbbr\times\bbbr^d\times\{\uparrow,\downarrow\}\times\{0,1\}$ 
viewed as momentum space&&beginning of \S\CHnewsectors&\cr
height2pt&\omit&&\omit&&\omit&\cr
&$\cB^\updownarrow$&&$\bbbr \times \bbbr^d \times \{\uparrow, \downarrow\}$
 viewed as position space&&Definition \NPsomespaces&\cr
height2pt&\omit&&\omit&&\omit&\cr
&$\cF_m(n;\Si)$&&functions on $\cB^m \times  \big( \cB \times\Si \big)^n$,
internal momenta in sectors&&Definition \defNPsectrepr.ii&\cr
height2pt&\omit&&\omit&&\omit&\cr
&$\cD_{\rm in}^{(j,\form)}$&&formal input data for scale $j$
&&Definition \defNPinputData&\cr
height2pt&\omit&&\omit&&\omit&\cr
&$\cD_{\rm out}^{(j,\form)}$&&formal output data for scale $j$
&&Definition \defNPoutputData&\cr
height2pt&\omit&&\omit&&\omit&\cr
&$\cD^{(j)}_{\rm in}$&&input data for scale $j$&&Definition \stepInputData&\cr
height2pt&\omit&&\omit&&\omit&\cr
&$\cD^{(j)}_{\rm out}$&&output data for scale $j$
&&Definition \stepOutputData&\cr
height4pt&\omit&&\omit&&\omit&\cr
}\hrule}}

\vfil
\goodbreak
\titleb{Other Notation}
\centerline{
\vbox{\offinterlineskip
\hrule
\halign{\vrule#&
         \strut\hskip0.05in\hfil#\hfil&
         \hskip0.05in\vrule#\hskip0.05in&
          #\hfil\hfil&
         \hskip0.05in\vrule#\hskip0.05in&
          #\hfil\hfil&
           \hskip0.05in\vrule#\cr
height2pt&\omit&&\omit&&\omit&\cr
&Not'n&&Description&&Reference&\cr
height2pt&\omit&&\omit&&\omit&\cr
\noalign{\hrule}
height2pt&\omit&&\omit&&\omit&\cr
&$r_0$&&number of $k_0$ derivatives tracked&&following (\eqnNPinteraction)&\cr
height2pt&\omit&&\omit&&\omit&\cr
&$r$&&number of $\k$ derivatives tracked&&following (\eqnNPinteraction)&\cr
height2pt&\omit&&\omit&&\omit&\cr
&$M$&&scale parameter, $M>1$&&before Definition \defNPscales&\cr
height2pt&\omit&&\omit&&\omit&\cr
&$\const$&&generic constant, independent of scale&& &\cr
height2pt&\omit&&\omit&&\omit&\cr
&$\abcst$&&generic constant, independent of scale and $M$&& &\cr
height2pt&\omit&&\omit&&\omit&\cr
&$\nu^{(j)}(k)$&&$j^{\rm th}$ scale function&&Definition \defNPscales&\cr
height2pt&\omit&&\omit&&\omit&\cr
&$\nu^{(\ge j)}(k)$&&$\smsum_{i\ge j}\nu^{(j)}(k)$&&Definition \defNPscales&\cr
height2pt&\omit&&\omit&&\omit&\cr
&$n_0$&&degree of asymmetry&&Definition \defNPstrongasymm&\cr
height2pt&\omit&&\omit&&\omit&\cr
&$J$&&particle/hole swap operator&&(\eqnNPjdef)&\cr
height2pt&\omit&&\omit&&\omit&\cr
&$\Om_S(\cW)(\phi,\psi)$
&&$\log\sfrac{1}{Z} \int  e^{\cW(\phi,\psi+\ze)}\,d\mu_{S}(\ze)$
&&Definition \defNPrengroupmap&\cr
height2pt&\omit&&\omit&&\omit&\cr
&$\tilde \Om_C(\cW)(\phi,\psi)$
&&$\log \sfrac{1}{Z}\int e^{\phi J\ze}\,e^{\cW(\phi,\psi +\ze)} d\mu_C(\ze)$
&&Definition \defNPrengroupmap&\cr
height2pt&\omit&&\omit&&\omit&\cr
&$\aleph$&&$\half<\aleph< \sfrac{2}{3}$&&following Definition \defNPsectrepr&\cr
height2pt&\omit&&\omit&&\omit&\cr
&$\la_0$&&maximum allowed ``coupling constant''&&Theorem \eqnNPnbndA&\cr
height2pt&\omit&&\omit&&\omit&\cr
&$\upsilon$&&$0<\upsilon< \sfrac{1}{4}$, power of $\la_0$ eaten by bounds&&Definition \defNPrhomn&\cr
height2pt&\omit&&\omit&&\omit&\cr
&$\rho_{m;n}(\la)$&&$\la^{-(1-\upsilon)\max\{m+n-2,2\}/2}$&&Definition \defNPrhomn&\cr
height2pt&\omit&&\omit&&\omit&\cr
&$\rho_{m;n}$&&$\rho_{m;n}(\la_0)\big\{\hbox{1 if $m=0$\ ;\ $\root{4}\of{\fl_j
M^j}$ if $m>0$}$&&Definition \defNPsectnorm.ii&\cr
height2pt&\omit&&\omit&&\omit&\cr
&$\fl_j$&&$=\sfrac{1}{M^{\aleph j}}$ = 
  length of sectors of scale $j$&&following Definition \defNPsectrepr&\cr
height2pt&\omit&&\omit&&\omit&\cr
&$\Si_j$&&the sectorization at scale $j$ of length $\fl_j$&&following Definition \defNPsectrepr&\cr
height2pt&\omit&&\omit&&\omit&\cr
&$\IB$&&$j$--independent constant&&Definition \defNPsectGrnorm&\cr
height2pt&\omit&&\omit&&\omit&\cr
&$\cb_j$&& $
=\sum_{|\bde|\le r\atop |\de_0|\le r_0}  M^{j|\de|}\,t^\de
+\sum_{|\bde|> r\atop {\rm or\ }|\de_0|> r_0}\infty\, t^\de
\in\fN_{d+1}
$&&Definition \defNPFancynormdomain&\cr
height2pt&\omit&&\omit&&\omit&\cr
&$\fe_j(X)$&& $= \sfrac{\cb_j}{1-M^j X}$&&Definition \defNPFancynormdomain&\cr
height2pt&\omit&&\omit&&\omit&\cr
&$f_{\rm ext}$&& extends $f(\x,\x')$ to 
$f_{\rm ext}\big((x_0,\x,\si,a),(x_0',\x',\si',a')\big)$&&
[Definition \defOSzeroext, FKTo4]&\cr
height2pt&\omit&&\omit&&\omit&\cr
&$*$&& convolution&&Definition \defNPconvol&\cr
height2pt&\omit&&\omit&&\omit&\cr
&$\bullet$&& ladder convolution&&Definition \deNPdefsectbubbleprop,&\cr
height2pt&\omit&&\omit&&\omit&\cr
&$\hat\mu$&&Fourier transform&&Notation \notNPfourierTI&\cr
height4pt&\omit&&\omit&&\omit&\cr
}\hrule}}

\end